# Magnetic phenomena in equiatomic ternary rare earth compounds

Sachin Gupta

Department of Physics, Bennett University, Greater Noida 201310, India

Email: sachin.gupta@bennett.edu.in; gsachin55@gmail.com

## Abstract:

This chapter discusses the structural and magnetic properties of equiatomic ternary *RTX* compounds, where *R* represents rare earth, *T* is $3d/4d$ transition metal, and *X* belongs to the *p*-block elements. *RTX* compounds exhibit a variety of crystal structures, which leads to a range of magnetic phenomenon ranging from long range antiferromagnetic/ferromagnetic ordering, unconventional superconductivity, magnetic frustration to spin ices. Coexistence of various magnetic phenomenon result in many exotic properties, which make these materials promising for next generation technological applications. Theoretical predictions suggested that certain compounds in this group possess topological properties as a result of bulk band inversion. Subsequently, experimental confirmation of these topological properties has been reported in some of these materials. The crystal structure of some of these compounds show structural modification from one crystal structure to another whereas few show isostructural transition as a function of temperatures. In *RTX* family of compounds, most transition metals, except manganese (Mn) do not contribute significantly to the magnetic moment and behave as non-magnetic. The lack of magnetic behaviour in the $3d/4d$ sublattices of these ternary compounds may be attributed to hybridization between *p* electron states of *X* atom and *d* electron states of the transition metal, resulting in the filling of the *d* band. The dominant interaction in these compounds is of Ruderman–Kittel–Kasuya–Yosida (RKKY) type due to the localized nature of the $4f$ electrons in the rare earths. The magnetic transition temperature in these materials varies from ultra-low to high temperatures, making them suitable for integration into devices operating at room temperature. The *RTX* series discussed here encompasses all rare earth elements, a range of transition metals ($3d/4d$), and various *p*-block elements such as Al, Ga, In, Si, Ge, Sn, Sb, and Bi. Most of the materials studied are arc-melted polycrystalline materials with some in the form of single crystal or thin films.

## Keywords:

# **<u>Abbreviations</u>**

ACS AC susceptibility

AFM antiferromagnetic

CEF crystalline electric field

FCC field cooled cooling

FCW field cooled warming

FM ferromagnetic

FP-LAPW full-potential linearized augmented plane wave

LTF low temperature form

HTF high temperature form

PPM Pauli paramagnetic

PM paramagnetic

RE rare earth

RKKY Ruderman–Kittel–Kasuya–Yosida

RSG re-entrant spin-glass

SC superconducting

SG spin glass

SOC spin orbit coupling

TMI thermomagnetic irreversibility

ZFCW zero field cooled warming

# 1. Introduction

Magnetic materials play a vital role in numerous daily life applications ranging from transportation to medicine and electronics, to energy harvesting and household appliances [1–5]. Various magnetic materials have been investigated to explore their magnetic properties and application potentials. Among them, rare earth (RE) based materials are extensively studied and still attract great interest of researchers due to their interesting properties [6–8]. Rare earth materials are used to produce some of the strongest permanent magnets available, called rare earth magnets [7]. The strong magnetic moment in rare earths is originated from localized unpaired $4f$ shell electrons [8]. The magnetic moment contribution from the delocalized $5d$ and $6s$ electrons is very small, however they play a crucial role in determining the magnetic properties by mediating the exchange interactions. On the other hand, the magnetism in transition metals is governed by their $3d$/$4d$ electrons. Compounds synthesized with combination of rare earths and transition metal were found to show exciting magnetic and related properties in terms of fundamental studies and technological applications. It has been observed that the magnetism in rare earths ($R$) - transition metal ($T$) intermetallics is generally governed by three types of interactions, viz. *R-R* (interaction between the magnetic moments within the rare earth sublattice), *T-T* (interaction between the magnetic moments within the transition metal sublattice) and *R-T* (interaction between rare earth and transition metal moments *i.e.* inter-sublattice interactions). While the direct exchange interaction is the predominant one in the case of *T-T*, it is the indirect Ruderman–Kittel–Kasuya–Yosida (RKKY) which dominates between *R-R*. The *R-R* interaction is weaker in these compounds while *T-T* interaction is the strongest. However, if the *T* element in these compounds is non-magnetic, the exchange interaction is of RKKY type.

There are many materials families comprising of the rare earth - transition metal intermetallic compounds. Among them equiatomic *RTX* family (*R*= rare earth, *T*= transition metal and *X*= *p*-block element of compounds has received huge attention due to their intriguing properties and great application potential. Many compounds of this family got renewed interest due to discovery of topological phases and their promising application in spintronics, magnetic refrigeration and quantum computation [6,9]. In 2014, Gupta and Suresh [9] extensively reviewed *RTX* compounds with an overview of crystal structure, magnetic, magnetocaloric, magneto-transport and hydrogenation. The present chapter discusses structural and magnetic properties of *RTX* (*T*= 3*d*/4*d* transition metal) family with updated results. Most of these materials are synthesized using induction melting technique and are in polycrystalline form. Very few are grown in the single crystal or thin films form. Depending on their composition these materials crystallize in different crystal structures. Some materials show structural transformation or isostructural transition with temperature. The magnetic properties of these materials vary depending on their crystal structure, encompassing a wide range of behaviours. These properties include long-range order with either antiferromagnetic or ferromagnetic ordering, as well as short-range order such as spin glass or spin ices. Additionally, some of these materials may even exhibit unconventional superconductivity. It has been observed that in these ternary compounds the 3*d*/4*d* sublattices do not carry any moment in most of the compounds except for Mn ones. The possible reason behind the non-magnetic behaviour of 3*d*/4*d* sublattice may be the hybridization of *p* electrons with *d* electron states of transition metal resulting in the filling of *d* band [10]. Therefore, due to the localized character of 4*f* electrons in rare earth, RKKY is the dominant interaction in these compounds. The magnetic transition temperature in these materials ranges from ultra-low to high temperature (above room temperature), which enables these materials to be integrated in devices for room temperature operation. Many compounds of *RTX* family are described as Heusler alloys showing tuneable electronic and magnetic properties and are very promising for spintronics applications [6,9]. Due to second order magnetic transition some of these materials have great potential for magnetic refrigeration, which is an alternative to conventional gas compression/expansion technology [9].

## 2. Structural and magnetic properties

In this section, the structural and magnetic properties of *RTX* compounds are discussed. Observations have indicated that compounds containing different *R* elements, but the same *T* and *X* elements typically exhibit similar crystal structures. However, in certain cases, both the crystal structure and space group can differ among these compounds. The crystal structures along with their space groups for different *RTX* compounds are summarized in Table 1. Given that the magnetic properties of materials are reliant on their crystal structure, it is logical to discuss both aspects together in the same section in order to enhance our understanding of them. The magnetic parameters determined from the magnetic properties such as nature of magnetic ground state, magnetic ordering temperature ($T_N$ or $T_C$), effective magnetic moment ($\mu_{eff}$) and paramagnetic Curie temperatures ($\theta_p$) for these compounds are summarized in Table 2.

## 2.1 $R$Sc$X$ compounds

Depending upon their annealing temperature, $R$ScSi and $R$ScGe compounds crystallize in Ti$_5$Ga$_4$ -type hexagonal and CeScSi (ordered variant of La$_2$Sb) or CeFeSi type tetragonal crystal structures [11–15]. CeScSi and CeFeSi type tetragonal structures are layered structure with layering sequence [Sc-X-R-X]$_2$ and [Sc-X-R-X] respectively as shown in Figure 1 [16]. These layered structures can allow for two-dimensional (2D) properties in these materials, which are yet to be explored [17–19].

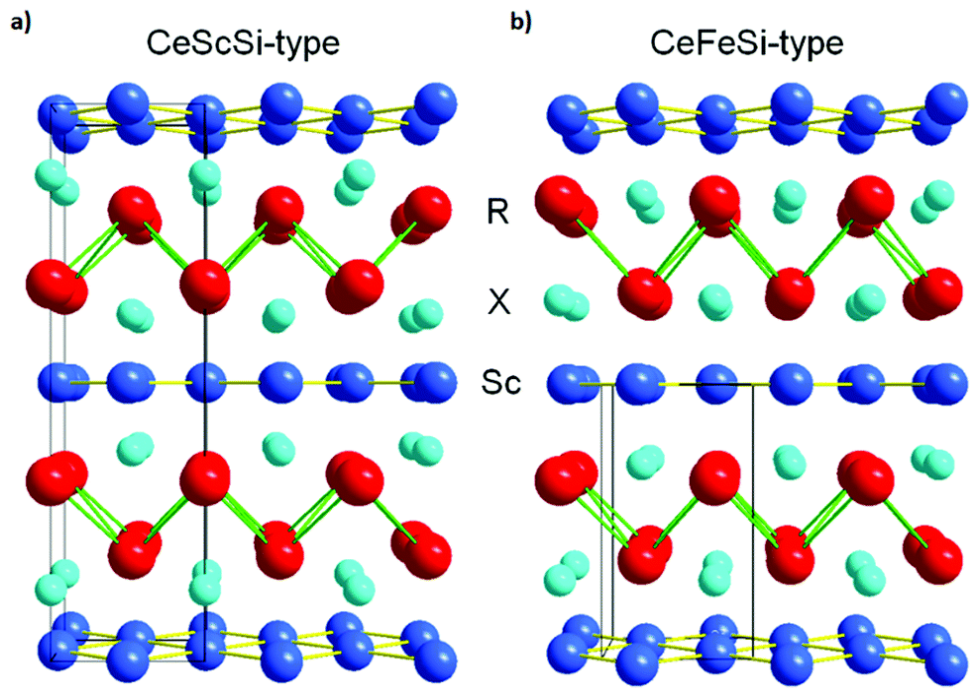

**FIG. 1** Crystal structure of layered (a) CeScSi and (b) CeFeSi – type tetragonal structure. Reproduced with permission from [16] © 2021, Royal Society of Chemistry.

Physical properties of $R$Sc$X$ are very sensitive to chemical variation. Magnetometry studies of these materials show antiferromagnetic (AFM) and ferromagnetic (FM) ordering with some of them showing relatively high magnetic transition temperatures. Uwatoko *et al.* [20] studied single crystalline samples of CeScSi and CeScGe prepared by Czochralski pulling method. Both the compounds show FM ordering. The magnetic properties of alloys with tetragonal structure of La$_2$Sb type are sensitive to lattice strain and the occupancy of lattice site.

It has been found that polycrystalline samples show changes in magnetic properties from AFM to field induced FM on annealing at different temperatures [20]. Later Singh *et al.* [12] reported that both the compounds order antiferromagnetically. It has been anticipated that different type magnetic orders in these samples may be originated due to parasitic phases present in the samples [12]. Nikitin *et al.* [10] reported magnetic properties of $R$ScSi and $R$ScGe ($R$= Gd -Er and Y) compounds. The authors reported FM behaviour in GdScSi, GdScGe and TbScGe with Gd compounds showing Curie temperature ($T_C$) above room temperature. The high Curie temperature in the materials is likely due to strong indirect interaction between $4f$ orbital of rare earth ion and hybridization of $3d$ orbital of Sc and $3p$ orbital of Si/Ge atoms. The effective magnetic moment ($\mu_{eff}$) for these compounds were found to be close to free rare earth ion ($R^{3+}$). Couillaud *et al* [11] studied Gd compounds i.e. GdScSi and GdScGe of this series for the application in magnetic refrigeration and determined a Curie temperature higher than reported in Ref. [10]. Singh *et al.* [13] reported magnetic properties of $R$ScX ($R$=Pr, Nd and Sm; $X$=Si and Ge) compounds. Magnetic measurements indicate FM ground states in all the compounds except PrScSi, which shows a spin-flop magnetic transition at the field of nearly 3T. The field dependence of magnetization for these compounds as a function of temperature is shown in Fig. 2. All the compounds show hysteresis in magnetization isotherms with Sm compounds exhibiting larger coercive fields.

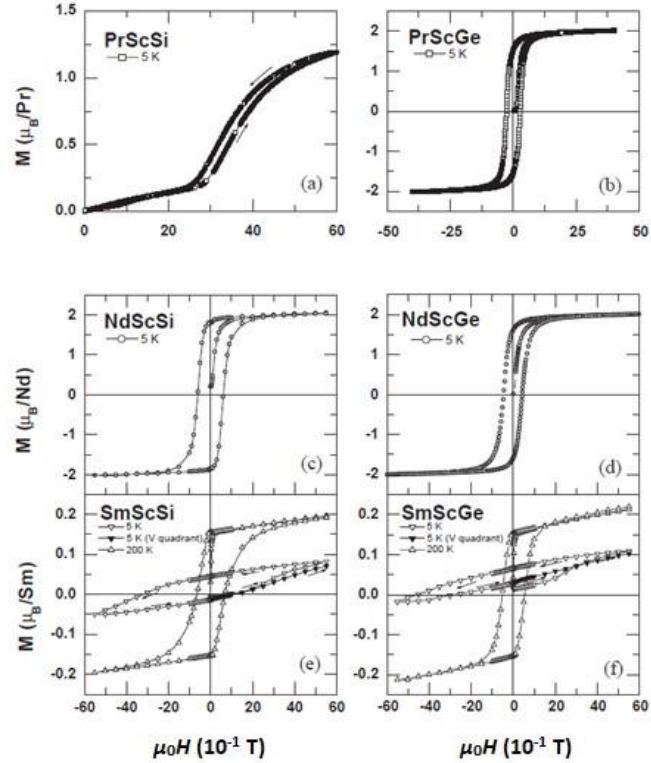

**FIG. 2** The magnetic field dependence of magnetization for $R$ScX ($R$=Pr, Nd and Sm; $X$=Si and Ge). Reproduced with permission from [13] © 2004, Elsevier B. V.

A thorough examination of the magnetic properties of $R$Sc$X$ compounds reveals a significant correlation between the magnetic nature of these materials and their crystal structure. Specifically, compounds such as $R$ScSi ($R$ = Ce, Sm, Gd) and $R$ScGe ($R$ = Ce, Pr, Sm, Gd, Tb), which possess a tetragonal crystal structure, exhibit ferromagnetic (FM) ordering. On the other hand, compounds like $R$ScSi ($R$ = Tb-Tm) and $R$ScGe ($R$ = Dy-Tm), which crystallize in the $Ti_5Ga_4$-type hexagonal structure, display paramagnetic (PM) or antiferromagnetic (AFM) behaviour [10]. It has been observed that Gd compounds in this series have their ordering temperature above room temperature. Also, it is worth to note that in contrast to the tetragonal polymorphs, $Ti_5Ga_4$ – type does not have layers of same atoms. This could result in smaller values of paramagnetic Curie temperature ($\theta_p$) and hence weaker exchange interactions in these compounds [10].

## 2.2 *R*TiX compounds

Bulk *R*TiSi (*R* = Y, Gd–Tm, Lu) [21,22], *R*TiGe (*R* = Y, La–Nd, Sm, Gd–Tm, Lu) [23,24] and GdTiSb [25] compounds in *R*Ti*X* series crystallize in CeFeSi- type tetragonal structure with space group *P*4/*nmm*. It has been observed that there is no significant change in lattice parameter *a* for Si and Ge compounds for a given *R* [22]. However, *c* parameter shows a little decrease on substituting Ge by Si. The high temperature form (HTF) of CeTiGe, GdTiGe and TbTiGe is found to form the CeScSi type tetragonal structure with space group *I*4/*mmm* [14,26].

Various crystal structure studies on *R*TiSi series reveal that only heavy rare earth compounds are stable. The temperature dependence of the magnetic susceptibility ($\chi$) for some of these compounds is shown in Fig. 3(a). The cusp like behaviour of susceptibility shows the antiferromagnetic nature of these compounds. The magnetic measurements show that all the existing compounds in *R*TiSi show AFM behaviour except YTiSi and LuTiSi, which are Pauli paramagnets (PPM) [22]. The magnetic transition temperatures of *R*TiSi (*R* =Gd–Tm) compounds are relatively high with GdTiSi ordering antiferromagnetically below 400 K. The experimental values of $\mu_{eff}$ for *R* = Gd-Tm is close to $R^{3+}$ ion, indicating no contribution of moment from Ti atom, which is further confirmed by neutron diffraction measurements. The neutron diffraction measurements carried out at low temperature (2 K) show additional peaks confirming antiferromagnetic ordering [22].

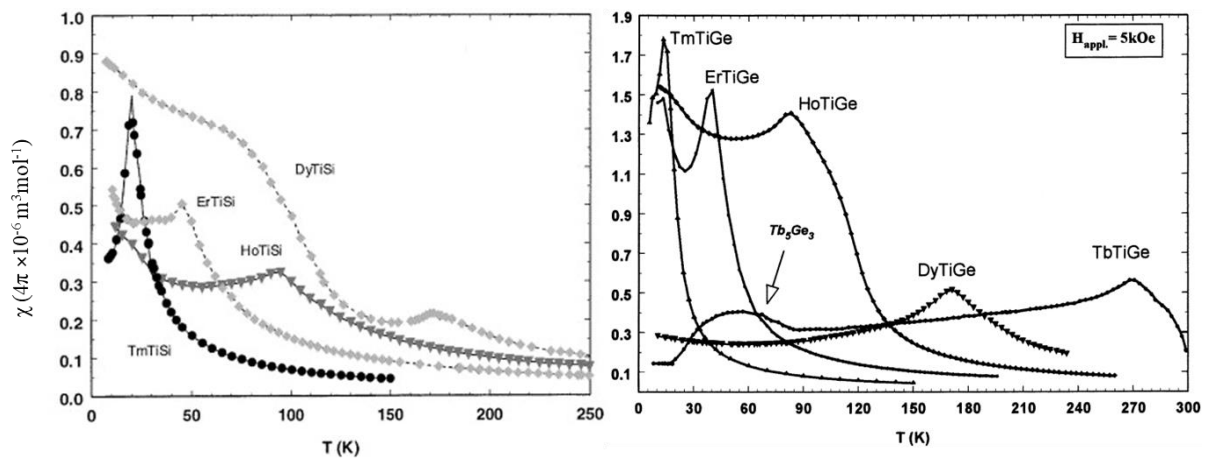

FIG. 3. (a) The temperature dependence of magnetic susceptibility for *R*TiSi (*R*= Dy-Tm) compounds at applied magnetic field of 0.2 T. (b) The temperature dependence of magnetic

susceptibility for $R$TiGe ($R$= Tb-Tm) compounds at applied magnetic field of 0.5 T. Reproduced with permission (a) from [22] © 2002, Elsevier B. V. (b) from [24] © 1999, Elsevier B. V.

Fig. 3(b) shows the magnetic susceptibility as a function of temperature for some of the $R$TiGe compounds. Similar to $R$TiSi compounds, CeFeSi-type $R$TiGe ($R$ = Nd, Sm, Gd–Tm) compounds show AFM ordering whereas compounds with $R$= Y, La and Lu are weak Pauli paramagnets [23,24,27]. It can be noted that HTF of GdTiGe and TbTiGe, which adopt CeScSi-type crystal structure order ferromagnetically [14,28]. Both low and high temperature forms of CeTiGe have been reported to be paramagnetic (PM) down to 0.4 K, despite of their different crystal structure types [24,26,29]. Deppe *et al.* [30] observed large steplike anomalies in the magnetization, magnetostriction and magnetoresistance results of polycrystalline CeTiGe compound at the critical field of 12 T. The result show hysteresis on sweeping magnetic field up and down, which suggests a first order phase transition. Analysis of these results suggest CeTiGe as a rare metamagnetic Kondo lattice system. Neutron diffraction studies confirm antiferromagnetic ordering in $R$TiGe ($R$= Pr, Nd, Tb-Er) compounds with PrTiGe and NdTiGe characterized by an easy plane sine modulated structure and $R$TiGe ($R$= Nd, Tb, Dy , Ho and Er) with commensurate structure [31]. It has been observed from the neutron diffraction studies that the values of the ordered magnetic moment at 2 K in all $R$TiGe ($R$= Pr, Nd, Tb-Er) compounds is close to the free-ion magnetic moment ($gJ$) values for the respective rare earth ions, indicating no contribution of moment from Ti.

A few interesting points can be noted from the structural and magnetic properties of $R$Ti$X$ compounds; (i) $R$TiSi and $R$TiGe crystallize in the same -type crystal structure and show very similar magnetic properties, (ii) for a fixed $R$, transition temperatures of these compounds are nearly the same (see Table 2), (iii) transition temperatures of these compounds are rather high despite of a non-magnetic behaviour of Ti and $X$ atoms. The observation of high transition temperatures in these compounds is correlated with its type of crystal structure. CeFeSi type crystal structure can be described as "BaAl$_4$ blocks" connected via $R$- $R$ contacts ("W blocks") as shown in Fig. 4. A possible reason for high transition temperatures is the close $R$-$R$ contact in the "W blocks" induced by segregation between $R$ and Ti atoms [22]. The reason for the almost same transition temperatures in both series despite of different radii of Si (1.319 Å) and Ge (1.369 Å) is that for heavy rare earths ($R$= Gd-Tb), the transition temperatures well obey the de Gennes scaling that results in same transition temperatures due to almost identical $R$-$R$ interatomic distances in $R$TiSi and $R$tiGe compounds [22].

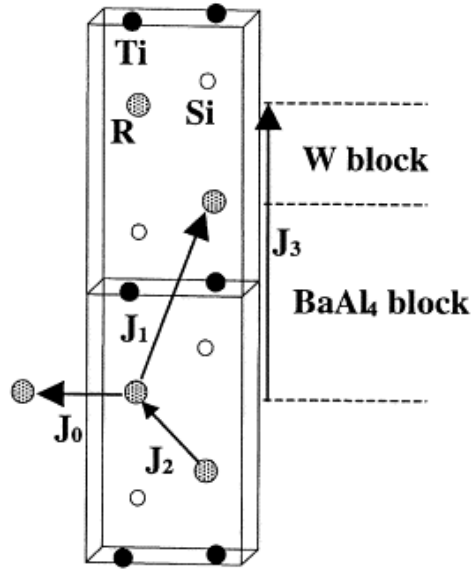

**FIG. 4** CeFeSi -type structure for $R$Ti$X$ compounds. Here $X$= Si is shown as a representative. Reproduced with permission from [22] © 2002, Elsevier B. V.

### 2.3 $R$Mn$X$ compounds

Compounds of type $R$MnAl ($R$= Ce, Nd, Gd, Tb), $R$MnGa ($R$= Ce, Nd, Pr, Gd-Dy) crystallize in the cubic MgCu$_2$- type C-15 Laves phase crystal structure, where Mn and $X$ (Al, Ga) atoms are randomly distributed [32–35]. $R$MnIn ($R$= Gd, Dy, Er, and Y) crystallize in closely related hexagonal C-14 type Laves phase crystal structure [33]. It has been observed that $R$MnSi compounds show two types of crystal structures; light rare earths ($R$= La-Sm, Gd) form in the CeFeSi type tetragonal structure while heavy rare earths ($R$= Tb-Er) compounds adopt the TiNiSi type orthorhombic crystal structure [36,37]. However, YMnSi, TbMnSi and DyMnSi were found to crystallize both in CeFeSi - tetragonal and TiNiSi – orthorhombic crystal structures depending on their heat treatment [38]. Usually, CeFeSi type structure is formed at low temperature while TiNiSi at high temperatures [37].

$R$MnGe compounds with light rare earths (($R$ = La –Nd) are reported to crystallize in CeFeSi type tetragonal crystal structure while heavy ones with ($R$ = Gd–Tm) crystallize in TiNiSi type orthorhombic crystal structure [39–41]. Since the size of Tm atomic radius lies on the boundary line for the TiNiSi $\rightarrow$ ZrNiAl structural transition, the crystal structure of TmMnGe changes with heat treatment and shows both TiNiSi (LT) and ZrNiAl (HT) type structures [40].

Until now we have seen that there was no magnetic moment contribution from the transition metal element. However, in *R*MnX compounds Mn contributes significantly to the total magnetic moment of the compounds. In these compounds, the Mn sublattice may order at relatively high temperatures.

The temperature dependence of magnetization of NdMnAl show broad peaks corresponding to magnetic transitions along with thermomagnetic irreversibility and hysteresis loops with high coercive field, indicating spin glass type behaviour [33]. The effective magnetic moment calculated from inverse-susceptibility data fits suggests contributions from both Nd and Mn ions. Similar behaviour was observed in GdMnIn and DyMnIn compounds. The magnitude of Mn moment in these compounds varies as it depends both on crystal symmetry and Mn-Mn bond lengths [33]. Chevalier *et al*. [42] studied structural and magnetic properties of unmilled and milled GdMnAl. The authors reported a AFM transition at 298 K in unmilled GdMnAl and spin-glass behaviour in milled GdMnAl with a freezing temperature of 53 K. Later, Klimczak and Talik [43] have investigated GdMnAl for magnetic refrigeration applications and reported that GdMnAl shows FM behaviour with a transition temperature of 274 K. Recently, Pasca *et al.* [44] studied DC and AC magnetization along with local magnetic hyperfine exchange interactions and summarized that GdMnIn shows spin glass behaviour with no long range magnetic ordering, which was attributed to a triangular spin frustration of magnetic ions.

The magnetometry and neutron diffraction measurements performed on polycrystalline *R*MnGa (*R*= Ce, Pr, Nd, Gd-Dy) samples reveal spin glass behaviour in these samples [35]. Among these, NdMnGa and DyMnGa show simple spin glass (SG) while PrMnGa, GdMnGa and TbMnGa show re-entrant spin glass (RSG) state [35]. A careful examination of crystal structure and magnetic properties suggest that almost all the compounds crystallizing in Laves phase cubic and hexagonal crystal structures show spin glass properties. The magnetic frustration among magnetic ions arises from the random distributions of Mn and *X* (Al, Ga, In) atoms at B sites (considering an $AB_2$ type crystal structure).

*R*MnSi compounds show high magnetic transition temperatures that are attributed to the stacking of AFM Mn layers [36]. Due to magnetic nature of Mn in this series, three main exchange interactions are taking place; interaction between rare earth sublattices ($R$–$R$ interaction), interaction between rare earth and Mn sublattices ($R$ - Mn interaction) and interaction between Mn sublattices (Mn- Mn interaction) [36]. A neutron diffraction study reveals that the magnetic order within the Mn planes depends on both Mn- Mn and Mn- $X$ contacts [37]. Accordingly, the interatomic distance between Mn-Mn decides the nature and strength of magnetic ordering within the magnetic layers. It has been reported that the in-plane Mn-Mn exchange coupling is FM for $d_{Mn-Mn} < 2.84$ Å and AFM for $d_{Mn-Mn} > 2.89$ Å and co-existence of FM and AFM components for intermediate inter-atomic distances [37]. Magnetization and neutron diffraction measurements reveal antiferromagnetic behaviour of $R$MnSi ($R$= La- Nd) compounds [36]. Magnetic structures of these compounds are characterized by a stacking of AFM (001) Mn layers around room temperature, which have not been detected in magnetization measurements [36]. Additional magnetic transitions were observed in PrMnSi and NdMnSi at low temperatures attributed to an antiferromagnetic ordering of rare earth sublattices followed by a spin reorientation process within the Mn sublattice. GdMnSi was reported to be FM below 310 K [36]. SmMnSi shows unusual magnetic behaviour with two magnetic transitions at 130 and 240 K (AFM) and a compensation point at 215 K [45]. This compound shows negative magnetization, which is usually shown in ferro and ferrimagnetic materials during the re-magnetization process when the magnetic field changes the direction [45]. Recently Ray *et al.* [46] studied magnetic and related properties of SmMnSi in detail and explained the unusual magnetic reversal / negative magnetization in terms of field induced switching of Sm orbital magnetic moment, which is coupled with its spin moment in the opposite direction, Mn-Mn and Mn-Sm exchange interactions and polarized conduction electron moment. Fig. 5 shows the temperature dependence of magnetization ($M$) in zero field cooled warming (ZFCW), field cooled cooling (FCC) and field cooled warming (FCW). It can be seen that magnetization shows AFM ordering around 250 K and a sign reversal between two compensation points T* and T**. The unusual magnetism of SmMnSi makes it a candidate for potential applications such as magnetization switching, magnetic memory, and clean energy harvesting [46–48].

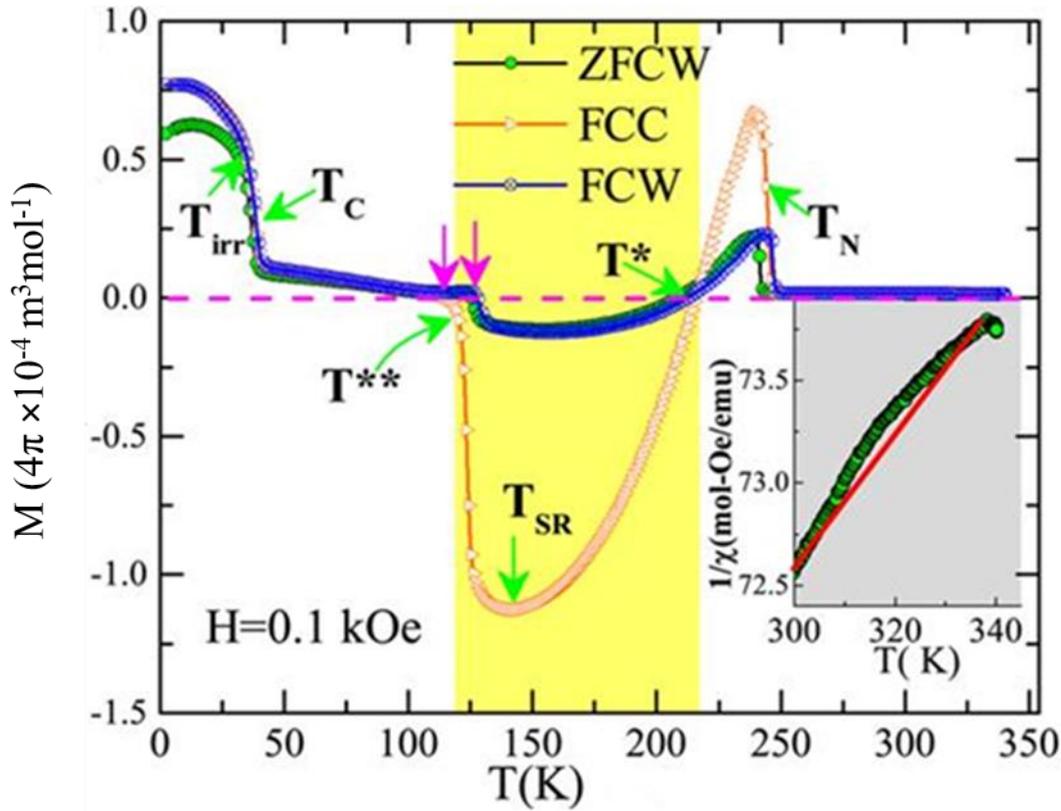

**FIG. 5.** The temperature dependence of magnetization, M at 0.01 T field under ZFCW, FCC, and FCW measurement configurations for SmMnSi. The inset shows inverse-magnetic susceptibility as a function of temperature. Reproduced with permission from [46] © 2020, AIP Publishing.

CeFeSi - type TbMnSi shows FM ordering below 260 K, while the high temperature form, TiNiSi - type TbMnSi and DyMnSi order antiferromagnetically below 410, and 400 K, respectively [37,49]. Observation of different magnetic ordering in polytypic forms of TbMnSi compounds clearly reveals the influence of Mn-Mn contacts on the Mn sublattice magnetic ordering [37]. Fig. 6 shows the evolution of Mn moment values for rare earth silicides. It can be noted that Mn moment values decrease as a function of rare earth size for both CeFeSi and TiNiSi structure types with almost parallel slopes. It is clear from Fig. 6 that Mn moment values are higher for TiNiSi type structure than CeFeSi type structure. The reason for higher value in TiNiSi type structure is that on structural transformation from 'CeNiSi → TiNiSi' Mn- $X$ hybridization is reduced due to relaxation in Mn- $X$ contacts, resulting in an enhancement of Mn moment [37]. In addition to this, there is a reduction in the number of direct Mn-Mn contacts from 4 to 2, which may affect the shape of the Mn $3d$ band [37].

HoMnSi was found to show multiple magnetic arrangements as a function of temperature; collinear AFM in the temperature range 55–300 K, canted AFM in the range 15-55 K and incommensurate AFM below 15 K [50].

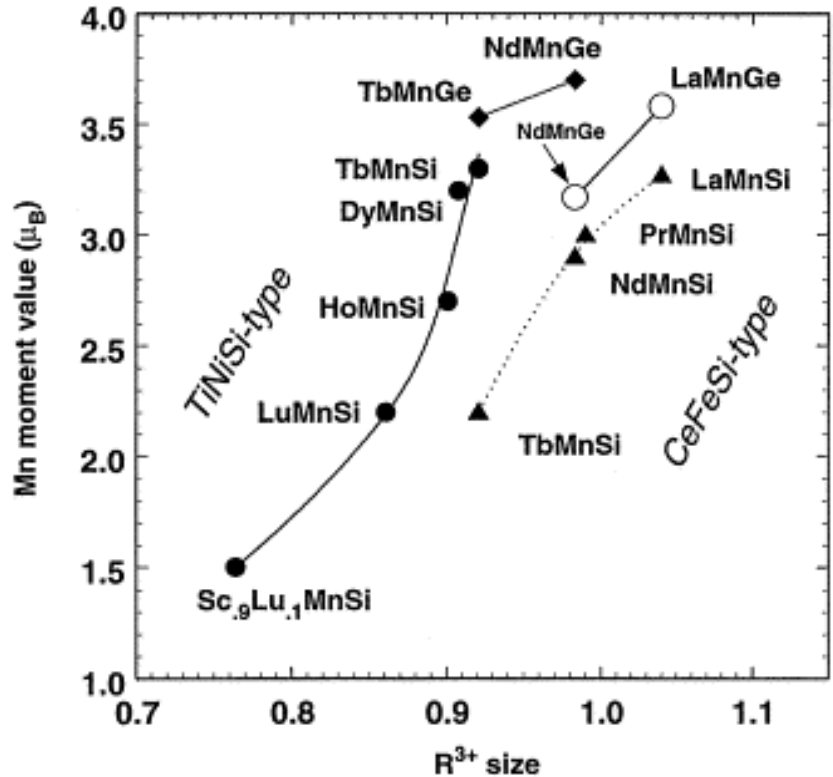

**FIG. 6.** Mn magnetic moment as a function of size of the rare earth ion ($R^{3+}$) for some CeFeSi and TiNiSi type structures. Reproduced with permission from [37] © 1999, Elsevier B. V.

Neutron diffraction measurements carried out on $R$MnGe ($R$ = La –Nd, Tb-Tm, Y) compounds show AFM ordering above room temperature with ordering exclusively on the Mn sublattice [39,40,51]. At low temperatures, $R$MnGe ($R$= La- Nd) compounds are characterized by FM ordering of rare earth layers, with different ordering along the stacking axis (001) while $R$ sublattices in $R$MnGe ($R$= Dy- Tm) orders antiferromagnetically [39,40]. It has been reported that at 150 K, PrMnGe orders ferromagnetically followed by a ferro-antiferromagnetic transition below 80 K due to a crystallographic phase transition from tetragonal to orthorhombic structure while NdMnGe shows FM order from 200 K down to 2 K [39]. At low temperatures, DyMnGe is characterized by a cycloid structure, HoMnGe by a conical structure, ErMnGe, YMnGe and TmMnGe order in a collinear AFM structure [40]. The magnetic moment values of Mn in these compounds is estimated to be in the range of 3.4 - 4.2 $\mu_B$ [40].

*R*MnGe compounds show magnetic properties similar to the *R*MnSi series as in both the series *R* and Mn sublattices show almost same ordering. However, there are some differences as well. The interlayer interactions between nearest neighbours in adjacent Mn planes are AFM in all rare earth germanides, whereas they are FM in some silicides such as LaMnSi and CeMnSi [39]. RKKY interactions possibly can explain various interlayer exchange interactions in these compounds. In *R*MnGe compounds, the Mn sublattice shows easy axis character while in *R*MnSi easy plain prevails. Moreover, exchange interactions between the Mn sublattices are stronger in germanides than in silicides despite of their larger unit cell parameters. [39]. Fig. 7 shows the evolution of Mn magnetic moments as a function of Mn-Mn interlayer distance for CeFeSi representatives. It can be noted that for $d_{Mn-Mn} > 2.86$ Å or Mn moments value greater than ~2.5 $\mu_B$ results in AFM behaviour in *R*Mn*X* above room temperature [39].

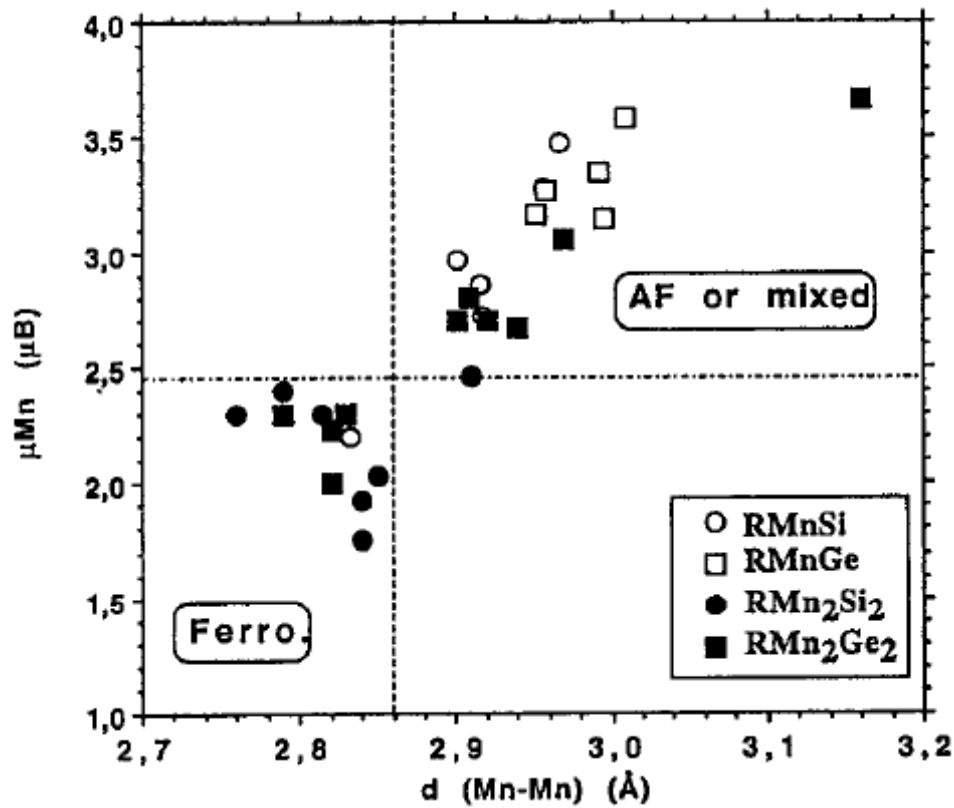

**FIG. 7** Mn-Mn interlayer exchange interaction as a function of Mn-Mn interlayer distance. Reproduced with permission from [39] © 1995, Elsevier B. V.

## 2.4 *RFeX* compounds

H. Oesterreicher [52] studied structural properties of *R*FeAl compounds. The author reported two types of crystal structures in *R*FeAl compounds. Light rare earths compounds (except LaFeAl) show a two-phase structure; out of which, one phase remains C15 type, the second phase is unidentified. Heavy rare earth compounds show C14 - MgZn$_2$ type hexagonal crystal structure with YbFeAl as an exception. *R*FeGa (*R*= Ho -Tm) compounds crystallize in MgZn$_2$ type hexagonal crystal structure [53]. *R*FeSi( *R*=La-Sm, Gd-Er) compounds crystallize in CeFeSi type tetragonal crystal structure [54,55].

Later H. Oesterreicher [56] reported magnetic properties of heavy rare earths *R*FeAl (*R*= Gd-Lu, Y). It has been observed that all the compounds show FM ordering. Kastil *et al.* [57] studied GdFeAl and TbFeAl for magnetic refrigeration applications. The authors reported FM in both the compounds with transition temperature values matching with the ones reported by H. Oesterreicher [56]. The saturation moment in FM GdFeAl is found to be 5.8 $\mu_B$/f.u., lower than the theoretical value of Gd$^{3+}$ ion [58], indicating no moment from Fe, however FM ordering in non-magnetic rare earth compounds such as YFeAl and LuFeAl suggest magnetic moment contribution from Fe atoms. A promising material for magnetic refrigeration application, TbFeAl was also reported to show large exchange bias, making it a multifunctional material [59]. The results suggest that the exchange bias in TbFeAl is induced by the inherent crystallographic disorder between Fe and Al and strong magneto-crystalline anisotropy of Tb$^{3+}$ [59]. These results pave the way for search of materials exhibiting exchange bias induced by atomic disorder. Fig. 8(a) shows the temperature dependence of magnetization measured in ZFC and FC modes in the field of 0.02 T. The results show a FM transition around 198 K followed by a second magnetic transition at 154 K. The isothermal magnetization shown in fig. 9 (b-c) features a jump in magnetization which disappears above 8 K. A loop shift in magnetization isotherms indicating exchange bias is also supported by the observation of training effects. A large coercive field of ~1.5 T at 2 K was shown by this material.

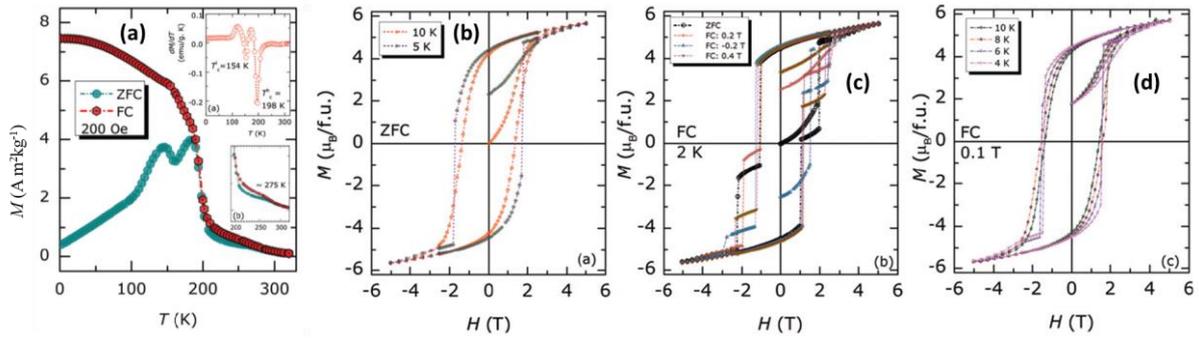

**FIG. 8.** (a) The temperature dependence of magnetization in a magnetic field of 0.02 T for TbFeAl. Magnetic isotherms as a function of temperature in (b) ZFC (c) FC and (d) FC modes. Reproduced with permission from [59] © 2016, IOP Publishing.

Zhang *et al.* [60] studied magnetic properties of ErFeAl and HoFeAl and observed that both the compounds show FM ordering with a second magnetic transition at low temperatures, attributed to spin reorientation. Mulders *et al.* [61] reported neutron diffraction on ErFeAl and TmFeAl. The authors observed no magnetic reflection from Fe atoms in these compounds possibly due to very small Fe moments i.e. 0.52 and 0.66 $\mu_B$ in TmFeAl and ErFeAl, respectively [61].

The temperature dependence of magnetization in ErFeGa shows FM behaviour below 77 K with a second magnetic transition at 58 K when applied field is 0.01 T [53]. The second magnetic transition shifts to 10 K on applying a magnetic field of 0.5 T. The authors discuss that the low temperature difference between ZFC and FC curves suggest the presence of large magneto-crystalline anisotropy or competing magnetic interactions. The effective magnetic moment determined from the fitting of Curie-Weiss law is found to be 9.59 $\mu_B$, slightly higher than the theoretical value for $Er^{3+}$ ion, suggesting a small contribution from Fe.

Susceptibility and neutron diffraction measurements show FM for $R$FeSi with $R$= Nd, Tb, Dy characterized by a collinear FM structure aligned along the *c*-axis while LaFeSi and CeFeSi are Pauli PM and PrFeSi is a Curie-Weiss paramagnet [55]. $R$FeSi ($R$= Tb and Dy) compounds studied by Zhang *et al.* [62] show different magnetic transition temperatures as reported previously, which arise due to the different heat treatment used in the synthesis of these materials. PPM behaviour of LaFeSi and neutron diffraction measurements for other magnetic $R$FeSi compounds confirm that Fe does not carry any moment [55]. Recently Chouhan *et al.* [63] performed DFT calculations using full-potential linearized augmented plane wave (FP-LAPW) method to examine *p* element substitution at Si in LaFeSi compound to induce Fe magnetic moments and study their magnetic properties. The authors demonstrated that either partial or complete substitution by several *p*-block atoms at the Si 2*c* - sites leads to non-zero Fe moments [63]. These results pave the way towards the manipulation of otherwise quenched Fe moments in La based compounds and can make them promising materials for magnetic sensor applications. HoFeSi was reported to show two successive magnetic transition temperatures; PM to FM at 29 K and FM to AFM/FIM at 20 K [64]. Magnetic properties of ErFeSi reveal that the compound orders ferromagnetically below 22 K [65].

## 2.5 *RCoX* compounds

*R*CoAl compounds with light rare earths (*R*= La-Sm) were reported to show multiple phases in their crystal structures while heavy rare earths (*R*= Gd- Lu) were reported to crystallize in C14 type hexagonal structures [66]. CeCoAl and CeCoGa compounds crystallize in a monoclinic structure [67,68]. Welter *et al.* reported that *R*CoSi (*R*=La-Sm, Gd, Tb) and *R*CoGe (*R*= La-Nd) compounds crystallize in CeFeSi type tetragonal structures [69,70]. *R*CoX (*R*= Y, Tb-Lu, *X*= Si, Ge, Sn) compounds were reported to crystallize in TiNiSi type orthorhombic crystal structure [71].

CeCoAl shows intermediate Ce-valent nature, which is in line with the short Ce–Co distances in this compound [67]. The heat capacity data of CeCoAl show a λ-anomaly at 271 K, indicating a structural phase transition [67]. Magnetic measurements performed on CeCoAl compound show an inflection point at the same temperature as that of the λ-anomaly, confirming the phase transition and two additional magnetic transition at 255 and 200 K, which are most likely due to superparamagnetic Co impurities forming on the grain boundaries [67]. The temperature dependence of magnetization shows FM behaviour for *R*CoAl (*R* = Gd, Tb, Dy, Ho, Tm) compounds [72,73]. It can be noted from Fig. 9 that except GdCoAl, all the compounds show significant thermomagnetic irreversibility (TMI) at low temperatures. The TMI have been observed in several *RTX* compounds, which can beattributed to the following reasons: (i) strong magnetocrystalline anisotropy of rare earth atoms/ions (ii) domain wall pinning effects (iii) magnetic frustration or spin glass behaviour and (iv) superparamagnetic behaviour.

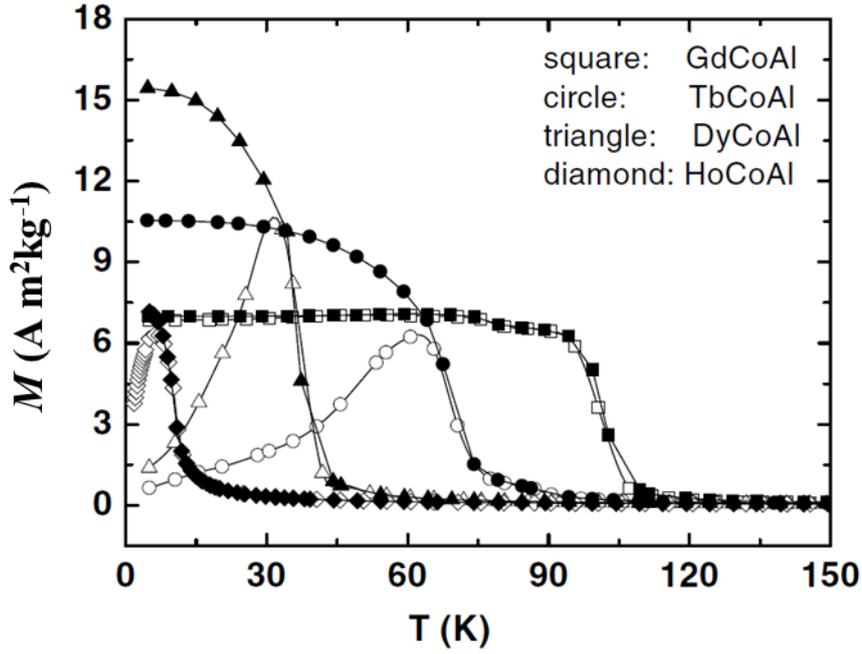

**FIG. 9** The temperature dependence of magnetization at applied field of 0.01 T for $R$CoAl ($R$ = Gd, Tb, Dy, and Ho) compounds. Reproduced with permission from [72] © 2001, IOP Publishing.

CeCoGa orders antiferromagnetically below 4.3 K ($T_N$) as confirmed by the temperature dependence of the magnetic susceptibility [68]. A Curie-Weiss fit to the susceptibility data gives $\mu_{eff}$ ~1.8 $\mu_B$ and $\theta_p$ ~ -82 K. The large value of $\theta_p$ suggests the presence of Kondo behaviour in this compound. The magnetic field dependence of magnetization shows two metamagnetic transitions and a shift in the hysteresis loop indicating the existence of exchange bias in the compound [68].

$R$CoSi ($R$ = Nd, Sm, Gd, Tb) were reported to be AFM below their respective magnetic transition temperatures (mentioned in Table 2) while LaCoSi, CeCoSi and PrCoSi were found to be PPM and Curie-Weiss paramagnets down to 1.6 K, respectively [69]. Later Chevalier *et al.* [74] reported AFM ordering in CeCoSi below 8.8 K. Neutron diffraction data in these compounds confirm no local moment at Co sites. Other than SmCoSi, all $R$CoSi ($R$ = Pr, Nd, Gd, Tb) compound follow Curie-Weiss law in the paramagnetic regime. Estimated $\mu_{eff}$ are in close agreement with theoretical values for respective rare earth ions, ruling out any moment from Co atoms. The positive values of $\theta_p$ suggests predominant FM interactions in these compounds [69]. Fig. 10 shows the field dependence of the magnetization (main plot) and the temperature dependence of the magnetic susceptibility (inset) along with the Curie-Weiss fit to the inverse susceptibility data for HoCoSi. A clear signature of onset of FM ordering can be noted. No hysteresis in magnetization data was observed while sweeping magnetic field up and down. The estimated value of effective magnetic moment from the inverse susceptibility was found to be close to the expected value for a free Ho ion [75,76]. The material shows significant change in magnetization around its magnetic phase transition temperature, which makes it promising for magnetic sensors and magnetic refrigeration applications. TmCoSi orders antiferromagnetically below 4.4 K, which is close to the boiling point of liquid Helium and therefore can be a good candidate for gas liquefaction applications [77]. The AFM ordering in TmCoSi transforms to FM ordering on field application of 1T indicating a metamagnetic transition in the material. Review of magnetic properties of $R$CoSi compounds confirms that among all compounds, only HoCoSi shows FM ordering.

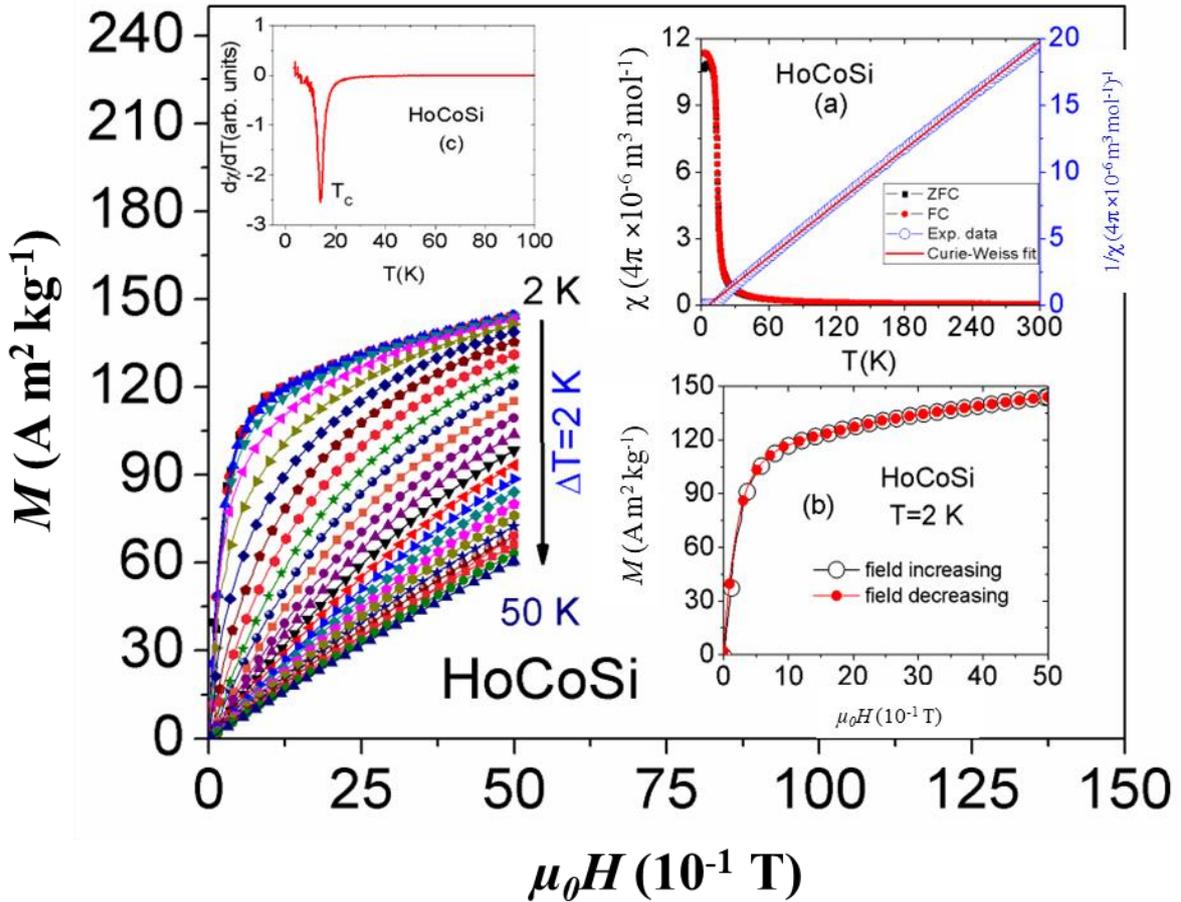

**FIG. 10** The field dependence of magnetization as a function of temperature. The inset shows derivative of magnetic susceptibility to check the magnetic transition temperature, the temperature dependence of magnetic susceptibility (at 0.05 T) and field dependence of magnetization in sweep up and down at 2 K for HoCoSi compound. Reproduced with permission from [75] © 2013, Elsevier B. V.

$R$CoGe compounds show similar magnetic properties as their silicide counterparts. $R$CoSi ($R$= La, Ce, Pr) compounds are PPMs and Curie-Weiss paramagnets down to 1.6 K while NdCoGe and TbCoGe order antiferromagnetically at low temperatures [70,78]. Chevalier *et al.* [79] reported CeCoGe to be AFM, which show spin fluctuation behaviour after hydrogenation without changing its crystal structure. Neutron diffraction data on TbCoGe suggest FM ordering [80]. DyCoGe shows complex magnetic properties with multiple magnetic phase transitions [78]. The compound shows FM ordering at 175 K, a FM to AFM transition at 10 K and spin-reorientation at low temperatures. No Co local moment was detected in these compounds. $R$CoSn ($R$ = Tb, Dy. Ho, and Er) compounds order antiferromagnetically at low temperatures and show field induced metamagnetic transitions [81,82]. The neutron diffraction data show modulated non-collinear magnetic structures for these compounds [82,83].

## 2.6 *RNiX* compounds

$R$NiAl ($R$= Ce-Nd, Gd-Lu) compounds crystallize in $Fe_2P$/ZrNiAl type hexagonal crystal structure, which is less densely packed than the Laves phases [84]. An isostructural transition with temperature has been seen in many of the compounds crystallizing in ZrNiAl type hexagonal structure [85–88]. Prchal *et al.* [85] summarised $RT$Al ($T$= Ni, Pd, Cu) compounds, which show structural discontinuity with temperature or composition. The general trend of isostructural transition in these compounds is that $c/a$ values from 0.565 to 0.575 are forbidden as shown in Fig. 11.

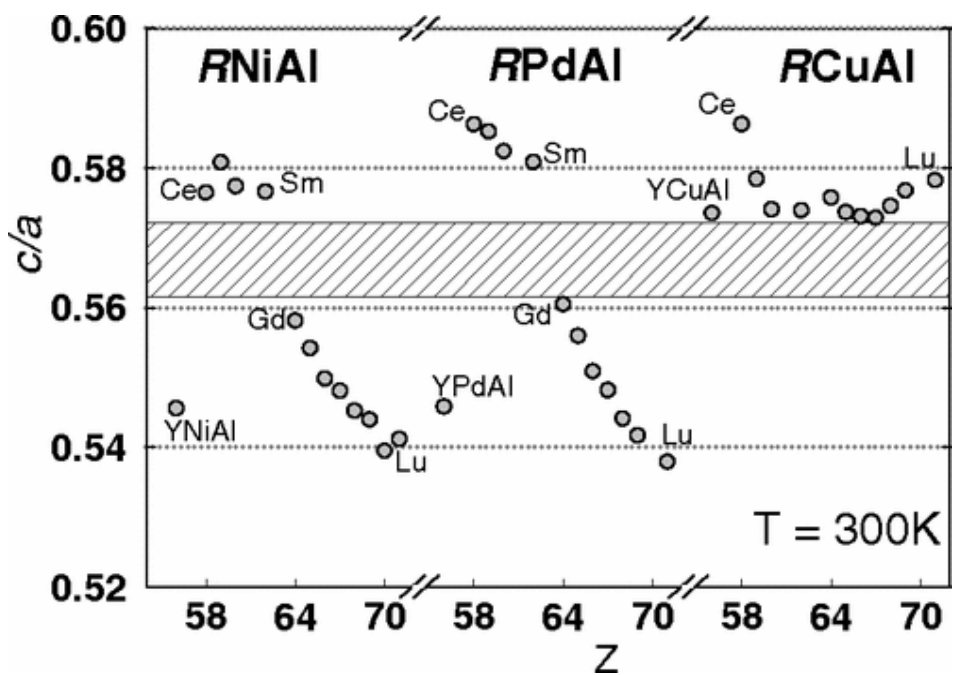

**FIG. 11.** Evaluation of $c/a$ values in $RT$Al ($T$ = Ni, Pd, Cu) compounds. Each point in the figure represents a rare earth element in the sequence Ce, Pr, Nd, Sm, Gd, Tb, Dy, Ho, Er, Tm, and Lu. Reproduced with permission from [85] © 2008, American Physical Society.

CeNiGa shows a crystallographic transformation with temperature. The high temperature form of CeNiGa crystallizes in TiNiSi-type orthorhombic whereas low temperature form adopts ZrNiAl type hexagonal structure [89]. $R$NiGa ($R$= Gd-Tm, Yb) compounds show TiNiSi -type orthorhombic crystal structure [90–96]. $R$NiIn ($R$= La, Ce, Nd, Gd-Tm) compounds crystallize in ZrNiAl type hexagonal structures [97–101]. $R$NiSi with light rare earths ($R$=La, Ce, Nd) compounds crystallize in LaPtSi-type tetragonal structure, whereas compounds with heavy rare earths ($R$=Gd-Lu) are characterized by TiNiSi-type orthorhombic crystal structures [76,102–106]. $R$NiGe ($R$=Ce, Gd-Tm, Y) compounds show TiNiSi type orthorhombic crystal structure [107,108]. $R$NiSn ($R$=La-Sm, Gd-Lu, Yb) compounds crystallize in TiNiSi- type orthorhombic crystal structure [109,110]. $R$NiSb ($R$=La-Nd, Sm) compounds crystallize in ZrBeSi type hexagonal crystal structure whereas compounds with $R$= Gd-Tm, Lu show MgAgAs-type cubic structure [111]. It has been reported that GdNiSb show two type of crystal structures.

$R$NiAl ($R$= Ce-Nd, Gd-Tm) compounds order ferromagnetically except PrNiAl and CeNiAl which show Curie-Weiss paramagnetic behaviour [84]. Later studies on CeNiAl show the evidence of possible mixed valence of Ce ion along with a high Kondo temperature [112].

Hydrogenation of CeNiAl induces an isostructural transition from ZrNiAl $\rightarrow$ AlB$_2$–type hexagonal structure and a valence transition for Ce ion [113]. A neutron diffraction study by Javorsky *et al.* [114] confirm AFM ordering in PrNiAl and NdNiAl below 6.5 and 2.4 K, respectively with field induced metamagnetic transitions in both the compounds. YbNiAl and LuNiAl do not show any magnetic ordering, suggesting no magnetic moments from Ni atoms. However, magnetic studies carried out by Schank *et al.* [115] show AFM ordering in YbNiAl and YbPtAl below 2.9 K and 5.9 K, respectively, which are reported to be the first Yb based heavy fermions showing AFM ordering. Ehlers and Maletta [116] reported frustrated magnetic moments in $R$NiAl ($R$ = Pr, Nd, Tb, Dy, Ho) compounds studied by neutron diffraction. It has been analysed that $R^{3+}$ nearest neighbours form a triangular lattice with antiferromagnetic coupling in the $a$-$b$ plane, which gives rise to frustration of one-third of $R^{3+}$ moments [116]. At low temperatures, magnetic structures of light rare earth compounds are incommensurate with the crystal structure while for heavy ones it is strictly commensurate. Neutron time of flight and spin-echo spectroscopy show both long range ordering and frustrated Tb spins in the AFM phase [117]. Lin He reported magnetic properties of SmNiAl and its hydride [118]. Field dependence of the magnetization as a function of temperature suggests paramagnetic nature of SmNiAl. After hydrogenation, SmNiAlH$_{1.17}$ was found to show FM behaviour at ~75 K and magnetic compensation at ~38 K. The magnetic compensation take place due to balancing of orbital magnetic moment by spin magnetic moment [118]. Strong diamagnetic effect was observed on field cooling in SmNiAlH$_{1.17}$ due to its strong magneto-crystalline anisotropy. An isostructural transition as a function of the temperature was observed in GdNiAl, which is reflected by a discontinuity in magnetic susceptibility and electrical resistivity data [86]. Single crystal studies on DyNiAl show FM ordering below 30K with the moments oriented along hexagonal $c$-axis followed by an FM $\rightarrow$AFM transition at 15 K where AFM component lies in the basal plane resulting in a non-collinear magnetic structure [119]. The high field magnetization measurements show very strong uniaxial magnetic anisotropy with an anisotropy field of 25 T and field induced transitions when the field is applied within the basal plane [119]. Single crystal studies on HoNiAl show similar magnetic properties as DyNiAl, however the magnetocrystalline anisotropy in HoNiAl is very weak [120].

Both low and high temperature forms of CeNiGa show intermediate valence nature however their Kondo temperature ($T_K$) are very different; $T_K$ >> 300 K for LTF and ~95 K for HTF. It has been reported that GdNiGa orders ferromagnetically below 30.5 K, whereas $R$NiGa ($R=$ Tb-Tm, Yb) are AFMs [90,94,96,121]. Neutron diffraction experiments show an incommensurate magnetic structure for $R$NiGa ($R=$ Dy-Tm) with no magnetic moment from the Ni atom [92,93,95,122].

Gondek *et al.* [123] studied $R$NiIn ($R=$ Ce, Pr, Nd, Tb-Er) compounds using magnetometric and neutron diffraction measurements and observed geometrical frustration of rare earths magnetic moments in AFM ordering. Because of magnetic frustration and presence of crystal field effects, a wide range of magnetic ordering was seen in these compounds. No magnetic ordering was observed in CeNiIn and PrNiIn compounds. $R$NiIn ($R=$Nd, Ho and Er) shows FM ordering whereas TbNiIn and DyNiIn are AFMs. At 1.5 K, both TbNiIn and DyNiIn exhibit coexistence of non-linear AFM phase and a phase described by propagation vector, $\boldsymbol{k}=$ (0.5,0,0.5) [123]. Zhang *et al.* [100] reported FM ordering in GdNiIn and ErNiIn and two successive magnetic phase transitions in $R$NiIn with $R=$ Tb, Dy and Ho.

LaNiSi is reported to be the first Ni based ternary superconductor, exhibits bulk superconductivity below 1.23 K, confirmed by ac susceptibility (diamagnetic behaviour) and heat capacity (sharp jump) measurements [104]. The temperature dependence of magnetic susceptibility of CeNiSi shows a broad maximum around 280 K and a sharp increase around 50 K [103]. NdNiSi was reported to show two magnetic phase transitions at 6.8 (AFM) and 2.8 K (spin reorientation) whereas GdNiSi is reported to be AFM below 11 K [105,124]. The broad maximum is the signature of intermediate valence Ce in this compound. Szytula *et al.* [102] studied magnetic properties of $R$NiSi ($R=$Tb- Er) compounds using magnetization and neutron diffraction measurements. The authors reported that all compounds show AFM ordering below their respective magnetic transition temperatures. At 1.5 K, HoNiSi and ErNiSi show collinear magnetic structures whereas TbNiSi and DyNiSi show square modulated magnetic structures [102]. Magnetic moments are oriented along the *b*-axis for $R=$ Tb, Dy and Ho compounds whereas it is localized in the *a-c* plane for the Er compound [102]. No magnetic moment was detected on Ni. Due to a strong magnetocrystalline anisotropy, a large rotating magnetocaloric effect was observed in single crystal DyNiSi, enabling its application in a new type of rotary magnetic refrigerator [125]. The temperature dependence of magnetic susceptibility at low fields (0.05 T) shows clear AFM ordering below 3.2 K, which transforms into FM with increasing the magnetic field strength, confirming the field induced metamagnetic transition in ErNiSi as shown in Fig. 12 [126]. It can be noted from the Fig. 12 that there is no difference between ZFC and FC, hence confirming the reversible behaviour of the compound with no thermal hysteresis.

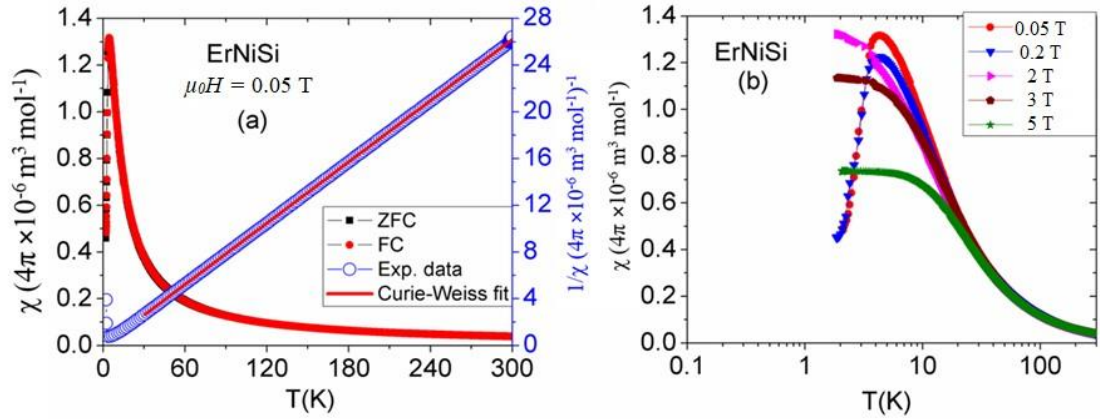

**FIG. 12.** (a) The temperature dependence of the magnetic susceptibility at (a) 0.05 T (b) 0.05 – 5 T fields. Reproduced with permission from [126] © 2014, AIP Publishing.

The temperature dependence of the magnetic susceptibility suggest PPM nature of CeNiGe [127]. Magnetic measurements for $R$NiGe ($R$= Gd-Tm, Y) compounds show AFM ordering in $R$= Gd, Tb, Dy and Er compounds at low temperatures, however no magnetic ordering was observed in Ho and Tm compounds down to 4.2 K [107,128]. Neutron diffraction measurements indicate square modulated magnetic structures for TbNiGe and DyNiGe at low temperature, which changes to a sinusoidal modulated structure for TbNiGe and a cycloidal spiral for DyNiGe on increasing the temperature while keeping the moment direction along the $c$-axis [128]. Neutron diffraction results on HoNiGe and ErNiGe show AFM ordering characterized by collinear magnetic structures, which gets modified to modulated structures at 2.5 and 2.3 K, respectively [129]. YNiGe and YNiSn show a temperature independent magnetic susceptibility [107,130].

A single crystal of LaNiSn exhibits superconductivity below 0.59 K as demonstrated by AC magnetic susceptibility and electrical resistivity results [131]. Routsi *et al.* studied magnetic properties of $R$NiSn compounds and observed that the compound CeNiSn shows Kondo or valence fluctuation behaviour, compounds with $R$= Sm, Gd-Dy order AFM at low temperatures whereas no magnetic ordering was observed for the compounds with $R$ = Pr, Nd, Ho-Tm down to 4.2 K [130,132,133]. Most of these compounds show field included metamagnetic transitions [130]. Szytula *et al* [134] reported AFM ordering below 3 K with the Nd moments forming an incommensurate sine-wave modulated structure [135]. There are some reports indicating a pseudo-gap opening in the Kondo state in CeNiSn at low temperatures [136,137]. Andoh *et al.* [138] studied $R$NiSn ($R$= Gd, Ho, Er) single crystals and observed that all compounds are AFM in nature and show strong anisotropy with the magnetization easy axis along the *b*-axis for Gd and Ho and along the *a* -axis for Er.

In the $R$NiSb series, Y, La and Lu compounds are PPM, indicating no magnetic moments on Ni atoms, SmNiSb shows van Vleck paramagnetic behaviour, NdNiSb shows FM nature, Ce, Tb-Ho show FM nature [111,139]. YbNiSb was interestingly found to be ordered below 0.8 K [140]. GdNiSb shows AFM ordering for both the crystal structures however, the magnetic transition temperature remains different [141].

## 2.7 *RCuX* compounds

Compounds in the $R$CuAl ($R$=Ce-Nd,Sm, Gd-Lu) series crystallize in the ZrNiAl type hexagonal crystal structures [142]. Some additional reflections were detected in XRD patterns due to some impurities [142]. In the $R$CuGa series, there are only reports on CeCoGa and EuCuGa compounds, which crystallize in the CeCu$_2$ type orthorhombic crystal structure [143,144]. $R$CuIn ($R$=La-Nd, Gd, Tb, Ho, Er, Lu) compounds were found to crystallize in the ZrNiAl type hexagonal crystal structures [145,146].

$R$CuSi ($R$=La-Nd, Sm, Gd-Lu, Yb, Y) compounds crystallize in two types of hexagonal crystal structures depending on their heat treatments [147–149]. The high temperature phase adopts AlB$_2$-type structure (P6/mmm), whereas the low-temperature phase crystallizes in the Ni$_2$In-type structure (P6$_3$/mmc). Crystal structures of $R$CuGe compounds show similarity with $R$CuSi as the crystal structure depends on the annealing treatment of the materials. As cast samples of $R$CuGe ($R$= La-Lu) crystallize in the AlB$_2$ type hexagonal crystal structure and after long time annealing the compounds with $R$= Tb – Lu adopt CaIn$_2$ type hexagonal crystal structures [150–155]. CaIn$_2$ and LiGaGe and CeCu$_2$ type hexagonal crystal structures were identified in $R$CuSn series of compounds. $R$CuSn ($R$=La, Ce- Nd, Lu, Y) [156,157] compounds crystallize in CaIn$_2$ type structure whereas $R$= Gd-Er in LiGaGe type structures [158,159] and EuCuSn in CeCu$_2$ type hexagonal structure [160]. EuCuAs, EuCuSb and YbCuBi crystallize in Ni$_2$In, ZrBeSi and LiGaGe type hexagonal crystal structures, respectively [161–163].

In the $R$CuAl series, the magnetic measurements indicate FM ordering in heavy rare earths $R=$ Gd-Tm compounds and AFM ordering in light rare earths Pr and Nd [142,164,165]. GdCuAl and DyCuAl show some additional magnetic phases. The experimental value of $\mu_{\text{eff}}$ is close to their respective theoretical $R^{3+}$ free ion values. CeCuAl was reported to show a large difference for experimental and expected values of $\mu_{\text{eff}}$, which may be due to the impurity phases present in the sample. Chevalier and Bobet reported CeCuAl to be AFM below 5.2 K [166]. Non-magnetic nature of YbCuAl and LaCuAl suggests only rare earth atoms contributed magnetism in these compounds [84]. The magnetic ordering temperatures in these materials except for PrCuAl and NdCuAl follow de Gennes scaling [84]. Dong *et al.* [167] studied magnetic properties of amorphous and crystalline GdCuAl ribbons. They observed that amorphous GdCuAl ribbon shows superparamagnetic behaviour while crystalline GdCuAl shows FM order. Magnetic susceptibility measurements suggests an intermediate valence state in CeCuGa and AFM in EuCuGa [143,168].

Magnetic measurements show $R$CuIn compounds with $R=$ La and Lu are PPMs, no magnetic ordering in $R=$ Ce and Pr down to 2 K, whereas NdCuIn and GdCuIn show AFM ordering [145,169]. Hybridization of Pr $4f$ and Cu $3d$ orbitals causes the absence of magnetic ordering in PrCuIn. Neutron diffraction measurements confirm AFM ordering in $R$CuIn ($R=$ Nd, Tb-Er) with non-collinear magnetic structures for $R=$ Nd, Tb, Ho and collinear for $R=$ Er [146]. The non-collinear magnetic structure in these compounds is attributed to the triangular arrangement of rare earth moments.

Studies on the magnetic properties of $R$CuSi compounds ($AlB_2$ type) show FM nature of compounds with $R=$ Pr, Gd and Tb [170,171]. CeCuSi shows FM ordering with moments oriented perpendicular to $c$-axis and the saturation moment value is lower than the expected $gJ$ values (where $g$ is the Lande's $g$ factor and $J$ is the total angular momentum quantum number), indicating the partial lifting of $f$-electron level degeneracy [171,172]. Saturation of magnetization in GdCuSi requires a relatively high magnetic field, suggesting anisotropic exchange interactions. It has been observed that $Ni_2In$ type PrCuSi and GdCuSi order antiferromagnetically, which transform into a FM phase under sizable field application whereas compounds with $AlB_2$ type structure show FM ordering [170,173–175]. The value of $\mu_{\text{eff}}$, estimated from susceptibility data is found to be 8.1 $\mu_B/Gd^{3+}$, higher than the expected free ion value [175]. The excess effective magnetic moment or saturation moment is observed in many other Gd ternary compounds and is mainly caused by the strong polarization of $5d$ conduction electrons via 4*f*-5*d* interactions [175–178]. Gupta *et al.*[179] carried out magnetometric and neutron diffraction measurements to study magnetic properties of NdCuSi. It has been observed that NdCuSi exhibits co-existence of AFM and FM phases at low temperatures, attributed to competing exchange and crystal field interactions. Fig. 13 shows the temperature dependence of the magnetic susceptibility and magnetic moments (estimated from zero-field neutron diffraction data). It can be clearly seen that at low field, the compound shows AFM ordering, which transforms to FM at higher magnetic fields. The temperature dependence of magnetic moments show that both AFM and FM phases are present below transition temperature in NdCuSi [179]. GdCuSi shows complex magnetic structure with an AFM transition at 14.2 K and another magnetic anomaly near 5 K [175]. Theoretical calculations suggest competing ferromagnetic and antiferromagnetic interactions leading to non-collinear magnetic structure [148]. Ni$_2$In type *R*CuSi with *R*= Dy-Tm compounds were reported to show AFM ordering with their moments forming sine-wave modulated structures [180–184]. Neutron diffraction studies show a crossover from AFM to FM phase in TmCuSi due to reorientation of Tm moments at low temperatures, which later was also confirmed by magnetization measurements [77,184].

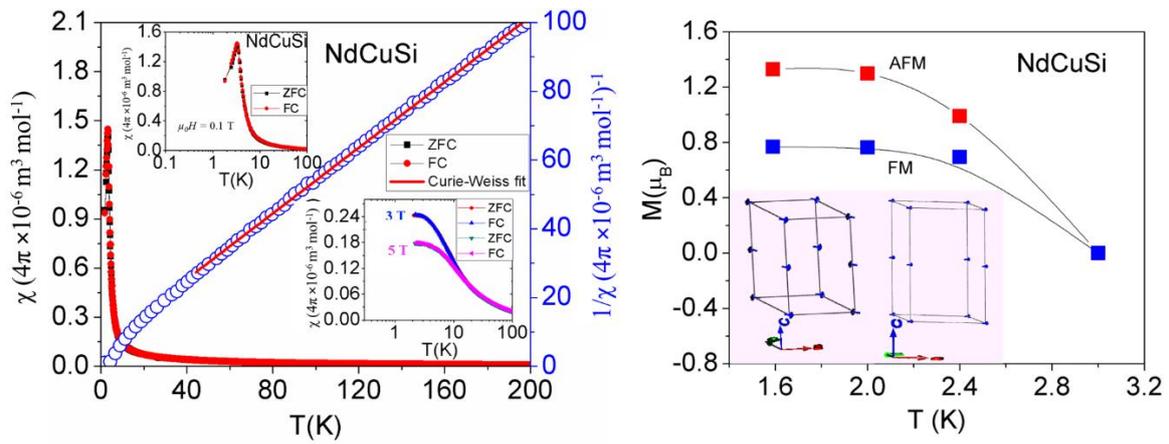

**FIG. 13.** The temperature dependence of (a) the magnetic susceptibility and (b) magnetic moments. Insets in (a) shows the temperature dependence of magnetic susceptibility at lower temperatures and at higher fields while in (b) it shows arrangement of magnetic moments from neutron diffraction measurements. Reproduced with permission from [179] © 2015, AIP Publishing.

In *R*CuGe series, CeCuGe shows FM ordering similar to CeCuSi at low temperatures [172]. *R*CuGe compounds with *R*= Pr, Nd, Gd-Er were found to show an AFM state at low temperatures, which undergoes a metamagnetic transition under field application [151]. All compounds except NdCuGe are characterized by sine modulate magnetic structures with the magnetic moments direction perpendicular to *c*-axis (for Tb and Dy), and parallel to *c*-axis (for Ho and Er) [151]. Collinear AFM structure was observed for NdCuGe compound. Gupta and Suresh [155] studied magnetic properties for *R*CuGe (*R*= Tb-Er) compounds in detail using DC and AC magnetization measurements. All these compounds show AFM ordering below their respective magnetic transition temperatures with field induced metamagnetic effects [155]. AC susceptibility (ACS) measurements were performed at constant AC field and a range of AC frequencies (*f*= 25-625 Hz). It has been observed that when magnetic structure of the material is in collinear AFM state, it displays a peak in the real part of ACS ($\chi'$) corresponding to AFM transition and no peak in the imaginary part of ACS ($\chi''$) [185]. Presence of a peak in the imaginary part of ACS signifies the energy losses associated with the ordered state, incurred in FM, FIM, canted or spin glass systems [186,187]. It can be noted from Fig. 14 that TbCuGe and HoCuGe show a peak in only the real part of ACS, whereas DyCuGe and ErCuGe show peaks in both real and imaginary part of ACS, indicating a collinear magnetic structure in the former and canted magnetic structures in the later compounds. The results for magnetic isotherms for these compounds show metamagnetic transitions with saturation trend for DyCuGe and ErCuGe, confirming the canted spins in these compounds as indicated in ACS susceptibility measurements [155].

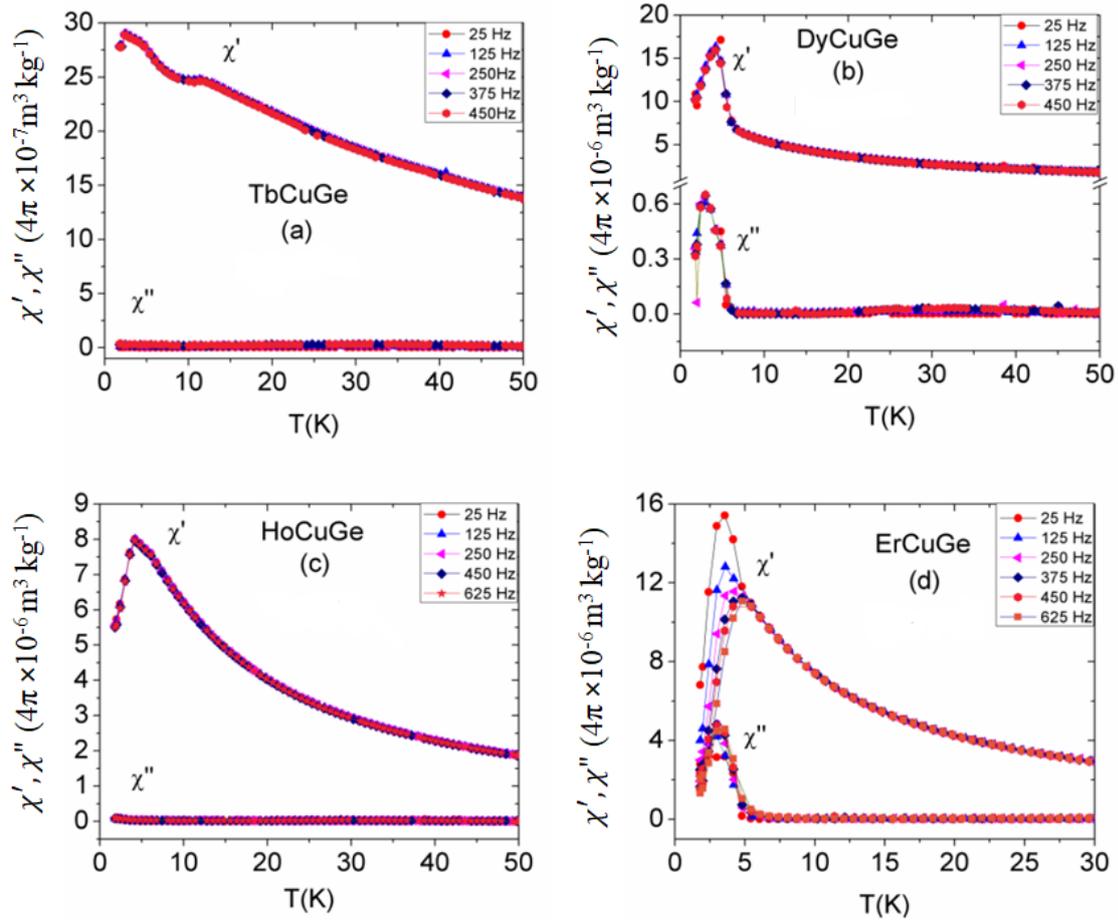

**FIG. 14.** The temperature dependence of AC magnetic susceptibility at different frequencies ($f$= 25-625 Hz) and constant AC magnetic field of $10^{-4}$ T for $R$CuGe ($R$=Tb-Er) compounds. Reproduced with permission from [155] © 2015, Elsevier B. V.

CeCuSn shows AFM ordering below 8.3 K with multiple metamagnetic transitions at 1.7 K suggesting a complex magnetic structure at low temperatures [188]. Neutron diffraction measurements show AFM ordering in PrCuSn and NdCuSn. PrCuSn exhibits a sine modulate magnetic structure while NdCuSn is characterized by collinear AFM structure at low temperatures [157]. In both the compounds the magnetic moments are oriented along $c$-axis. $R$CuSn compounds with $R$= Gd-Er studied by neutron diffraction and magnetization measurements were found to AFM at low temperatures with Tb, Dy and Ho compounds characterized by collinear magnetic structures at 1.5 K, which transformed to incommensurate structures for Tb and Dy on increasing temperatures [159]. The field dependence of magnetization shows metamagnetic transitions in these compounds. EuCuSn does not show any magnetic ordering down to 4.2 K unlike other Eu compounds such as EuCuSb, EuCuAs and EuCuBi, which are found to be AFM in nature [161,189,190].

## 2.8 *RRuX* compounds

Similar to 3*d* transition metal compounds, 4*d* compounds were also reported to be crystallized in a variety of crystal structures. Single crystal CeRuAl and CeRhAl compounds crystallize in the LaNiAl type orthorhombic crystal structure [191]. It has been reported that compounds with light rare earths in *R*RuSi and *R*RuGe crystallize in CeFeSi type tetragonal crystal structure, whereas heavy rare earth compounds adopt Co$_2$Si and TiNiSi type orthorhombic crystal structures [15,192,193]. Only CeRuSn has been reported in the *R*RuSn series, which crystallizes in CeCoAl type monoclinic crystal structure [194].

Magnetic measurements show a PPM nature of LaRuSi and LaRhSi, Curie-Weiss PM behaviour of CeRuSi, CeRuSn and CeRhSi, a FM ground sate in GdRuSi and Sm silicides and germanides and AFM nature of Pr and Nd silicides and germanides [192,194,195]. Neutron diffraction measurements show a collinear AFM structure in NdRuSi. HoRuSi shows two magnetic transitions with ferromagnetic ordering dominant in nature [196]. Magnetic and neutron diffraction measurements show that ErRuSi orders ferromagnetically below 8 K with the magnetic moments aligned in *ab* plane [197,198]. No significant moment was detected on Ru sites resulting in a saturation moment of the compounds less than expected from the *gJ* value for Er$^{3+}$ free ion. Moreover, the AC susceptibility shows peaks in both real and imaginary parts, confirming the FM nature.

*R*RuGe (*R*=Gd-Er) compounds order ferromagnetically below their transition temperatures with the Gd compound exhibiting the highest transition temperature in this series [199,200]. Neutron diffraction data show that Tb, Ho and Er compounds show a non-collinear FM structure at low temperatures [199,201]. Fig. 15 shows the temperature dependence of the magnetic susceptibility for *R*RuGe (*R*=Gd-Er) compounds in ZFC and FC measurement protocols. It can be noted that all the compounds show a huge difference in ZFC and FC at low temperatures, suggesting the possibility of a domain wall pinning effect [199]. The value of $\mu_{\text{eff}}$, estimated from susceptibility data is smaller than the expected free ion value in their respective compound, suggesting no contribution from the Ru moment. TmRuGe was reported to show no magnetic ordering down to 1.7 K [202].

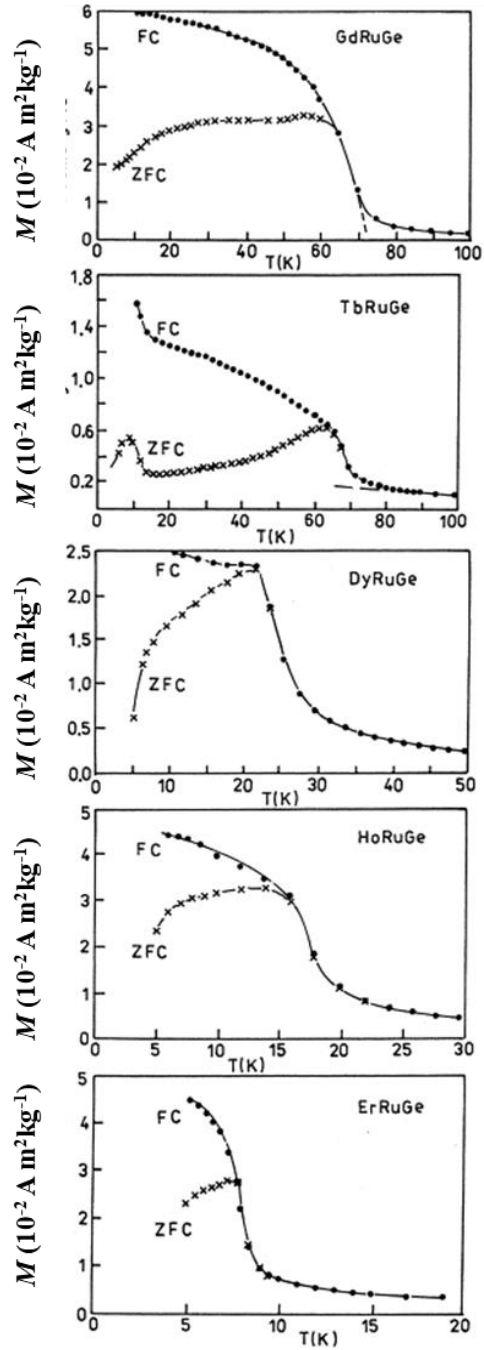

**FIG. 15.** Temperature dependence of magnetization for $R$RuGe compounds. Reproduced with permission from [199] © 1998, Elsevier B. V.

## 2.9 $R$Rh$X$ compounds

$R$RhAl ($R$=Y, Pr, Nd, Gd, Ho, Tm) compounds crystallize in TiNiSi type orthorhombic crystal structures whereas LaRhAl and CeRhAl were found to crystallize in Pd(Pd, Mn)Ge type structure with Ce showing mixed valency of 3 and 4, at room temperature [203]. All *R*RhGa (*R*=La-Nd, Gd, Tb, Ho-Lu, Yb, Y) compounds crystallize in TiNiSi type orthorhombic crystal structures [204,205]. *R*RhIn (R=La-Nd, Sm, Gd-Tm) compounds show ZrNiAl type hexagonal crystal structure, whereas YbRhIn, EuRhIn LuRhIn crystallize in TiNiSi type orthorhombic crystal structure [206–212]. *R*RhSi (*R*=Y, Gd-Er) compounds form in the TiNiSi type orthorhombic crystal structure and LaRhSi crystallizes in a cubic structure [213–215]. Crystallographic results show that *R*RhGe (*R*=Ce-Nd, Sm, Gd-Yb, Y) compounds crystallize in TiNiSi type orthorhombic crystal structure [216–221].

*R*RhSn (*R*=La-Nd, Sm, Gd-Lu) compounds crystallize in ZrNiAl type hexagonal structure [178,222–226]. GdRhSn in this series was found to show a first order iso-structural transition around 245 K as shown in Fig. 16 [178]. The high temperature phase shows a decease in *a* lattice parameter and increase in *c* lattice parameter, resulting in a sharp change in *c/a* ratio at the transition temperature.

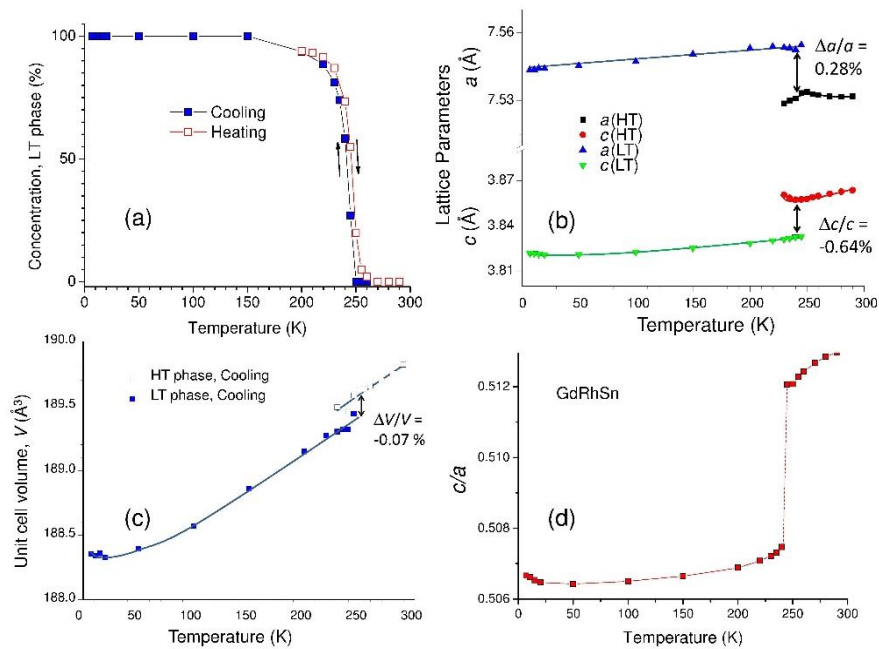

**FIG. 16**. The temperature dependence of (a) low temperature phase concentration, (b) lattice parameters (b) volume of unit cell (d) *c/a* ratio obtained from Rietveld refinement of XRD data for GdRhSn.  Reproduced with permission from [178] © 2014, Elsevier B. V.

*R*RhSb (*R*=La, Ce, Pr, Sm, Gd-Tm) compounds crystallize in the TiNiSi orthorhombic crystal structure [227–229]. *R*RhBi (*R*= La, Ce) compounds crystallize in TiNiSi type orthorhombic crystal structure [230].

Dong *et al.* [231] reported the non-magnetic nature of CeRhAl down to 2 K and a superconducting transition at 2.4 K in LaRhAl [232]. Ce in CeRhAl shows mixed valent character, confirmed by XRD and susceptibility data whereas La in LaRhAl remains in trivalent state [231]. Later, CeRhAl was reported to show AFM ordering below 3.8 K [233]. Compounds with $R=$ Pr, Nd and Gd order ferromagnetically below their magnetic transition temperatures [232]. All FM compounds show positive $\theta_p$ except in PrRhAl. It has been observed that YRhAl shows superconductivity at 0.9 K despite of a different crystal structure from LaRhAl, which is also a superconductor [234]. These results suggest that crystal structures of these compounds are not crucial in observation of superconductivity. It can be noted that the superconductivity in YRhAl is observed in resistivity data but not in magnetization data, due to its lowest temperature limit in the magnetization measurements. Both resistivity and the heat capacity data show down and up jump, respectively at the superconducting transition temperature, which gets suppressed by the application of a magnetic field [234].

Magnetic measurements carried out on $R$RhGa compounds show that CeRhGa is a valence stable compounds for Ce $4f$ shell and does not show any magnetic ordering, NdRhGa and GdRhGa are FMs, whereas compounds with $R=$ Tb, Ho, Er and Tm are AFMs below their magnetic transition temperatures [205,235–237]. Neutron diffraction measurements reveal that compounds with $R=$ Tb, Ho, Er exhibit collinear magnetic structures [205]. CeRhIn compound shows valence fluctuations and a weak temperature dependence of the magnetic susceptibility at higher temperature [238]. EuRhIn and GdRhIn order ferromagnetically below 22 and 34 K, respectively with Eu showing a divalent nature [211,212].

LaRhSi crystallizes in a cubic crystal structure unlike other compounds of this series and shows superconductivity at 4.3 K [215]. Chevalier *et al.* [214] studied structural and magnetic properties of $R$RhSi ($R=$ Y, Gd-Er) compounds and found that compounds with $R=$ Gd, Tb and Dy order ferromagnetically and $R=$ Ho and Er order antiferromagnetically below their magnetic transition temperatures. Magnetometric and neutron diffraction measurements show that $R$RhSi compounds with $R=$ Tb- Ho show AFM ordering with HoRhSi characterized by a stable commensurate magnetic structure whereas TbRhSi and DyRhSi show a change in magnetic structures below $T_N$ [239,240]. Bazela *et al.* [240] reported a double flat spiral structure in TbRhSi and ErRhSi whereas a collinear magnetic structure with two directions of magnetic moments in HoRhSi. Recent studies of the magnetic properties show AFM ordering at low temperatures for $R$RhSi ($R=$ Tb, Dy, Ho) with field induced metamagnetic transitions in DyRhSi and HoRhSi compounds (Fig. 17) [241]. No significant moment contribution was observed from the Rh atom, which is also supported by theoretical calculations [241].

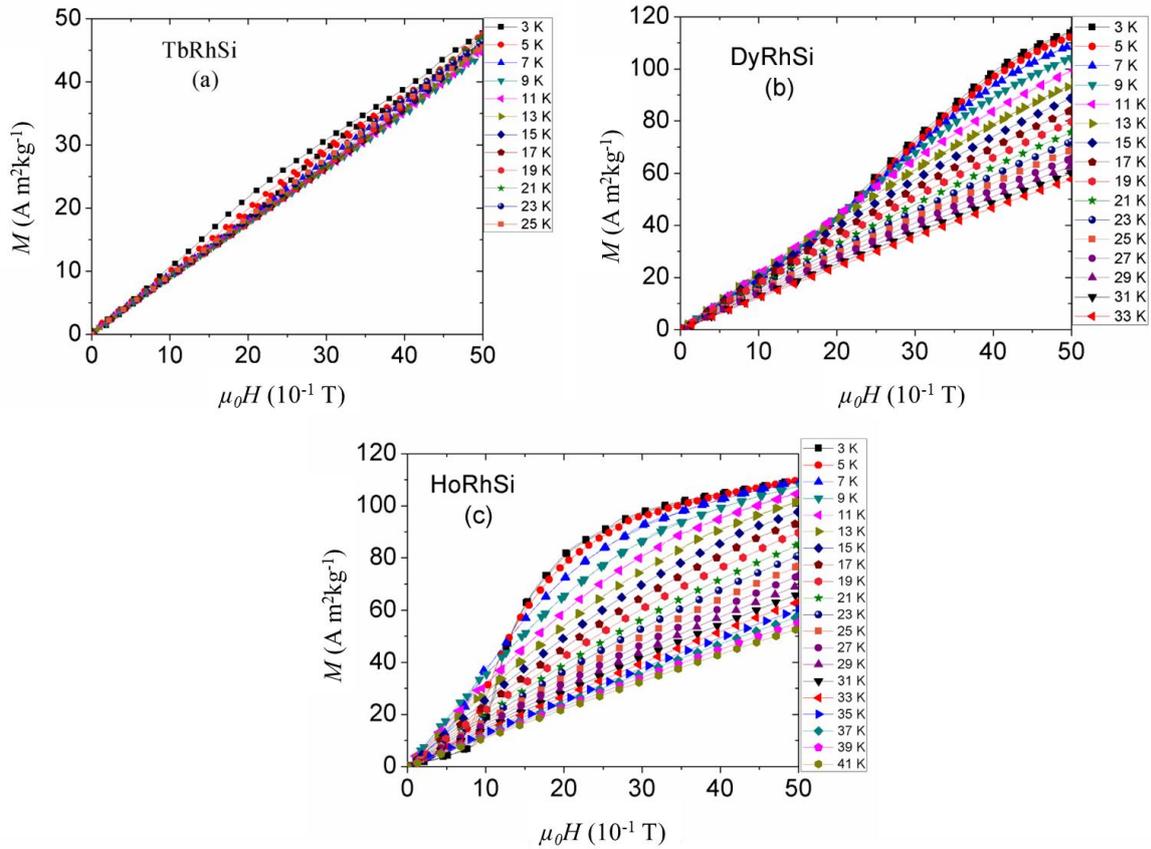

**FIG. 17.** The field dependence of magnetization as a function of temperature for $R$RhSi (Tb, Dy, Ho) compounds. Reproduced with permission from [241] © 2021, Elsevier B. V.

Single crystal CeRhGe shows magnetic easy axis along the $a$-axis with large magnetic anisotropy associated with orthorhombic crystal structure and an incommensurate antiferromagnetic structure derived from neutron diffraction [242]. Bazela et al. [243] studied structural, magnetometric and neutron diffraction measurements of CeRhGe and NdRhGe polycrystalline samples. Both the compounds were found antiferromagnetically ordered below their respective $T_N$ with CeRhGe characterized by a collinear (C mode) magnetic structure and NdRhGe is described by a wavevector, $k = (0.5, 0, 0.5)$ with their magnetic moments aligned along the $b$-axis [243]. Magnetization measurements show that GdRhGe show multiple magnetic transitions; at 31.8 ($T_1$) & 24 K ($T_2$) [244]. The peak at $T_2$ becomes sharp on the application of a magnetic field (2 T) while the transition temperature corresponding to $T_1$ decreases with increasing the magnetic field, indicating that at $T_1$ there is the AFM transition [244]. The value of $\mu_{eff}$ (8.5 $\mu_B$), estimated from susceptibility data is higher than the expected $Gd^{3+}$ free ion value (7.4 $\mu_B$) in their respective compound, suggesting either contribution from Rh and/or due to strong polarization of conduction electrons by Gd [244,245]. The ACS measurements also suggest that the magnetic transition at $T_1$ is AFM while the magnetic transition at $T_2$ is a complex one [244]. Neutron diffraction and magnetic measurements show AFM ordering characterized by collinear magnetic structure at 1.6 K in HoRhGe and ErRhGe, which with increasing temperature changes to an incommensurate sine modulate structure for ErRhGe at 5 K [219,246]. SmRhGe was reported to show FM ordering followed by a metamagnetic transition at low temperatures [218].

Reports suggest that TiNiSi type orthorhombic YbRhGe is an AFM with $T_N$= 7 K, whereas YbRhGe with LiGaGe- hexagonal structure show FM ordering [217]. Magnetic and neutron diffraction measurements show AFM ordering in DyRhGe and TmRhGe [247]. TmRhGe is characterized by a collinear magnetic structure with the magnetic cell doubled along the $b$ axis with respect to crystal unit cell. Magnetic measurements carried out on $R$RhGe ($R$= Tb, Dy, Er, Tm) show that all the compounds undergo AFM ordering with some of them showing spin reorientation at low temperatures [248]. $\mu_{eff}$ estimated from the magnetic susceptibility data is close to their respective rare earth free ion and theoretical calculated value [248,249]. The negative values of $\theta_p$ indicate AFM ordering in these compounds. TbRhGe among these compounds has the highest magnetic transition temperature similar to TbRhSn in RRhSn series reflecting the similarities between them. Slight TMI was observed in TbRhGe at low temperatures and field, which disappears at the field of 2T as can be seen from Fig. 18. Moreover, the compounds with $R$ = Tb, Dy and Er show an upturn in magnetic susceptibility data at low temperatures, which may arise due to spin-reorientation and have been discussed further by performing ACS measurements as shown in Fig. 19 [248]. ACS measurements show a single peak followed by an upturn at low temperatures in $\chi'(T)$ data. This single peak in all the compounds coincides with the AFM transition temperatures determined from the DCS measurements. Absence of a peak in $\chi''(T)$ corresponding to an AFM transition is another confirmation of collinear antiferromagnetic nature of these compounds [248]. In case of DyRhGe and ErRhGe the trend is different. The $\chi''(T)$ shows small peaks at low temperatures, reflecting energy losses in the magnetic systems, which may arise from extra or intradomain effects in ferro, ferri, spin glass or canted systems [248,250]. Absence of a peak in $\chi''(T)$ of TbRhGe may possibly be due to the weak canting of the moments.

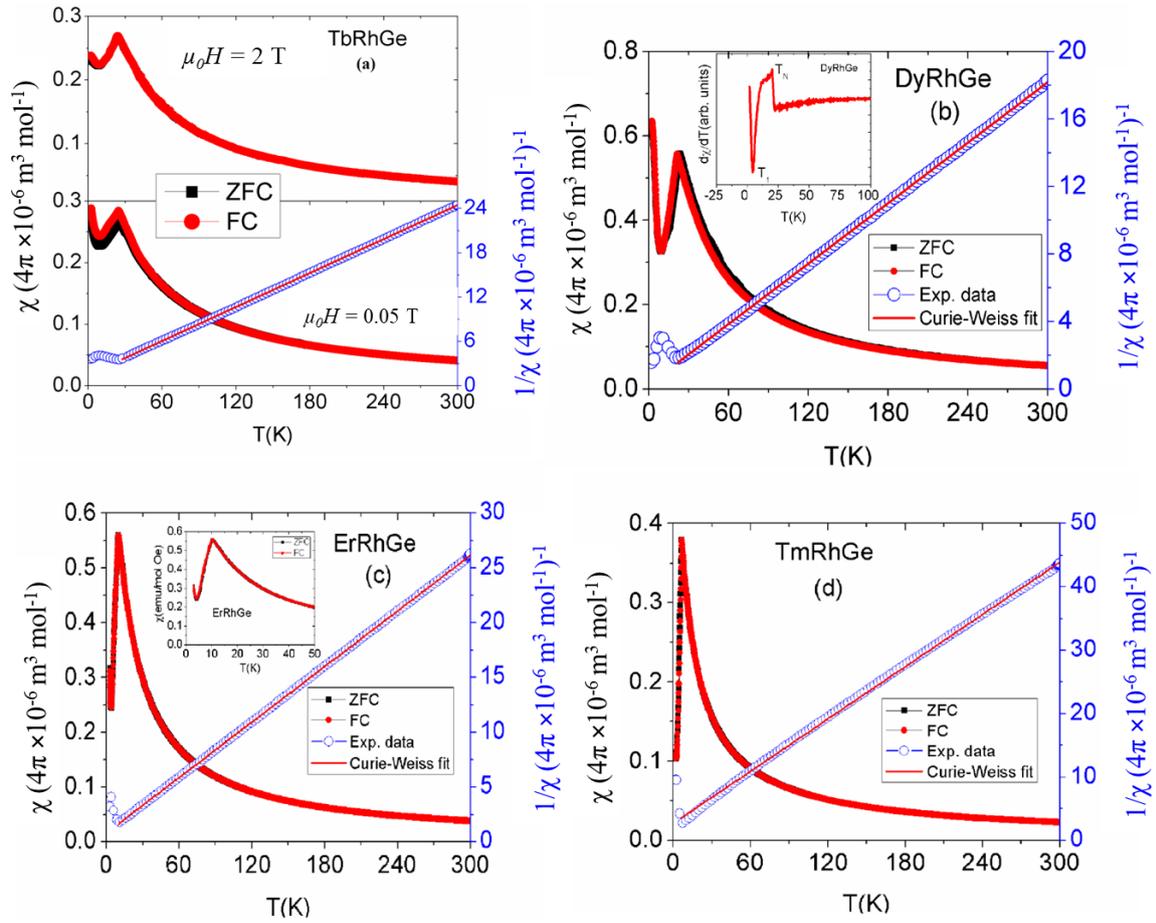

**FIG. 18**. The temperature dependence of the magnetic susceptibility and inverse magnetic susceptibility for $R$RhGe ($R$= Tb, Dy, Er, Tm) compounds along with Curie-Weiss fit. Reproduced with permission from [248] © 2015, Elsevier B. V.

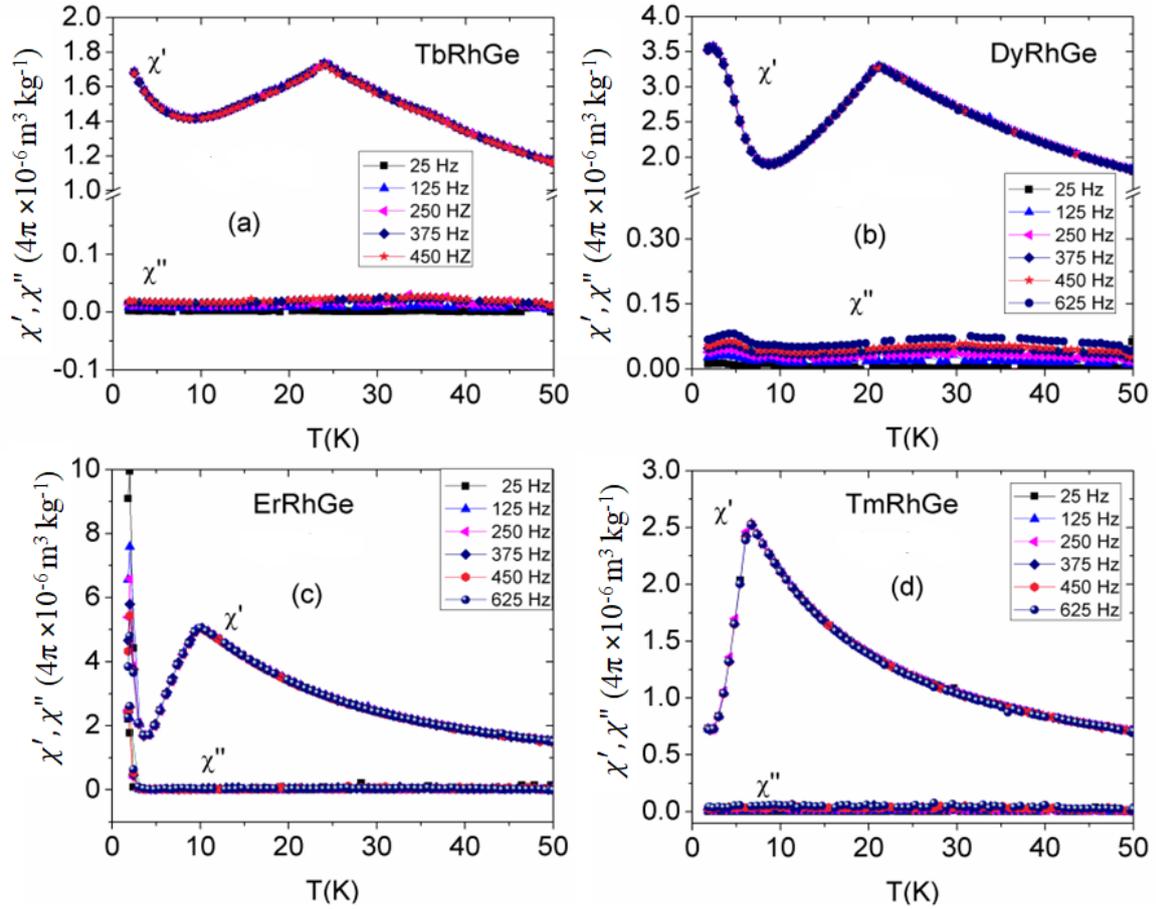

**FIG. 19**. The temperature dependence of the ac magnetic susceptibility as a function of frequency and constant AC field of $10^{-4}$ T for $R$RhGe ($R$= Tb, Dy, Er, Tm) compounds. Reproduced with permission from [248] © 2015, Elsevier B. V.

Magnetic and Mössbauer ([119]Sn and [155]Gd sources) studies reveal that LaRhSn is non-magnetic, CeRhSn is PM down to 2 K, PrRhSn and NRhSn are FMs while GdRhSn is AFM [222]. Tin (Sn) atoms do not possess any moment so there is no splitting for Sn spectra recorded for LaRhSn. The observed splitting in Sn spectra due to magnetic hyperfine field at tin nucleus is induced solely by rare earth magnetic moments and hence it can be a good measure of net magnetic moment in the vicinity of tin [222,251]. Gupta *et al.* [178,185] performed magnetic measurements for $R$RhSn ($R$= Gd-Tm) compounds and observed that all these compounds (except HoRhSn) show AFM ordering at low temperatures. TbRhSn and DyRhSn were found to show multiple magnetic transitions and field induced metamagnetic transitions. GdRhSn compound shows first order iso-structural transition around 240 K, which was evidenced in temperature dependence XRD, magnetic, and electrical resistivity data [178]. SmRhSn shows FM ordering below 14.9 K with additional transition at low temperatures [225]. [119]Sn

Mössbauer spectra hint some complex magnetic structure in this compound. YbRhSn was found to show AFM ordering below 2 K [252]. Recently single crystal LaRhSn has been reported to show nodeless superconductivity probed by the specific heat, muon spin relaxation and tunnel diode oscillator techniques [226]. Non-centrosymmetric materials such as LaRhSn induces antisymmetric spin-orbit coupling which helps to lift two-fold spin degeneracy of the electronic bands enabling the material to host unconventional superconducting properties [226]. Fig. 20 shows the signature of superconductivity in LaRhSn as evidenced by a sharp jump in the temperature dependence of the electrical resistivity and electronic heat capacity at the same temperature.

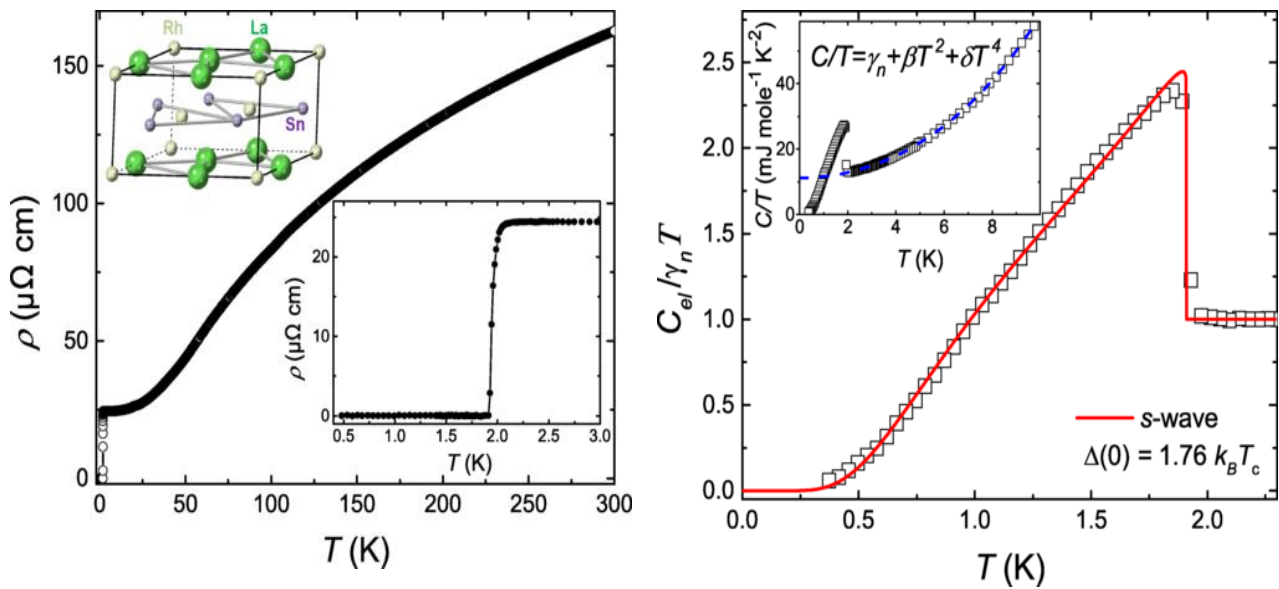

**FIG. 20.** The temperature dependence of the electrical resistivity and the electronic heat capacity for LaRhSn compound. Reproduced with permission from [226] © 2022, American Physical Society.

Malik *et al.* [228] reported LaRhSb as a superconductor below 2.1 K and PrRhSb to be AFM at 18 K with a spin glass transition at 6 K. Later, superconductivity at 2 K was reported in single crystal LaRhSb [253]. LaRhBi was reported to show superconducting nature below 2.4 K while CeRhBi an AFM Kondo coupling at low temperatures [230].

## 2.10 *R*Pd*X* compounds

Compounds of *R*PdAl show two types of crystal structures; TiNiSi type orthorhombic and ZrNiAl type hexagonal. High temperature forms of these materials adopt ZrNiAl type hexagonal crystal structure while at low temperature TiNiSi type orthorhombic structure is preferred [254]. GdPdAl undergoes an iso-structural hexagonal transformation at 180 K [255]. On cooling there is an increase in $c$ parameter while decrease in $a$ parameter. $R$PdGa ($R$=La-Nd, Sm, Eu, Gd-Tm, Lu, Y) compounds crystallize in TiNiSi type orthorhombic crystal structure [168,213,256]. $R$PdIn ($R$=La-Sm, Gd-Lu, Y) compounds adopt ZrNiAl type hexagonal crystal structure whereas EuPdIn crystallizes in TiNiSi type orthorhombic crystal structure [168,257–263]. $R$PdSi ($R$=La-Nd, Sm-Lu, Y) compounds were found to crystallize in two different superstructures of the orthorhombic unit cell with space groups *Pnma* and *Pmmn* [213,264,265]. $R$PdGe ($R$=La-Nd, Sm, Gd-Tm, Y) compounds crystallize in the CeCu$_2$ type orthorhombic structure [266–269]. As cast $R$PdSn ($R$= La, Ce-Nd, Sm-Yb) compounds crystallize in the TiNiSi type orthorhombic crystal structure, however, some of these compounds on annealing at 950˚C transform into Fe$_2$P type hexagonal crystal structure [270,271]. Compounds of $R$PdSb show formation into multiple crystal structures; compounds with $R$=Ce-Nd, Sm, Gd-Dy crystallize in the CaIn$_2$ hexagonal crystal structure, EuPdSb crystallizes in TiNiSi type orthorhombic crystal structure and compounds with $R$=Ho, Er, Tm crystallize in MgAgAs type cubic crystal structure [272,273]. $R$PdBi (R=La-Nd, Sm Gd-Lu) compounds crystallize in MgAgAs type cubic crystal structure [274–276].

Geometrical frustration in materials can lead to many interesting properties (such as partially ordered states) at low temperatures because of existence of energetically degenerate states [277]. Coexistence of localized moments instability and geometrical frustration near the magnetic phase transition may lead to complicated magnetic structure including partially ordered state [278]. 4$f$ electron compounds exhibiting Kondo effect (e.g. Ce and Yb compounds) are examples of instability of magnetic moments, because there is an competition between the Kondo effect, which tends to cancel out moments and RKKY interaction, which tend to order moments. Some of compounds of $R$PdAl series show geometrical frustration of rare earth moments due to triangular symmetry of rare earth atoms characteristic for ZrNiAl type structure. CePdAl was found to show AFM heavy fermion character at 2.7 K. Neutron diffraction studies reveal that two-thirds of Ce moment order antiferromagnetically below 2.7 K, characterized by a propagation vector $\mathbf{k}$= (0.5,0,0.35) and one-third remains paramagnetic [279,280] and thus there is coexistence of magnetically ordered and frustrated Ce moments. NMR studies suggest spin fluctuation in this compound below 2.7 K [280]. Very recently, Lucas *et al.* [281] studied entropy evaluation in the magnetic phases of CePdAl by applying magnetic fields to supress Kondo screening. The authors estimated the frustration using the magnetic entropy as a frustration parameter – when the Kondo screening is suppressed, it maximizes the magnetic entropy and enhances the frustration. The obtained results can be promising in the search of quantum spin liquids [281]. Similar to CePdAl, recently, PrPdAl has also been reported to show frustrated antiferromagnetism and heavy-fermion like behaviour [282]. AFM character is described by two-thirds of fully ordered Pr moments and one-third with strongly frustrated and reduced moments. Fig. 21 shows the field dependence of magnetization along *a* and *c* axis. It can be noted that the magnetization easy axis is along the *c*-axis and the S shape behaviour along *a* axis exhibits a metamagnetic transition, indicating that ordered Pr moments lie in the basal plane and hence reflecting the important effect of frustration [282].

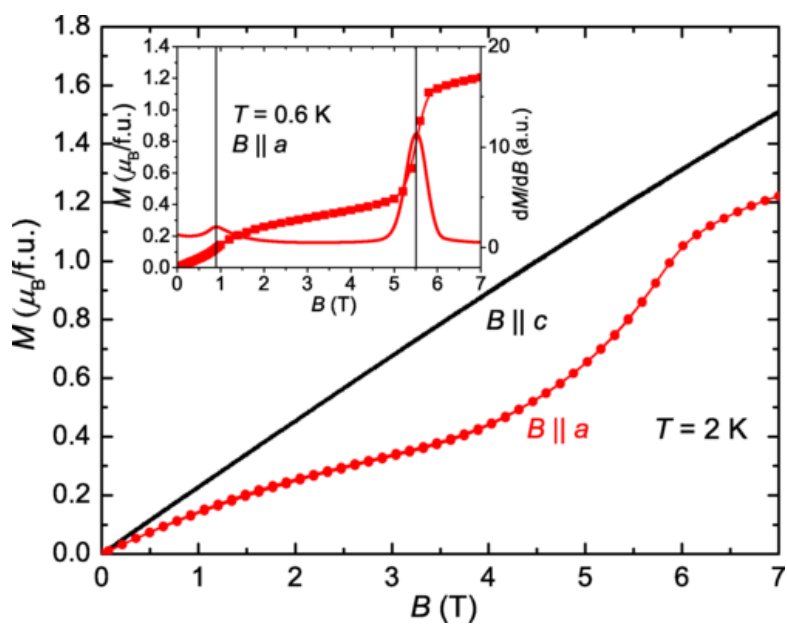

**FIG. 21.** The field dependence of magnetization for PrPdAl compound at 2 K. The field is applied along *a* and *c* axis. Reproduced with permission from [282] © 2022, American Physical Society.

Hexagonal NdPdAl and TbPdAl undergo AFM transition while DyPdAl and HoPdAl order ferromagnetically at low temperatures [254,283–287]. Talik *et al.* reported magnetic properties of single crystal HoPdAl and found that the compound shows AFM character along the *a*-axis and FM character along the *c*-axis [287]. NdPdAl shows a second magnetic transition below $T_N$. Single crystal GdPdAl show two magnetic transitions; at 48 and 20 K and a slope change at 180 K due to an iso-structural transition.

Magnetometric and neutron diffraction measurements show that $R$PdGa compounds with $R=$ Gd-Er show AFM ordering at low temperatures except DyPdGa, which is a FM below 20 K [256]. Compounds which order antiferromagnetically show a change in their magnetic structures with temperature, associated with additional magnetic transitions at low temperatures [256]. EuPdGa shows a FM transition at low temperatures [168].

Similar to LaRhSn [288], fully gapped superconductivity below 1.6 K was also observed in non-centrosymmetric LaPdIn compound [289]. CePdIn shows an AFM behaviour with $T_N$ below 1.7 K [257,290]. The value of $\mu_{eff}$ (2.58 $\mu_B$) was determined to be close to the Ce$^{3+}$ free ion value and the $\theta_p$ (-52 K) is negative and has the magnitude larger than the $T_N$, hinting at strong Kondo interactions in this compound. The heat capacity data suggest heavy fermion nature in this compound. PrPdIn does not show any ordering down to 1.7 K, which may result from strong hybridization between Pr 4$f$ and Pd 4$d$ electron states and the presence of a singlet as the crystalline electric field ground state [257]. In contrast to these, NdPdIn was reported to be FM below 26 K. Later Li $et$ $al.$ [291] studied PrPdIn and NdPdIn compounds and reported that both the compounds order ferromagnetically at 11.2 and 34.3 K, respectively, which is in contrast with a previous report [257], in which PrPdIn does not show any magnetic ordering and FM ordering in NdPdIn was reported at a lower Curie temperature. Gondek $et$ $al.$ studied magnetic properties of $R$PdIn compounds with $R=$ Nd, Ho, Er by means of neutron diffraction measurements [258]. It has been observed that all the compounds order ferromagnetically at low temperatures with a change in the orientation of Nd moments from basal plane to along the $c$-axis. Balanda $et$ $al.$ [259] reported AC and DC magnetic susceptibility studies on $R$PdIn compounds with $R=$ Gd- Er. The authors observed that in contrast to GdPdIn, which orders ferromagnetically, other compounds show complex ferrimagnetic behaviour with additional magnetic transitions below their Curie temperatures [259]. The analysis shows that ferromagnetic interactions are dominant at higher temperatures whereas at low temperatures, AFM interactions play a role. Li $et$ $al.$ [288] reported FM ordering in TbPdIn and DyPdIn compounds. A second magnetic phase transition at low temperatures was reported in DyPdIn due to antiferromagnetic coupling of moments. It was observed that unlike other members of $R$PdIn compounds, GdPdIn and TmPdIn (which shows AFM transition) were found to show single phase magnetic transitions [260]. Single crystal SmPdIn was found to show FM ordering at 54 K with magnetization easy direction along the $a$-axis [261].

Magnetic measurements show LaPdSi is non-magnetic down to low temperatures while CePdSi and PrPdSi show FM nature [292]. The value of $\mu_{eff}$ calculated from magnetic susceptibility data is close to the expected $Ce^{3+}$ and $Pr^{3+}$ free ion value. The non-magnetic behaviour of LaPdSi and the value of $\mu_{eff}$ in case of CePdSi and NdPdSi indicate that Pd does not carry moment in these compounds. Magnetic properties of $R$PdSi ($R$ = Y, Gd–Er) compounds, described as half-Heusler alloys reveal AFM nature of all the compounds except YPdSi, which is a PPM [264]. All magnetically ordered materials show field induced metamagnetism. YbPdSi was found to show FM heavy fermion behaviour below 8 K, indicating that Kondo and RKKY interactions are competing in this compound [293].

$R$PdGe compounds with $R$= Ce, Pr and Tb were studied by means of magnetometric and neutron diffraction measurements. PrPdGe does not show any magnetic ordering down to 2 K, CePdGe orders antiferromagnetically at $T_N$= 3.4 K, characterized by a sine modulated magnetic structure at 2K while TbPdGe undergoes AFM ordering below 35 K, with non-collinear magnetic structure [266]. $R$PdGe compounds with $R$= Gd, Dy, Er are AFMs while HoPdGe is a PM [267]. DyPdGe is characterized by non-collinear magnetic structure which is stable with temperature while ErPdGe shows a complicated magnetic structure which changes with increasing temperature [267]. Kotsanidis *et al.* [269] reported magnetic properties of $R$PdGe ($R$= Y, Pr, Nd, Tb, Dy, Er) compounds. The authors observed from the magnetic measurements results that except YPdGe, all compounds exhibit AFM ordering below their magnetic transition temperatures. At 1.8 K, NdPdGe is described by a sine wave modulated magnetic structure with propagation vector, $\boldsymbol{k}$ = (0.0, 0.0, 0.01). The magnetic measurements reveal FM character of YbPdGe compound below 11.4 K [294].

Magnetic measurements were carried out on orthorhombic $R$PdSn ($R$= Ce-Yb) compounds, which reveal that compounds with $R$= Ce, Sm, Eu, Gd, Tb, Dy and Er order antiferromagnetially below their respective $T_N$ and compounds with $R$= Pr, Nd, Ho and Tm are PM down to 4.2 K [270]. YbPdSn follows Curie-Weiss law at temperatures higher than 150 K and deviates below that. The smaller value of $\mu_{eff}$ than the expected value suggests instability of Yb moments or the divalent state. The magnetic transition temperatures in these compounds do not follow de Gennes scaling, which can be explained on the basis of CEF [270]. Later Zygmunt and Szytula [295] studied magnetic properties of $R$PdSn ($R$= Ce-Nd, Tb-Tm) and observed that except TmPdSn, all the compounds show magnetic ordering at low temperatures with negative paramagnetic Curie temperatures. Neutron diffraction measurements reveal that AFM PrPdSn and NdPdSn exhibit collinear and sine modulated magnetic structures, respectively while DyPdSn shows highly anisotropic magnetic properties with its easy magnetization axis along *b*-axis, hard axis along *a*-axis which changes to *c*-axis below 30 K [296,297].

Magnetic properties studies of *R*PdSb (*R*= La – Tm) compounds reveal that CePdSb undergoes FM ordering, NdPdSb, SmPdSb, EuPdSb and GdPdSb order antiferromagnetically while other compounds are PMs within the investigated temperature range [272,298]. Similar to other compounds in the family, magnetic transition temperatures do not follow de Gennes scaling. Mukhopadhyay *et al.* [299,300] studied magnetic properties of rare earth based Heusler alloys; *R*PdSb (*R*= Gd, Tb, Er, Ho). Magnetic measurements show all compounds except ErPdSb show AFM ordering at low temperature whereas ErPdSb does not show any magnetic ordering down to 1.8 K. The linear magnetoresistance behaviour in ErPdSb resembles that of the Weyl semimetals such as TaAs and NbAs [301].

Rare earth compounds containing bismuthides *R*PdBi with Pd/Pt are an excellent group of half Heusler materials as some of them have been reported to show coexisting magnetic states and non-trivial topological behaviour. Combinations of intriguing physical properties could make these materials promising for quantum computation [302,303]. Some of these materials show unconventional superconductivity characterized by mixed spin singlet/triplet pairing symmetry [304].

Nowak *et al.* reported [305] band inversion in non-trivial topological half Heusler LuPdBi, LuPtBi and YPtBi single crystals by means of [209]Bi NMR measurements. NMR as a local probe is sensitive to the electronic bands near the Fermi level. As in trivial topological systems, the conduction band possess predominantly *s*-character while in non-trivial topological systems, due to band inversion, conduction band is characterised by *p* or *d* symmetry [305]. Thus in NMR spectroscopy the positive frequency shifts reflects Bi-*s* interactions whereas the negative one suggests the chemical shielding contribution rather than Bi-*p* core polarization [305]. It has been observed that [209]Bi NMR frequency shifts are positive for topological trivial ScPdBi and YPdBi compounds whereas strongly negative for topologically non-trivial LuPdBi, LuPtBi and YPtBi associated with band inversion near the Fermi level [305,306]. These results create a new tool for studying band inversion in topological insulators by means of NMR spectroscopy [305].

Pollycrystalline CePdBi reported to show an AFM transition below 2 K and onset of superconductivity below 1.3 K [307]. The superconducting phase in CePdBi is possibly originated from atomic disorder, which is a typical feature of Heusler alloys. Single crystal half Heusler $R$PdBi ($R$= Y, Gd- Er) were studied for their magnetic and transport properties [274,275,308]. HoPdBi and LuPdBi were found to show superconductivity below 0.7 and 1.8 K, whereas compounds with $R$= Gd- Er undergo AFM ordering below $T_N$= 12.8, 5.3, 3.7, 1.9 and 1.2 K, respectively. It is interesting to note that HoPdBi shows coexistence of superconducting and AFM states, making it a very intriguing material [274,309]. Neutron diffraction measurements reveal that the magnetic structure in TbPdBi can be described by $k$ = (0.5,0.5,0.5) which is in agreement with the theory developed for AFM topological insulators [275]. The real part of ACS in ErPdBi becomes negative below 1.6 K and corresponding to this the imaginary part of ACS exhibit a maximum, suggesting the onset of superconducting state in this compound [308,310]. On the application of a magnetic field, TbPdBi shows Shubnikov-de Haas oscillations fitted with the parameters similar to archetypal topological insulators [308]. YPdBi and TbPdBi were reported to show unconventional superconductivity at ~1 and 1.7 K, respectively with an additional AFM transition at 5.5 K in TbPdBi half Heusler [311–313]. Polycrystalline sample of DyPdBi exhibits AFM ordering at $T_N$ = 4 K, with field induced spin flip transitions below $T_N$ [314] whereas thin film of DyPdBi (110) grown by using pulsed laser deposition gives topologically non-trivial signature evidenced by weak antilocalization effect and Shubnikov-de Hass quantum oscillations [315,316]. Recently, Radmanesh *et al.* [313] reported non-trivial superconducting paired states in half Heusler TbPdBi single crystal using the temperature dependence of the magnetic penetration depth analysis down to 40 mK. It has been observed that TbPdBi shows $T^3$ power law behaviour of penetration depth confirming the nodeless topological superconductivity characterized by an anisotropic gap structure.

## 2.11 $R$Ag$X$ compounds

$R$AgAl ($R$=La-Nd, Sm, Gd- Er, Y) [317,318] and $R$AgGa ($R$=La- Nd, Eu, Gd-Lu, Y) compounds crystallize in CeCu$_2$ type orthorhombic structure while YbAgAl crystallizes in the MgZn$_2$ type structure [319]. $R$AgSi ($R$=Sm, Gd- Tm, Yb, Lu, Y) compounds crystallize in ZrNiAl type hexagonal crystal structure [320,321]. RAgGe compounds adopt three type of crystal structures; $R$AgGe ($R$=La, Ce) [322] and RAgGe (R= Gd-Lu) compounds [323] crystalize in CaIn$_2$ type and ZrNiAl type hexagonal crystal structures, respectively while EuAgGe crystallizes in CeCu$_2$ orthorhombic structure [324]. Similar to $R$AgGe, $R$AgSn compounds crystallize in three crystal structures; $R$AgSn ($R$=La, Ce, Sm, Gd-Er, Yb) [325] and $R$AgSn ($R$=Ce-Nd, Gd-Er) compounds [326,327] crystallize in CaIn$_2$ type and LiGaGe type hexagonal crystal structures, respectively while EuAgSn forms in CeCu$_2$ orthorhombic structure [189].

$R$AgAl ($R$=La, Ce, Pr, Nd, Tb and Er) compounds have been studied for their structural and magnetic properties [328]. Except LaAgAl and CeAgAl, all the compounds show long range as well as short range magnetic ordering, possibly due to the magnetic frustration arising from the random distribution of non-magnetic atoms. Magnetically ordered compounds show strong TMI indicating complex magnetic states in these compounds. The paramagnetic Curie temperature was found to be negative for Ce and Pr compounds, while it is positive for the others, suggesting the predominant FM character [328]. Later Slebarski *et al.* reported that CeAgAl exhibits a signature of weak FM with its $\mu_{eff}$ close to the expected free Ce ion. [329]. GdAgAl was reported to show spin glass behaviour with strong TMI and slightly higher value of $\mu_{eff}$ than the expected free Gd$^{3+}$ ion value [330]. $R$AgAl ($R$ =Dy, Ho, Er) compounds show AFM ordering at low temperatures [331]. The magnetic susceptibility shows strong TMI in these compounds, suggesting spin-glass signature. The estimated value of $\mu_{eff}$ from the magnetic susceptibility data fit found to be slightly higher than the expected for their respective free rare earth ion values. The excess in the $\mu_{eff}$ values may arise due to polarisation of 4d Ag electrons. The positive values of paramagnetic Curie temperature in these compounds suggest predominant FM character [331]. Almost all the compounds in $R$AgAl were found to show TMI and coexistence of AFM and FM interactions, TmAgAl show different behaviour. TmAgAl exhibits single phase FM transition with no TMI [332].

CeAgGa shows FM ordering with the signature of a spin glass phase [333,334]. The magnetic properties of $R$AgGa compounds reveal that the compounds with $R$= Gd-Tm compounds order magnetically at low temperatures; Gd, Ho and Er compounds order ferromagnetically, Tb orders antiferromagnetically while Dy compound show metamagnetic behaviour [335–337]. Zygmunt *et al.* [338] reported magnetic properties of $R$AgGa with $R$ = Tb, Dy, Ho compounds, showing spin glass behaviour in TbAgGa and FM ordering in DyAgGa and HoAgGa compounds. Recently, França *et al.* [339] studied detailed structural and magnetic properties of of GdAgGa compounds. The authors report FM ordering at 30 K and a field dependent anomaly below 14 K in GdAgGa compound. Time dependence of magnetization relaxation, the heat capacity, ACS and TMI results indicate the coexistence of re-entrant spin-glass (RSG) phase with long range FM ordering. Similar to $R$AgAl compounds, the RSG behaviour in GdAgGa arises due to a random distribution of non-magnetic atoms resulting in disorder, inducing the non-uniform FM phase [339]. Fig. 22 shows the temperature dependence of the magnetic susceptibility for GdAgGa, where both the FM and field dependent transition can be clearly seen from the temperature derivative of the magnetic susceptibility. Fig. 24 (c) shows the time dependence of the remanent magnetization measured after cooling the sample from room temperature to 5 K followed by application of magnetic field and switching it off before recording the data. The experimental data was fitted by a stretched exponential equation and different parameters were determined, indicating coexistence of SG and FM components [339].

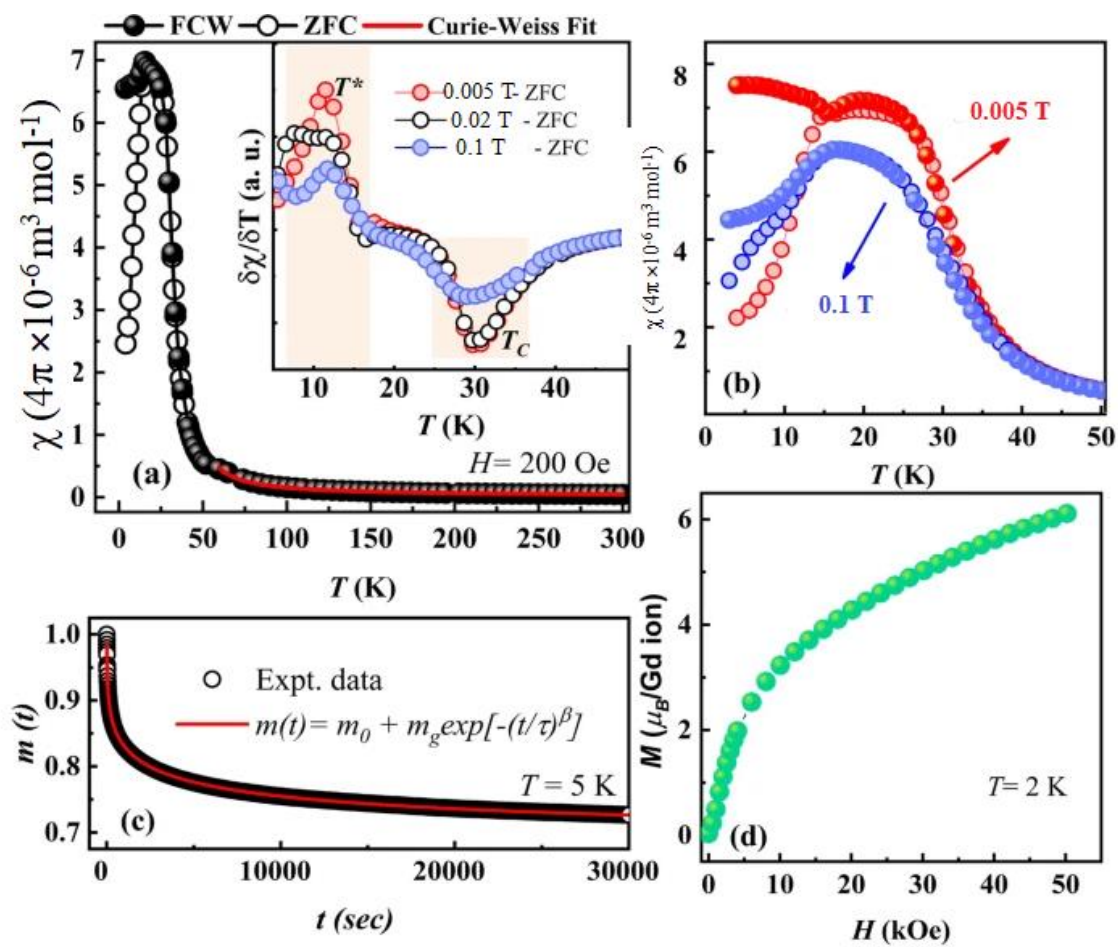

**FIG. 22.** (a) The temperature dependence of magnetic susceptibility for GdAgGa (inset show the derivative) (b) magnetic susceptibility at smaller temperature regime. (c) the time dependence of normalized remanent magnetization at 5 K (d) the field dependence of magnetization at 2 K. Reproduced with permission from [339] © 2023, Springer Nature.

Magnetometric and neutron diffraction measurements show that $R$AgSi ($R$=Gd-Tm) compounds order antiferromagnetically at low temperatures with TbAgSi exhibiting additional magnetic transitions at low temperatures [321,340]. The magnetic ordering in TmAgSi is characterised by a non-collinear magnetic structure with moments lying in the basal plane [340]. The temperature dependence of the magnetic susceptibility data in NdAgSi shows FM transition at 44 K with large TMI on decreasing temperature and an additional magnetic transition at 11 K [341].

CeAgGe shows second order AFM ordering with an effective magnetic moment close to the expected trivalent Ce ion [322]. Polycrystalline $R$AgGe ($R$= Gd - Er) compounds were investigated using magnetization and neutron diffraction measurements [342]. GdAgGe and ErAgGe show AFM ordering at 15.6 and 3.6 K, respectively. Neutron diffraction measurements show a complicated temperature dependent sine modulate magnetic structure for TbAgGe and square modulated magnetic structures for DyAgGe and HoAgGe compounds [342]. Magnetic measurements show multiple magnetic transition for TbAgGe and one of DyAgGe compounds along with the field induced metamagnetic transitions. The TMI in these compounds may arise from magnetic domains and motion of domain walls. Single crystal $R$AgGe ($R$=Tb–Lu) compounds show AFM ordering for TbAgGe and YbAgGe at low temperatures [323]. The anisotropic measurements show that more magnetic phase transitions are associated with TbAgGe when the field direction is along the $c$-axis than when it is perpendicular to $c$-axis, compounds with $R$= Ho, Er and Tm show magnetic transitions while DyAgGe shows magnetic phase transition in both the field directions [323]. TmAgGe shows AFM ordering with spins confined to distorted Kagome like planes and magnetization is strongly anisotropic [343,344]. Recently, scientists have experimentally realized Kagome spin ice state in HoAgGe, which has attracted significant attention in the field of condensed matter physics due to its unique properties and potential for studying exotic magnetic phenomena [345,346]. Spin ices are unique states of matter manifested by frustrated spins following local "ice rules," similar to how electric dipoles behave in water ice [346]. In case of a two-dimensional system, these ice rules can be applied to Ising-like spins, sitting on a Kagome lattice within the plane, provided that each triangular unit must possess a single monopole, resulting in different ordering and excitations [346]. Kagome spin ice state has only been experimentally realized in artificial spin ice systems in which ferromagnetic nano rods are arranged in honeycomb structure. Zhao *et al.* [346] demonstrated naturally occurring Kagome spin ice state in HoAgGe in its fully ordered ground state by means of various experiments such as magnetometric, thermodynamic and neutron diffraction and theoretical approaches. Fig. 23 shows structural and magnetic properties of single crystal HoAgGe. The temperature dependence of magnetic susceptibility shows multiple magnetic transitions when the field is applied along *b*- axis and monotonic increase in susceptibility with decrease in temperature for field applied along *c*-axis. The field (*H*) dependence (*H* ∥ *b*) of magnetization shows step like field induced metamagnetic transitions at low temperatures, however no such behaviour was observed when field was along the *c* axis. These results suggests that Ho moments in HoAgGe lie in *ab* plane and possess additional in-plane magnetic anisotropies. The neutron diffraction data (at 10 K) reveal that among three non-equivalent Ho sites viz. $Ho_1$, $Ho_2$ and $Ho_3$; $Ho_1$, and $Ho_2$ exhibit ordered moments with same magnitude (but in opposite direction) whereas $Ho_3$ moment fluctuate without ordering as shown in Fig. 24 (c). Below 7 K, $Ho_3$ moments show long range ordering as shown in inset of Fig. 24 (a). At 4 K temperature, it gets into a fully ordered ground state as shown in Fig. 24(f), which is exactly the same as the classical Kagome spin ice predicted theoretically.

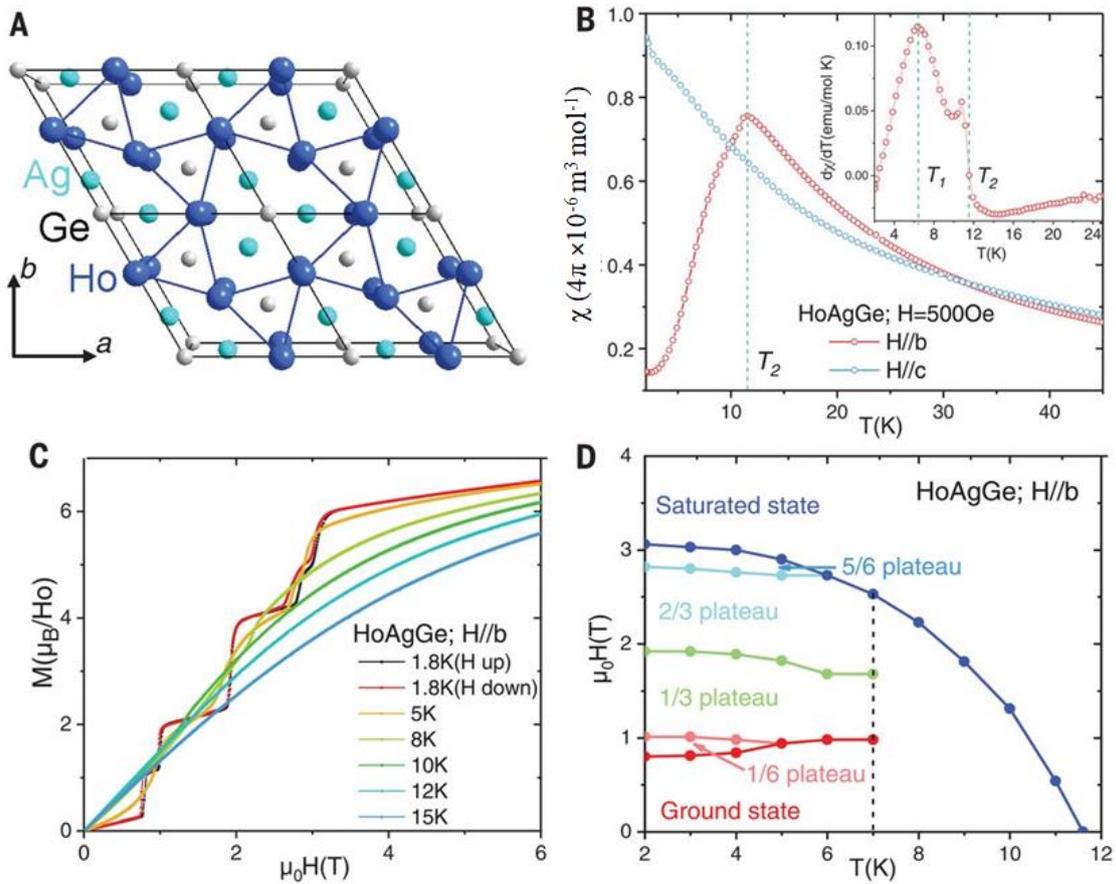

**FIG. 23.** (a) Schematic of HoAgGe crystal structure, (b) the temperature dependence of susceptibility for field of 0.05 T along *b* and *c*-axis, (c) In-plane field dependence of magnetization, (d) the temperature dependence of metamagnetic transitions. Reproduced with permission from [346] © 2020, Science.

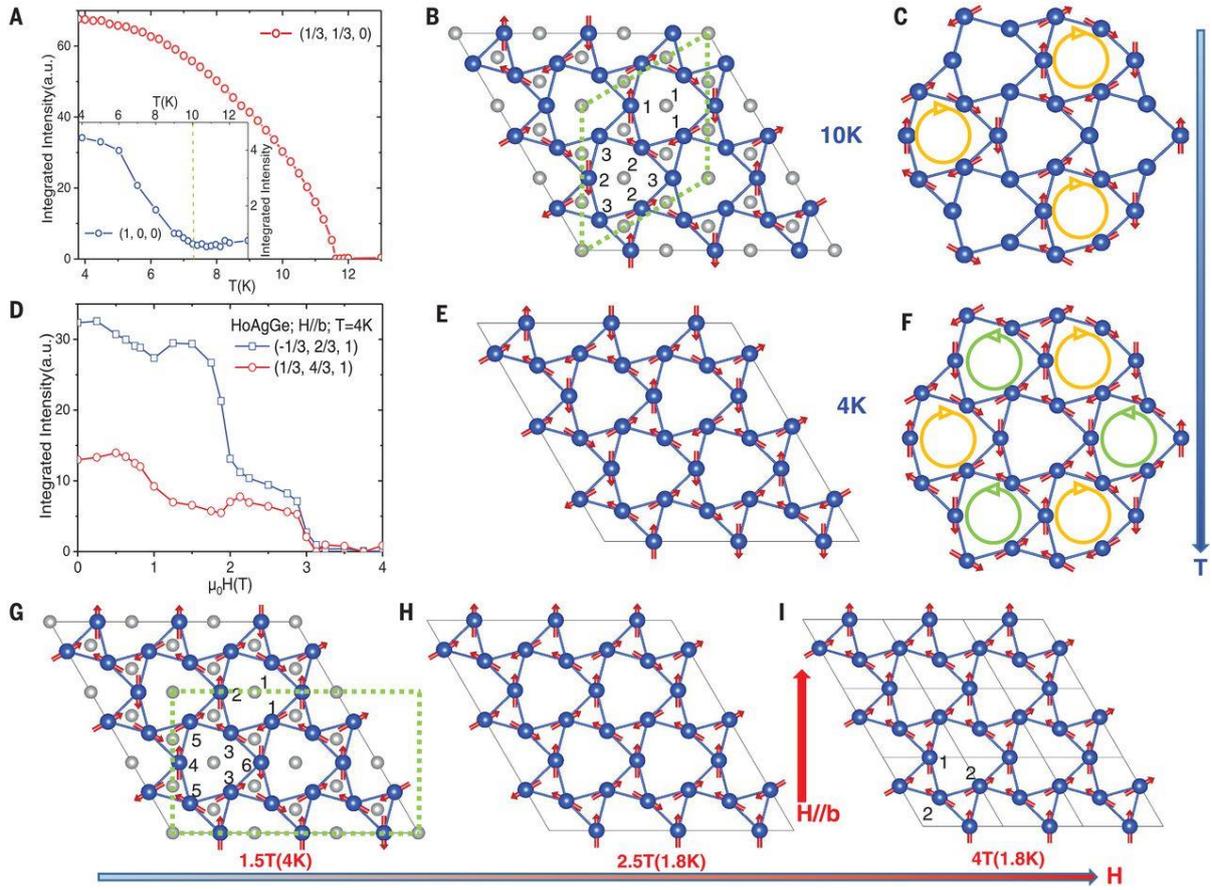

**FIG. 24**. (a) The temperature dependence of the integrated intensities from the neutron diffraction data, (b) refined magnetic structure at 10 K for HoAgGe compound. Three inequivalent Ho sites are represented as $Ho_1$, $Ho_2$ and $Ho_3$ (c) Partially ordered structure of HoAgGe at 10 K, exhibiting counterclockwise hexagons of spins (d) the field dependence of (-1/3, 2/3, 1) and (1/3, 4/3, 1) integrated peaks (e) magnetic structure at 4 K (f) clockwise and counterclockwise hexagons of spins at 4 K, representing Kagome spin ice ground state (g) magnetic structure at $H$= 1.5 T and $T$= 4 K. (H) magnetic structure at $H$= 2.5 T and $T$= 1.8 K (I) magnetic structure at $H$= 4T and $T$= 1.8 K. Reproduced with permission from [346] © 2020, Science.

Guillot *et al.* [336] reported AFM ordering associated with metamagnetic transitions in $R$AgSn with $R$= Nd, Tb, Ho compounds. Neutron diffraction show a collinear AFM structure in these compounds [347]. Baran *et al*. [327] carried out neutron diffraction measurements for $R$AgSn compounds with $R$= Ce-Nd, Gd-Er and observed that at 1.6 K, most of compounds show a collinear AFM structure described by propagation wave vector, $k$ = (0.5,0,0) with magnetic moments aligned normal to the *c*-axis in CeAgSn, at the inclination of angle of 40° to the *c*-axis in DyAgSn and parallel to the *c*-axis in the rest of the studied compounds. At high temperature (below $T_N$) sine modulated magnetic structures were identified in some of these compounds. [155]Gd Mossbauer spectroscopy and neutron diffraction measurements confirm AFM ordering in GdAgSn compound with moments aligned along the hexagonal *c*-axis [348]. The neutron diffraction carried out on TmAgSn sample show an incommensurate magnetic structure with magnetic moments lying in the basal plane and show triangular geometry [349,350].

## 3. Conclusion and outlook

In conclusion, the *RTX* family of compounds exhibits a diverse range of crystal structures, including layered structures that offer opportunities for studying two-dimensional properties through exfoliation. The hexagonal structure prevalent in these magnetic materials introduces frustration coexisting with long-range order, resulting in intriguing phenomena such as spin ices, which hold promise for various applications. Moreover, these materials demonstrate a wide spectrum of magnetic properties, spanning from antiferromagnetism and ferromagnetism to superconductivity. Additionally, some members of this family exhibit non-trivial topological properties, making them potential candidates for quantum computation applications. The broad range of magnetic properties and their phase transitions across different temperature regimes further highlight the potential for diverse applications across a wide temperature range. Moving forward, exploring and harnessing the unique characteristics of the *RTX* compounds hold great promises.

## Acknowledgement
The author would like to thank Prof. K. G. Suresh, IIT Bombay for fruitful discussion on rare earth intermetallics.

## Table 1

List of crystal structures with their structure types and space groups for *RTX* compounds. The table is adopted with permission from [9] © 2015, Elsevier B. V.

| Compound | Structure type | Crystal structure | Space group | References |
|---|---|---|---|---|
| *R*ScSi (*R*=La-Nd, Sm, Gd) *R*ScSi (*R*=Tb-Tm) | La$_2$Sb/CeScSi Ti$_5$Ga$_4$ | tetragonal hexagonal | *I4/mmm* *P6$_3$/mcm* | [11–15] |
| *R*ScGe (*R*= Dy-Tm) *R*ScGe (*R*=La-Nd, Sm, Eu Gd,Tb) | Ti$_5$Ga$_4$ La$_2$Sb/CeScSi | hexagonal tetragonal | *P6$_3$/mcm* *I4/mmm* | [11–15] |
| *R*TiSi (*R*=Y, Gd-Tm, Lu) | CeFeSi | tetragonal | *P4/nmm* | [21,22] |
| *R*TiGe (*R*=Y, La-Nd, Sm, Gd-Tm, Lu) CeTiGe (HTM), GdTiGe, TbTiGe | CeFeSi CeScSi | tetragonal tetragonal | *P4/nmm* *I4/mmm* | [23,24] [14,26] |
| GdTiSb | CeFeSi | tetragonal | *P4/nmm* | [25] |
| RMnAl (R=Ce, Nd, Gd) | MgCu$_2$ | cubic | *Fd$\bar{3}$m* | [32,33] |
| RMnGa (R=Ce, Pr, Nd, Gd - Dy) | MgCu$_2$ | cubic | *Fd$\bar{3}$m* | [35] |
| RMnIn (R=Gd, Dy, Er, Y) | MgZn$_2$ | hexagonal | *P6$_3$/mmc* | [33] |
| *R*MnSi (*R*=La-Sm, Gd) | CeFeSi | tetragonal | *P4/nmm* | [36] |
| *R*MnSi (*R*=Tb-Ho) | TiNiSi | orthorhombic | *Pnma* | [37] |
| RMnGe (La-Nd) | CeFeSi | tetragonal | *P4/nmm* | [39] |
| RMnGe (Gd-Er, Y) | TiNiSi | orthorhombic | *Pnma* | [40] |
| TmMnGe** | TiNiSi | orthorhombic | *Pnma* | [40] |
| TmMnGe*** | ZrNiAl | hexagonal | *P$\bar{6}$2m* | [40] |
| RFeAl (Ce-Nd, Sm) | MgCu$_2$+* | cubic | | [52] |
| RFeAl (Gd-Lu) | MgZn$_2$ | hexagonal | *P6$_3$/mmc* | [52] |
| RFeGa (Ho, Er, Tm) | MgZn$_2$ | hexagonal | *P6$_3$/mmc* | [53] |
| RFeSi( R=La-Sm, Gd-Er) | CeFeSi | tetragonal | *P4/nmm* | [54] |
| CeCoAl | CeCoAl | monoclinic | *C2/m* | [67] |
| RCoAl (R=Gd-Lu) | MgZn$_2$ | hexagonal | *P6$_3$/mmc* | [66] |
| CeCoGa | CeCoAl | monoclinic | *C2/m* | [68] |
| RCoSi (R=La-Sm, Gd, Tb) | CeFeSi | tetragonal | *P4/nmm* | [69] |
| TbCoSi | TiNiSi | orthorhombic | *Pnma* | [71] |
| RCoSi (R=Ho-Lu) | TiNiSi | orthorhombic | *Pnma* | [71] |
| RCoGe (R=La-Nd) | CeFeSi | tetragonal | *P4/nmm* | [70] |
| RCoGe (R=Gd-Lu, Y) | TiNiSi | orthorhombic | *Pnma* | [71] |
| RCoSn (R=Tb-Lu, Y) | TiNiSi | orthorhombic | *Pnma* | [71] |
| RNiAl (R=Ce-Nd,Sm, Gd-Lu) | ZrNiAl | hexagonal | *P$\bar{6}$2m* | [84] |
| CeNiGa (LTP) | ZrNiAl | hexagonal | *P$\bar{6}$2m* | [89] |
| CeNiGa (HTP) | TiNiSi | orthorhombic | *Pnma* | |
| RNiGa (R=Gd-Tm, Yb) | TiNiSi | orthorhombic | *Pnma* | [90–96] |
| RNiIn (R=La-Sm, Gd-Tm) | ZrNiAl | hexagonal | *P$\bar{6}$2m* | [97–101] |
| RNiSi (R=La, Ce, Nd) | LaPtSi | tetragonal | *I4$_1$md* | [102–105]. |
| RNiSi (R=Gd-Lu) | TiNiSi | orthorhombic | *Pnma* | [102–105]. |
| RNiGe (R=Ce, Gd-Tm, Y) | TiNiSi | orthorhombic | *Pnma* | [107,108] |
| RNiSn (R=La-Sm, Gd-Lu, Yb) | TiNiSi | orthorhombic | *Pnma* | [109] |
| *R*NiSb (*R*=La-Nd, Sm) | ZrBeSi | hexagonal | *P6$_3$/mmc* | [111] |
| RNiSb (R=Gd-Lu) | MgAgAs | cubic | *F$\bar{4}$3m* | [111] |
| RCuAl (R=Ce-Nd,Sm, Gd-Lu) | ZrNiAl | hexagonal | *P$\bar{6}$2m* | [142] |
| RCuGa (R= Ce, Eu) | CeCu$_2$ | orthorhombic | *Imma* | [143,168] and Refs. therein |
| $R$CuIn ($R$=La-Nd, Gd, Tb, Ho, Er, Lu) | ZrNiAl | hexagonal | $P\bar{6}2m$ | [145] and Refs. therein |
| $R$CuSi ($R$=La-Nd, Sm, Gd-Lu, Yb, Y) | Ni$_2$In** AlB$_2$*** | hexagonal hexagonal | $P6_3/mmc$ $P6/mmm$ | [147–149] |
| RCuGe (R=La-Lu) (R=La-Nd, Sm, Gd) (R=Tb-Lu, Y) | AlB$_2$[a*] AlB$_2$[a**] CaIn$_2$[a**] | hexagonal hexagonal hexagonal | $P6/mmm$ $P6/mmm$ $P6_3/mmc$ | [150–152] |
| RCuSn (R=La, Ce- Nd, Lu, Y) | CaIn$_2$ | hexagonal | $P6_3/mmc$ | [156,157] |
| EuCuSn | CeCu$_2$ | orthorhombic | $Imma$ | [160] |
| RCuSn (Gd-Er) | LiGaGe | hexagonal | $P6_3mc$ | [158,159] |
| EuCuAs | Ni$_2$In | hexagonal | $P6_3/mmc$ | [162] |
| EuCuSb | ZrBeSi | hexagonal | $P6_3/mmc$ | [161] |
| YbCuBi | LiGaGe | hexagonal | $P6_3mc$ | [163] |
| CeRuAl | LaNiAl | orthorhombic | $Pnma$ | [191] |
| CeRhAl | LaNiAl | orthorhombic | $Pnma$ | [191] |
| $R$RuSi ($R$=La-Sm,Gd) | CeFeSi | tetragonal | $P4/nmm$ | [192] |
| $R$RuSi ($R$=Y, Tb-Tm) | Co$_2$Si | orthorhombic | $Pnma$ | [193] |
| $R$RuGe ($R$=La-Sm) | CeFeSi | tetragonal | $P4/nmm$ | [15,192,193] |
| $R$RuGe ($R$=Sm, Gd-Tm) | TiNiSi | orthorhombic | $Pnma$ | [15] |
| CeRuSn | CeCoAl | monoclinic | $C2/m$ | [194] |
| RRhAl (R=Y, Pr, Nd, Gd, Ho, Tm) | TiNiSi | orthorhombic | $Pnma$ | [203] |
| RRhAl (R=La, Ce) | Pd$_2$(Mn,Pd)Ge$_2$ | orthorhombic | $Pnma$ | [203,234] |
| RRhGa (R=La-Nd, Gd, Tb, Ho-Lu, Yb, Y] | TiNiSi | orthorhombic | $Pnma$ | [204,205] |
| RRhIn (R=La-Nd, Sm, Gd-Tm) | ZrNiAl | hexagonal | $P\bar{6}2m$ | [206–210] |
| $R$RhIn ($R$=Eu, Yb, Lu) | TiNiSi | orthorhombic | $Pnma$ | [206,211] |
| $R$RhSi ($R$=Y, Gd-Er) | TiNiSi | orthorhombic | $Pnma$ | [213–215] |
| LaRhSi | ZrOS | cubic | $P2_13$ | [215] |
| $R$RhGe ($R$=Ce-Nd, Sm, Gd-Yb, Y) | TiNiSi | orthorhombic | $Pnma$ | [216–219] |
| RRhSn (R=La-Nd, Sm, Gd-Lu) | ZrNiAl | hexagonal | $P\bar{6}2m$ | [178,222–225] |
| RRhSb (R=La,Ce Pr, Sm, Gd-Tm) | TiNiSi | orthorhombic | $Pnma$ | [227–229] |
| RRhBi (R= La, Ce) | TiNiSi | orthorhombic | $Pnma$ | [230] |
| RPdAl (R=Sm, Gd-Tm, Y) | TiNiSi | orthorhombic | $Pnma$ | [254] |
| RPdAl (R= Ce-Nd, Sm, Gd-Tm, Lu, Y)[a***] | ZrNiAl | hexagonal | $P\bar{6}2m$ | [254] |
| RPdGa (R=La-Nd, Sm, Eu, Gd-Tm, Lu, Y) | TiNiSi | orthorhombic | $Pnma$ | [168,213,256] |
| RPdIn (R=La-Sm, Gd-Lu, Y) | ZrNiAl | hexagonal | $P\bar{6}2m$ | [168,257–261] |
| EuPdIn | TiNiSi | orthorhombic | $Pnma$ | [168] |
| RPdSi (R=La-Nd, Sm-Lu, Y) | TiNiSi | orthorhombic | $Pnma$ | [213,264,265] |
| $R$PdGe ($R$= La-Nd, Sm, Gd-Tm, Y) | CeCu$_2$ | orthorhombic | $Imma$ | [266–269] |
| EuPdGe | EuNiGe | monoclinic | $P2_1/n$ | [168] |
| $R$PdSn ($R$=La, Ce-Nd, Sm-Yb) | TiNiSi | orthorhombic | $Pnma$ | [270,271] |
| $R$PdSn ($R$=Er, Tm) | Fe$_2$P | hexagonal | $P\bar{6}2m$ | [270,271] |
| EuPdAs | Ni$_2$In | hexagonal | $P6_3/mmc$ | [168] |
| $R$PdSb ($R$=Ce-Nd, Sm, Gd-Dy) | CaIn$_2$ | hexagonal | $P6_3/mmc$ | [272] |
| EuPdSb | TiNiSi | orthorhombic | $Pnma$ | [272] |
| $R$PdSb ($R$=Ho, Er, Tm) | MgAgAs | cubic | $F\bar{4}3m$ | [272] |
| $R$PdBi (R=La-Nd, Sm Gd-Lu) | MgAgAs | cubic | $F\bar{4}3m$ | [274–276] |

| $R$AgAl ($R$=La-Nd, Sm, Gd- Er, Y) | CeCu$_2$ | orthorhombic | *Imma* | [317,318] |
|---|---|---|---|---|
| $R$AgGa ($R$=La- Nd, Eu, Gd-Lu, Y) | CeCu$_2$ | orthorhombic | *Imma* | [319] |
| $R$AgSi (R=Sm, Gd- Tm, Yb, Lu, Y) | ZrNiAl | hexagonal | $P\bar{6}2m$ | [320,321] |
| RAgGe (R=La, Ce) | CaIn$_2$ | hexagonal | *P6$_3$/mmc* | [322] |
| EuAgGe | CeCu$_2$ | orthorhombic | *Imma* | [324] |
| RAgGe (R=Gd-Lu) | ZrNiAl | hexagonal | $P\bar{6}2m$ | [323] |
| RAgSn (R=La, Ce, Sm, Gd-Er, Yb) | CaIn$_2$ | hexagonal | *P6$_3$/mmc* | [325] |
| EuAgSn | CeCu$_2$ | orthorhombic | *Imma* | [189] |
| RAgSn (R=Ce-Nd, Gd-Er) | LiGaGe | hexagonal | *P6$_3$mc* | [326,327] |
| EuAgSb | ZrBeSi | hexagonal | *P6$_3$/mmc* | [351] |

## Table 2

List of $RTX$ compounds with their magnetic ground state, magnetic transition temperatures ($T_C$ or $T_N$), paramagnetic Curie temperature ($\theta_p$) and experimental effective magnetic moment ($\mu_{eff}$). The table is adopted with permission from [9] © 2015, Elsevier B. V.

| Compound | Type of ordering | Magnetic phase transition(s) | $\theta_p$ (K) | $\mu_{eff}$ ($\mu_B$/f.u.) | References |
|---|---|---|---|---|---|
| CeScSi | FM | 46 | | | [20] |
| | AFM | ~26 | | | [12] |
| PrScSi | AFM | 145 | 99 | 2.65 | [13] |
| NdScSi | FM | 175 | 178 | 2.96 | [13] |
| SmScSi | FM | 270 | | | [13] |
| GdScSi | FM | 318 | 338 | 7.9 | [10] |
| | FM | 352 | | | [11] |
| TbScSi | AFM | 165 | 80 | 10.1 | [10] |
| DyScSi | | | 34 | 11.4 | [10] |
| HoScSi | | | 13 | 11.4 | [10] |
| ErScSi | | | 5 | 10.3 | [10] |
| TmScSi | | | -11 | 8.4 | [10] |
| CeScGe | FM | 35 | | | [20] |
| | AFM | 46 | | | [12] |
| PrScGe | FM, AFM | 80, 88, 140 | 125 | 2.88 | [13] |
| NdScGe | FM | 200 | 195 | 2.65 | [13] |
| SmScGe | FM | 270 | | | [13] |
| GdScGe | FM | 320 | 332 | 7.8 | [10] |
| | FM | 348 | | | [11] |
| TbScGe | FM | 216 | 215 | 9.9 | [10] |
| DyScGe | | | 37 | 11.3 | [10] |
| HoScGe | | | 10 | 11.2 | [10] |
| ErScGe | | | -5 | 10.6 | [10] |
| TmScGe | | | 10 | 8.2 | [10] |
| YTiSi | PPM | | | | [22] |
| GdTiSi | AFM | 400 | 369 | 7.3 | [22] |
| TbTiSi | AFM | 286 | 186 | 10.14 | [22] |
| DyTiSi | AFM | 170 | 110 | 10.43 | [22] |
| HoTiSi | AFM | 95 | 50 | 11 | [22] |
| ErTiSi | AFM | 50 | 25 | 9.7 | [22] |
| TmTiSi | AFM | 20 | 11 | 7 | [22] |
| LuTiSi | PPM | | | | [22] |
| YTiGe | PPM | | | | [27] |
| LaTiGe | PPM | | | | [24] |
| CeTiGe | | | -20 | 2.7 | [24] |
| CeTiGe[HTM] | | | -59 | 2.6 | [26] |
| PrTiGe | AFM | 70, 55 | 26 | 3.5 | [24] |
| NdTiGe | AFM | 150, 90 | 93 | 3.6 | [24] |
| SmTiGe | AFM | 260, 215 | NCW | NCW | [24] |
| GdTiGe[#] | AFM | 412 | 317 | 8.3 | [27] |
| GdTiGe[##] | FM | 400 | 460 | 7.8 | [24] |
| GdTiGe[##] | FM | 376 | 416 | 8.39 | [14] |
| TbTiGe[#] | AFM | 270 | 263 | 9.9 | [24] |
| TbTiGe[#] | AFM | 288 | 248 | 9.4 | [27] |
| TbTiGe[##] | FM | 300 | 285 | 10.41 | [28] |
| DyTiGe | AFM | 170 | 155 | 10.7 | [24] |
| | AFM | 175 | 120 | 10.7 | [27] |
| HoTiGe | AFM | 115, 85 | 103 | 10.7 | [24] |
| | AFM | 115 | 101 | 10.7 | [27] |
| ErTiGe | AFM | 41 | 34 | 10 | [24] |
| TmTiGe | AFM | 15 | 1 | 7.6 | [24] |
| LuTiGe | PPM | | | | [24] |
| GdTiSb | FM | 268 | 350 | 7.5 | [25] |
| NdMnAl | SG | - | | 7.02 (includes Mn moments) | [33] |
| GdMnAl | AFM | 298 | | | [42] |
| | FM | 274 | | | [43] |
| TbMnAl | AFM | 34 | | | [34] |
| CeMnGa | SG | | -144 | 3.1 | [35] |
| PrMnGa | RSG | 90, 50$^r$ | -17 | 3.7 | [35] |
| NdMnGa | SG | 10$^g$ | -11 | 3.4 | [35] |
| GdMnGa | RSG | 220, 80$^r$ | - | - | [35] |
| TbMnGa | RSG | 120, 85$^r$ | -6 | 2.7 | [35] |
| DyMnGa | SG | 40$^g$ | 5 | 3.5 | [35] |
| GdMnIn | SG | 118 | −3.8 | 9.92 | [44] |
| LaMnSi | AFM | 310$^{**}$ | -350 | 3.07 | [36] |
| CeMnSi | AFM | 240$^{**}$ | -50 | 3.80 | [36] |
| PrMnSi | AFM | 130, 80, 265** | 75 | 4.69 | [36] |
| NdMnSi | AFM | 175, 80, 280** | 110 | 4.6 | [36] |
| SmMnSi | AFM | 235, 220, 110 | NCW | | [36] |
| GdMnSi | FM | 310 | 265 | 8.52 | [36] |
| TbMnSi$^{###}$ | AFM | 410, 140 | 44 | 10.2 | [37] |
| | FM | 260 | 164 | 10.37 | [49] |
| DyMnSi | AFM | 400, 55 | 22 | 11.1 | [37] |
| HoMnSi | AFM | 15 | -58 | 12.35 | [50] |
| LuMnSi | AFM | 255 | -201 | 3.17 | [50] |
| LaMnGe | AFM | 420[**] | | | [39] |
| CeMnGe | AFM | 415[**]<br>313, 41 | | | [39] |
| PrMnGe | AFM<br>FM | 415[**]<br>150 | | | [39] |
| NdMnGe[###] | AFM<br><br>FM | 430<br>410[**]<br>190 | -31 | 5.3 | [37]<br>[39]<br>[39] |
| GdMnGe | AFM<br>AFM | 490, 200, 60<br>350 | 64<br>80 | 8.65<br>8.6 | [40]<br>[41] |
| TbMnGe | AFM | 510, 186, 70 | 39 | 10.6 | [51] |
| DyMnGe | | 70 | 24 | 11.2 | [40] |
| HoMnGe | | 12 | 11 | 11.2 | [40] |
| ErMnGe | | | -34 | 10.39 | [40] |
| TmMnGe | | 17 | -36 | 8.7 | [40] |
| YMnGe | | | -78 | 3.6 | [40] |
| YFeAl | FM | 38 | | | [56] |
| GdFeAl | FM | 260 | | | [56] |
| TbFeAl | FM | 195 | | | [56] |
| DyFeAl | FM | 128 | 143 | 10.76 | [56] |
| HoFeAl | FM | 92 | | | [56] |

|  |  | 80 |  |  | [60] |
|---|---|---|---|---|---|
| ErFeAl | FM | 56 |  |  | [56] |
| TmFeAl | FM | 38 |  |  | [56] |
| LuFeAl | FM | 39 |  |  | [56] |
| ErFeGa | FM | 77 | 27 | 9.59 | [53] |
| LaFeSi | PPM |  |  |  | [55] |
| CeFeSi | IV |  |  |  | [55] |
| PrFeSi |  |  | 35 | 3.62 | [55] |
| NdFeSi | FM | 25 | 20 | 3.90 | [55] |
| SmFeSi | FM | 40 |  |  | [55] |
| GdFeSi | FM | 135 | 165 | 8.09 | [55] |
| TbFeSi | FM | 125 | 110 | 9.62 | [55] |
|  | FM | 110 | 90 | 10.28 | [62] |
| DyFeSi | FM | 110 | 75 | 10.57 | [55] |
|  | FM | 70 | 69 | 11.41 | [62] |
| HoFeSi | FM, AFM/FIM | 29, 20 | 24.8 | 11.25 | [64] |
| ErFeSi | FM | 22 | 19.7 | 9.84 | [65] |
| GdCoAl | FM | 100 |  |  | [72] |
| TbCoAl | FM | 70 |  |  | [72] |
| DyCoAl | FM | 37 | 44 | 10.73 | [72] |
| HoCoAl | FM | 10 |  |  | [72] |
| TmCoAl | FM | 6 | 4 | 5.93 | [73] |
| CeCoGa | AFM | 4.3 | -82 | 1.8 | [68] |
| LaCoSi | PPM | | | | [69] |
| CeCoSi | | | -53 | 2.8 | [69] |
| | AFM | 8.8 | -55 | 2.71 | [74] |
| PrCoSi | | | -10 | 3.9 | [69] |
| NdCoSi | AFM | 7 | 30 | 3.9 | [69] |
| SmCoSi | AFM | 15 | NCW | | [69] |
| GdCoSi | AFM | 175 | 70 | 8.6 | [69] |
| TbCoSi | AFM | 140 | 70 | 9.6 | [69] |
| HoCoSi | FM | 14 | 6 | 10.9 | [60] |
| TmCoSi | AFM | 4.4 | | 7.5 | [77] |
| LaCoGe | PPM | | | | [70] |
| CeCoGe | | | 15 | 2.6 | [70] |
| | AFM | 5 | -39 | 2.6 | [79] |
| PrCoGe | | | -7 | 3.9 | [70] |
| NdCoGe | AFM | 8 | -6 | 4 | [70] |
| TbCoGe | FM | 16 | 17.5 | 9.48 | [80] |
| | AFM | 17 | | | [78] |
| DyCoGe | AFM, FM | 10, 170 | | | [78] |
| TbCoSn | AFM | 20.5 | 15 | 9.8 | [81] |
| DyCoSn | AFM | 10 | 9 | 10.5 | [81] |
| HoCoSn | AFM | 7.8 | 6.5 | 11 | [81] |
| ErCoSn | AFM | 5 | 0 | 9.8 | [81] |
| RNiAl (R=Yb, Lu) | PPM | | | | [84] |
| PrNiAl | | | -10 | 3.73 | [84] |
| | AFM | 6.5 | | | [114] |
| | TSW | 6.9 | -23 | 3.7 | [116] |
| NdNiAl | FM | 15 | 5 | 3.84 | [84] |
| | AFM | 2.4 | | | [114] |
| | TSW | 2.7 | -7.5 | 3.8 | [116] |
| SmNiAl | PM | | | | [118] |
| GdNiAl | FM | 66 | 56 | 8.5 | [84] |
| | FM, AFM | 57, 31 | | | [116] |
| TbNiAl | FM | 57 | 45 | 10.1 | [84] |
| | AFM | 47, 23 | 30 | 10.3 | [116] |
| DyNiAl | FM | 39 | 30 | 11.1 | [84] |
| | FM, AFM | 31, 15 | 17 | 10.9 | [116] |
| HoNiAl | FM | 25 | 11 | 10.8 | [84] |
| | FM, AFM | 14.5, 12.5, 5.5 | 7.3 | 10.7 | [116] |
| ErNiAl | FM | 15 | -1 | 9.8 | [84] |
| TmNiAl | FM, AFM | 12, 4.2 | -11 | 7.8 | [84] |
| YbNiAl | AFM | 2.9 | | 4.4 | [115] |
| GdNiGa | FM | 30.5 | 32 | 7.85 | [90] |
| TbNiGa | AFM | 23 | | | [91] |
| DyNiGa | AFM | 17 | -12.5 | 10 | [121] |
| HoNiGa | AFM | 10 | 13.3 | 10.43 | [94] |
| ErNiGa | AFM | 8 | | | [93] |
| TmNiGa | AFM | 5.5 | -49 | 7.79 | [95] |
| YbNiGa | AFM | 1.9, 1.7 | -30 | 4.4 | [96] |
| NdNiIn | FM | 14.5 | | | [123] |
| GdNiIn | FM | 98 | 101 | 7.96 | [100] |
| TbNiIn | FM, AFM | 71, 12 | 47 | 9.8 | [100] |
| | AFM | 68 | | | [123] |
| DyNiIn | FM, AFM | 30, 14 | 31 | 10.68 | [100] |
| | AFM | 32 | | | [123] |
| HoNiIn | FM | 20, 7 | 18 | 10.61 | [100] |
| ErNiIn | FM | 9 | 8 | 9.64 | [100] |
| TmNiIn | AFM | 2.5 | 0.4 | 7.5 | [101] |
| NdNiSi | AFM | 6.8, 2.8 | | | [105] |
| GdNiSi | AFM | 11 | | | [124] |
| TbNiSi | AFM | 16 | -6.1 | 9.7 | [102] |
| DyNiSi | AFM | 8.8 | -7.5 | 10.5 | [102] |
| HoNiSi | AFM | 4.1 | -5 | 10.3 | [102] |
| | AFM | 4.2 | -10.1 | 10.9 | [106] |
| ErNiSi | AFM | 3.3 | 0 | 8.9 | [102] |
| | AFM | 3.2 | 0 | 9.57 | [126] |
| CeNiGe | PPM | | | | [127] |
| GdNiGe | AFM | 11 | -10 | 7.85 | [107] |
| TbNiGe | AFM | 7 | -9 | 9.45 | [107] |
| | AFM | 18.5 | 4.1 | 8.97 | [128] |
| DyNiGe | AFM | 9 | -6 | 10.40 | [107] |
| | AFM | 4.7 | -8.2 | 10.42 | [128] |
| HoNiGe | AFM | | -3 | 10.43 | [107] |
| | AFM | 5 | -2.6 | 11.2 | [129] |
| ErNiGe | AFM | 6 | -3 | 9.28 | [107] |
| | AFM | 2.9 | -1.4 | 9.19 | [129] |
| TmNiGe | AFM | | -4 | 7.23 | [107] |
| YbNiGe | | | 3 | 4.31 | [107] |
| LaNiSn | SC | 0.59 | | | [131] |
| NdNiSn | AFM | 3 | | 3.32 | [134,135] |
| SmNiSn | AFM | 9 | -40 | 1.37 | [132] |
| GdNiSn | AFM | 10.5 | -3 | 8.8 | [130] |
| TbNiSn | AFM | 7.2 | 6 | 11.27 | [130] |
| DyNiSn | AFM | 8.2 | -1 | 11.06 | [130] |
| HoNiSn | AFM | - | -2 | 10.75 | [130] |
| | AFM | 3 | | 8.6 | [138] |
| ErNiSn | AFM | - | 6 | 9.85 | [130] |
| | | 4 | | 9 | [138] |
| TmNiSn | AFM | - | -4 | 7.65 | [130] |
| YbNiSn | FM | 5.5 | -43 | 4.3 | [110] |
| YbNiSb | AFM | 0.8 | -13 | 4.6 | [140] |
| RNiSb (R=Y, La) | PPM | | | | [139] |
| CeNiSb | FM+AFM | 3.5 | -27 | 2.9 | [139] |
| PrNiSb | PM | | -0.7 | 3.8 | [139] |
| NdNiSb | FM | 23 | 13 | 3.7 | [139] |
| SmNiSb | VVP | | | 1.58 | [111] |
| GdNiSb$^{cubic}$ | | | -15 | 8.1 | [111] |
| GdNiSb$^{cubic}$ | AFM | 9.5 | | | [141] |
| GdNiSb$^{hex}$ | AFM | 3.5 | | | [141] |
| TbNiSb | AFM | 5.5 | -17 | 9.7 | [139] |
| DyNiSb | AFM | 3.5 | -9.8 | 10.9 | [139] |
| HoNiSb | AFM | 2 | -10.8 | 10.7 | [139] |
| CeCuAl | AFM | 5.2 | -37 | 2.53 | [166] |
| PrCuAl | AFM | 7.9 | -0.7 | 3.54 | [142] |
| NdCuAl | AFM | 18 | 20.3 | 3.29 | [142] |
| SmCuAl | | 47 | 29.2 | 0.4 | [142] |
| GdCuAl | FM | 82 | 77.2 | 8.21 | [142] |
| TbCuAl | FM | 49 | 47.3 | 9.97 | [142] |
| DyCuAl | FM | 28 | 25.9 | 10.65 | [142] |
| HoCuAl | FM | 12.1 | 11.5 | 10.6 | [142] |
| ErCuAl | FM | 6.8 | 4.2 | 9.55 | [142] |
| TmCuAl | FM | 2.8 | 9 | 7.45 | [165] |
| CeCuGa | IV | | -92 | 1.92 | [143] |
| EuCuGa | AFM | 12 | | | [168] |
| CeCuIn | | | -15 | 2.4 | [145] |
| PrCuIn | | | -3 | 3.65 | [145] |
| NdCuIn | AFM | 4.9 | -7.2 | 2.96 | [145] |
| GdCuIn | AFM | 20 | 20 | 7.9 | [169] |
| TbCuIn | AFM | 14.5 | -5.4 | 9.44 | [146] |
| HoCuIn | AFM | 5 | -10 | 10.51 | [146] |
| ErCuIn | AFM | 3.1 | -12.3 | 9.93 | [146] |
| CeCuSi | FM | 15.5 | -2 | 2.54 | [171] |
| PrCuSi | FM | 14 | 8 | 3.39 | [170] |
| | AFM | 5.1 | -11 | 3.88 | [173] |
| NdCuSi | AFM | 3.1 | -11 | 3.87 | [179] |
| GdCuSi | FM | 49 | 58 | 8.32 | [170] |
| | AFM | 14.2 | 11.7 | 8.1 | [175] |
| TbCuSi | FM | 47 | 52 | 9.62 | [170] |
| DyCuSi | AFM | 11 | 20 | 10.4 | [180] |
| HoCuSi | AFM | 9 | 15 | 10.7 | [180] |
| | AFM | 7 | 8 | 10.62 | [182] |
| ErCuSi | AFM | 6.8 | | | [183] |
| TmCuSi | AFM | 6.5/6.2 | | 7.2 | [77,184] |
| CeCuGe | FM | 10 | 2 | 2.56 | [172] |
| PrCuGe | AFM | 1.8 | 1 | 3.56 | [151] |
| NdCuGe | AFM | 3.5 | -7.5 | 3.62 | [151] |
| GdCuGe | AFM | 17 | -0.3 | 7.89 | [151] |
| TbCuGe | AFM | 11.6 | -21 | 9.8 | [151] |
| | | 11.8 | -30.6 | 10.1 | [155] |
| DyCuGe | AFM | 6 | -12.9 | 10.73 | [151] |
| | | 5.2 | -14.4 | 10.8 | [155] |
| HoCuGe | AFM | 6.1 | -11.6 | 10.58 | [151] |
| | | 4.7 | -5.4 | 10.7 | [155] |
| ErCuGe | AFM | 4.3 | -14 | 9.75 | [151] |
| | | 4.1 | -10.1 | 9.8 | [155] |
| CeCuSn | AFM | 8.3 | | | [188] |
| PrCuSn | AFM | 3 | -3.4 | 3.84 | [157] |
| NdCuSn | AFM | 10 | -19.5 | 3.84 | [157] |
| EuCuSn | PM | | -16 | 7.75 | [189] |
| GdCuSn | AFM | 26 | -36 | 7.8 | [159] |
| TbCuSn | AFM | 17.8 | -18 | 9.8 | [159] |
| | AFM | 15.3 | -38 | 9.82 | [158] |
| DyCuSn | AFM | 5 | -15 | 10.3 | [159] |
| HoCuSn | AFM | 7.8 | -3 | 10.3 | [159] |
| ErCuSn | AFM | 4.8** | 0 | 9.2 | [159] |
| YbCuSn | | | -4.4 | 0.2 | [273] |
| EuCuAs | AFM | 18 | 28 | 7.67 | [190] |
| EuCuSb | AFM | 10 | 3 | 7.84 | [161] |
| EuCuBi | AFM | 18 | -13 | 7.65 | [190] |
| LaRuSi | PPM | | | | [192] |
| CeRuSi | | | -52 | 2.56 | [192] |
| PrRuSi | AFM | 73 | 7 | 3.52 | [192] |
| NdRuSi | AFM | 74 | 29 | 3.46 | [192] |
| SmRuSi | FM | 65 | NCW | | [192] |
| GdRuSi | FM | 85 | 78 | 8.58 | [192] |
| ErRuSi | FM | 8 | 8.1 | 9.48 | [197,198] |
| LaRuGe | | | | | [192] |
| CeRuGe | | | -73 | 2.55 | [192] |
| PrRuGe | AFM | 62 | 0 | 3.82 | [192] |
| NdRuGe | AFM | 65 | 4 | 3.91 | [192] |
| SmRuGe | FM | 45 | NCW | | [192] |
| GdRuGe | FM | 72 | 67 | 8.19 | [199] |
| TbRuGe | FM | 70 | 25 | 9.92 | [199] |
| DyRuGe | FM | 26 | 24 | 10.82 | [199] |
| HoRuGe | FM | 18 | 24 | 10.82 | [199] |
| ErRuGe | FM | 8 | 7 | 9.82 | [199] |
| TmRuGe | PM | | 0 | 7.52 | [202] |
| YRhAl | SC | 0.9 | | 0.1 | [234] |
| LaRhAl | SC | 2.4 | | | [231,232] |
| CeRhAl | | | -24.8 | 1.17 | [232] |
| | AFM | 3.8 | -11.3 | 1.57 | [233] |
| PrRhAl | FM | 4.7 | -5.4 | 3.49 | [232] |
| NdRhAl | FM | 10.5 | 6.2 | 3.37 | [232] |
| GdRhAl | FM | 29.8 | 32.9 | 7.82 | [232] |
| CeRhGa | | | 8 | 1.5 | [237] |
| TbRhGa | AFM | 22 | 23 | 10.1 | [205] |
| HoRhGa | AFM | 5.4 | 2 | 10.6 | [205] |
| ErRhGa | AFM | 4.8 | 9 | 9.37 | [205] |
| TmRhGa | AFM | 3.9 | 5.6 | 7.67 | [236] |
| EuRhIn | FM | 22 | 34 | 7.9 | [211] |
| GdRhIn | FM | 34 | 24.3 | 8 | [212] |
| LaRhSi | SC | 4.3 | | | [215] |
| GdRhSi | FM | 100 | 94 | 7.95 | [214] |
| TbRhSi | FM | 55 | 48 | 9.92 | [214] |
| | AFM | 26 | -13.7 | 9.8 | [241] |
| DyRhSi | FM | 25 | 11.5 | 10.31 | [214] |
| | AFM | 12.5 | -8.1 | 10.7 | [241] |
| HoRhSi | AFM | 8 | 10.5 | 10.71 | [214] |
| | AFM | 8.7 | -14 | 10.5 | [241] |
| ErRhSi | AFM | 7.5 | -3 | 9.54 | [214] |
| CeRhGe | AFM | 10.5 | -56 | 2.3 | [243] |
| NdRhGe | AFM | 14 | -10 | 3.73 | [243] |
| SmRhGe | FM | 56 | NCW | | [218] |
| GdRhGe | AFM | 31.8 | -4.9 | 8.5 | [244] |
| TbRhGe | AFM | 24.3 | -18.9 | 10.2 | [248] |
| DyRhGe | AFM | 20.3 | -3 | 10.9 | [248] |
| HoRhGe | AFM | 5.5 | -0.9 | 10.9 | [219] |
| ErRhGe | AFM | 10.2 | -9.6 | 9.7 | [248] |
| TmRhGe | AFM | 6.8 | -4.4 | 7.6 | [248] |
| YbRhGe | AFM | 7 | -16.4 | 4.48 | [217] |
| LaRhSn | SC | 2 | | | [226] |
| CeRhSn | PM | | -70 | 1.30 | [222] |
| PrRhSn | FM | 3 | 2.9 | 3.64 | [222] |
| NdRhSn | FM | 10.3 | -6.7 | 3.50 | [222] |
| SmRhSn | FM | 14.9 | | | [225] |
| GdRhSn | AFM | 16 | 19 | 7.58 | [222] |
| | AFM | 16.2 | 13.9 | 7.91 | [178] |
| TbRhSn | AFM | 18.3 | -13.2 | 9.8 | [185] |
| DyRhSn | AFM | 7.2 | -19.9 | 10.63 | [185] |
| HoRhSn | FM | 6.2 | 4 | 10.65 | [185] |
| ErRhSn | AFM | | -6.6 | 9.58 | [185] |
| TmRhSn | AFM | 3.1 | -47.1 | 7.01 | [185] |
| YbRhSn | AFM | 2 | | | [252] |
| LaRhSb | SC | 2.1, 2.4 | | | [228,253] |
| PrRhSb | FM, AFM | 6,18 | -1.3 | 3.5 | [228] |
| LaRhBi | SC | 2.4 | | | [230] |
| CeRhBi | | | -77.6 | 2.68 | [230] |
| CePdAl | AFM | 2.7 | | | [279] |
| PrPdAl | AFM | 4 | | | [282] |
| NdPdAl | AFM | 5 | | | [283] |
| GdPdAl | FM | 48 | 49*** | 8.35 | [255] |
| | | | 67**** | 7.94 | |
| TbPdAl | AFM | 43 | 38 | 9.66 | [285] |
| DyPdAl | FM | 25 | 49.3 | 10.6 | [286] |
| HoPdAl | AFM | 16 | 28 | 10.5 | [287] |
| EuPdGa | FM | 38 | 17 | 7.86 | [168] |
| GdPdGa | AFM | 5.1 | | | [256] |
| TbPdGa | AFM | 34 | | 8.14 | [256] |
| DyPdGa | FM | 20 | | | [256] |
| HoPdGa | AFM | 6.8 | | 7.15 | [256] |
| ErPdGa | AFM | 5 | | 9.2 | [256] |
| LaPdIn | SC | 1.6 | | | [289] |
| CePdIn | AFM | <1.7 | -52.5 | 2.58 | [257] |
| EuPdIn | | | 40 | 7.99 | [262] |
| | AFM | 13 | 13 | 7.6 | [263] |
| PrPdIn | | | -8.8 | 3.57 | [257] |
| | FM | 11.2 | | | [291] |
| NdPdIn | FM | 30** | 2 | 3.59 | [257,258] |
| | FM | 34.3 | | | [291] |
| SmPdIn | FM | 54 | NCW | | [261] |
| EuPdIn | AFM | 13 | 13 | 7.6 | [168] |
| GdPdIn | FM | 102 | 96.5 | 12 | [259] |
| | FM | 101.5 | 88.3 | 7.99 | [260] |
| TbPdIn | FIM | 70 | 6 | 10.4 | [259] |
| | FM | 74 | 60 | 9.73 | [288] |
| DyPdIn | FIM | 34 | 5.2 | 11 | [259] |
| | FM | 35 | 29 | 10.5 | [288] |
| HoPdIn | FIM | 25 | 7 | 10.8 | [259] |
| | FM | 22** | | | [258] |
| ErPdIn | FIM | 12.3 | 1.6 | 9.7 | [259] |
| | FM | 11** | | | [258] |
| TmPdIn | AFM | 2.7 | -8.1 | 7.47 | [260] |
| YPdSi | PPM | | | | [264] |
| CePdSi | FM | 7 | -34 | 2.58 | [292] |
| PrPdSi | FM | 5 | -7 | 3.65 | [292] |
| GdPdSi | AFM | 18 | -52 | 8.37 | [264] |
| TbPdSi | AFM | 14 | -32 | 9.65 | [264] |
| DyPdSi | AFM | 7.4 | -17 | 11.14 | [264] |
| HoPdSi | AFM | 3.8 | -10.6 | 10.62 | [264] |
| ErPdSi | AFM | 3 | -12.7 | 9.47 | [264] |
| EuPdSi | | | 9 | 7.45 | [168] |
| YbPdSi | FM | 8 | -41.2 | 4.7 | [293] |
| CePdGe | | | -25 | 2.4 | [269] |
| | AFM | 3.4 | -17 | 2.23 | [266] |
| PrPdGe | AFM | 8 | -8 | 3.48 | [269] |
| | | | -12 | 3.7 | [266] |
| NdPdGe | AFM | 4 | -4 | 3.61 | [269] |
| EuPdGe | AFM | 8.5 | 12 | 8 | [168] |
| GdPdGe | | | -32 | 7.92 | [269] |
| | AFM | 17 | -36 | 7.46 | [267] |
| TbPdGe | AFM | 18 | -22 | 9.7 | [269] |
| | AFM | 32 | -24 | 9.69 | [266] |
| DyPdGe | AFM | 6 | -10 | 10.68 | [269] |
| | AFM | 8.4 | -10 | 10.4 | [267] |
| HoPdGe | | | -8 | 10.56 | [267] |
| ErPdGe | AFM | 4 | -6 | 9.3 | [267] |
| TmPdGe | | | -4 | 7.38 | [269] |
| YbPdGe | FM | 11.4 | | | [294] |
| CePdSn | AFM | 7.5 | -68 | 2.67 | [270] |
| | AFM | 6 | -63 | 2.7 | [295] |
| PrPdSn | | | -2 | 3.6 | [270] |
| | AFM | 4.3 | -5.5 | 3.51 | [295] |
| NdPdSn | | | -8 | 4.93 | [270] |
| | AFM | 2.4 | -11 | 3.68 | [295] |
| SmPdSn | AFM | 11 | NCW | | [270] |
| EuPdSn | AFM | 15.5, 6 | 13 | 7.78 | [168] |
| EuPdSn | AFM | 13 | 5 | 8.27 | [270] |
| GdPdSn | AFM | 14.5 | -27 | 8.16 | [270] |
| TbPdSn | AFM | 23.5 | -16 | 10.17 | [270] |
| | AFM | 19 | -11 | 10.1 | [295] |
| DyPdSn | AFM | 11.4 | -2 | 11.1 | [270] |
| | AFM | 10 | -7 | 10.5 | [295] |
| HoPdSn | | | -7 | 11.07 | [270] |
| | AFM | 3.7 | -7.5 | 10.7 | [295] |
| ErPdSn | AFM | 5.6 | -0.3 | 9.51 | [270] |
| | AFM | 5.2 | 3 | 9.62 | [295] |
| TmPdSn (hex) | | | -0.1 | 7.98 | [270] |
| YbPdSn | | | -5 | 1.45 | [270] |
| CePdSb | FM | 16.5 | 11 | 2.6 | [295] |
| PrPdSb | FM | 9 | 3 | 3.6 | [295] |
| NdPdSb | | | 8 | 3.6 | [295] |
| | AFM | 10 | 9 | 2.95 | [298] |
| EuPdSb | AFM | 13 | -35 | 8.19 | [168] |
| SmPdSb | FM | 3 | | | [295] |
| GdPdSb | AFM | 16.5 | -21 | 8.13 | [295] |
| TbPdSb | AFM | 2.2 | -10 | 9.8 | [295] |
| DyPdSb | AFM | 4.6 | -6 | 10.6 | [295] |
| HoPdSb | AFM | 2.2 | -4 | 10.6 | [295] |
| YbPdSb | | | -9 | 4.39 | [273] |
| YPdBi | SC | ~1 | | | [312] |
| CePdBi | SC, SG | 1.4, 2.5 | 0.5 | 2.23 | [307] |
| GdPdBi | AFM | 13.5 | -36.5 | 8 | [276] |
| TbPdBi | SC, AFM | 5.3 | -25.5 | 9.17 | [275,311] |
| DyPdBi | AFM | 3.5 | -11.9 | 10.7 | [276] |
| HoPdBi | SC, AFM | 0.7, 2.2 | -6.1 | 10.59 | [274,276] |
| ErPdBi | SC, AFM | 1.22, 1.06 | | | [310] |
| LuPdBi | SC | 1.8 | | | [274] |
| CeAgAl | SG | | -9.6 | 2.6 | [328] |
| | FM | 2.9 | -18 | 2.65 | [329] |
| PrAgAl | SG | 10 | -0.9 | 3.5 | [328] |
| NdAgAl | SG, FM | 16 | 6.7 | 3.4 | [328] |
| GdAgAl | SG | 40 | 53 | 8.3 | [330] |
| TbAgAl | SG, FM | 64 | 45 | 9.9 | [328] |
| ErAgAl | SG, FM | 10 | 1.9 | 9.4 | [328] |
| DyAgAl | SG, FM | 43.8, 34.7[a*] | 37 | 10.69 | [331] |
| HoAgAl | SG, FM | 23.7, 18.2[a*] | 16.4 | 10.79 | [331] |
| ErAgAl | SG, FM | 15.3, 12.5[a*] | 8.1 | 9.9 | [331] |
| TmAgAl | FM | 3.3 | 4.3 | 7.71 | [332] |
| CeAgGa | FM | 5.5 | | 2.49 | [334] |
| | SG, FM | 5.1, 3.6 | -43 | 2.5 | [333] |
| PrAgGa | | | 31 | 3.18 | [335] |
| NdAgGa | | | 4 | 3.65 | [335] |
| GdAgGa | FM | 27 | 52 | 7.95 | [335] |
| | FM, SG | 30, 14 | 42 | 7.9 | [339] |
| TbAgGa | AFM | 18 | 20 | 10.03 | [335] |
| | AFM | 34 | 39 | 9.1 | [336] |
| | SG | 50 | 27 | 9.6 | [338] |
| DyAgGa | | | 17 | 10.6 | [335] |
| | FM | 20 | 2.6 | 10.9 | [338] |
| HoAgGa | FM | 4.7 | 14 | 10.43 | [335] |
| | FM | 18 | 3 | 10.6 | [338] |
| | FM | 7.2 | | 10.9 | [337] |
| ErAgGa | FM | 3 | 12 | 9.43 | [335] |
| TmAgGa | | | 9 | 7.38 | [335] |
| NdAgSi | FM | 44 | 43.8 | 2.2 | [341] |
| GdAgSi | AFM | 13.4 | -13 | 8.26 | [321] |
| TbAgSi | AFM | 21.5, 16 | -23.4 | 9.8 | [321] |
| DyAgSi | AFM | 11 | -10.3 | 10.9 | [321] |
| HoAgSi | AFM | 10 | -8 | 10.7 | [321] |
| ErAgSi | AFM | 2.5 | -1.5 | 9.8 | [321] |
| TmAgSi | AFM | 3.3 | -0.8 | 7.45 | [340] |
| CeAgGe | AFM | 4.8 | -9.3 | 2.61 | [322] |
| GdAgGe | AFM | 15.6 | -31.4 | 7.88 | [342] |
| TbAgGe | AFM | 25, 20 | -14.6 | 9.97 | [342] |
| DyAgGe | AFM | 14.5, 11 | 0 | 10.86 | [342] |

| HoAgGe | AFM | 10.3 | 0 | 10.73 | [342] |
|--------|-----|------|-----|-------|-------|
| ErAgGe | AFM | 3.6** | 6.4 | 9.54 | [342] |
| TmAgGe | AFM | 4.1 | -14.4 | 7.9 | [323] |
| CeAgSn | AFM | 3.6 | -17 | 2.46 | [327] |
| PrAgSn | AFM | 3.8 | -7.5 | 3.56 | [327] |
| NdAgSn | AFM | 11.5 | -20 | 3.95 | [327,336] |
| EuAgSn | AFM | 6 | -31 | 7.96 | [189] |
| GdAgSn | AFM | 34 | -26 | 7.65 | [327] |
| TbAgSn | AFM | 35 | -42 | 9.33 | [327,336] |
| DyAgSn | AFM | 10.6 | -13 | 10.5 | [327] |
| HoAgSn | AFM | 10.8 | -11 | 10.7 | [327,336] |
| ErAgSn | AFM | 6.2 | -2 | 9 | [327] |
| TmAgSn | AFM | 4.1 | 17 | 7.49 | [349,350] |
| YbAgSn |  |  | -4.5 | 0.2 | [273] |
| EuAgAs | AFM | 11 | 19 | 7.45 | [168] |
| EuAgSb | AFM | 8 | 2 | 7.62 | [168] |
| EuAgBi | AFM | 10 | -4 | 7.39 | [168] |

**Symbols meaning:** g=spin glass temperature, r= re-entrant spin glass temperature,

# = CeFeSi-type, ## = CeScSi-type, ### = TiNiSi-type, *=along *a*-axis, ** = estimated from neutron diffraction data, *** = HTF II, **** = HTF I,  a* = $T_{max}$ ZFC

## References

[1]   M. Ohring, Magnetic properties of materials, in: Engineering Materials Science, Elsevier, 1995: pp. 711–746. https://doi.org/10.1016/b978-012524995-9/50038-6.

[2]   J. Mohapatra, J.P. Liu, Rare-earth-free permanent magnets: The past and future, in: Handbook of Magnetic Materials, Elsevier, 2018: pp. 1–57. https://doi.org/10.1016/bs.hmm.2018.08.001.

[3]   M. Staňo, O. Fruchart, Magnetic Nanowires and Nanotubes, in: Handbook of Magnetic Materials, Elsevier, 2018: pp. 155–267. https://doi.org/10.1016/bs.hmm.2018.08.002.

[4]   V. Krizakova, M. Perumkunnil, S. Couet, P. Gambardella, K. Garello, Spin-orbit torque switching of magnetic tunnel junctions for memory applications, in: Handbook of Magnetic Materials, Elsevier, 2022: pp. 1–53. https://doi.org/10.1016/bs.hmm.2022.10.001.

[5]   A.C. Lima, N. Pereira, P. Martins, S. Lanceros-Mendez, Magnetic materials for magnetoelectric coupling: An unexpected journey, in: Handbook of Magnetic Materials, Elsevier, 2020: pp. 57–110. https://doi.org/10.1016/bs.hmm.2020.09.002.

[6]   S. Gupta, Exotic rare earth-based materials for emerging spintronic technology, in: Including Actinides, Elsevier, 2023: pp. 99–140. https://doi.org/10.1016/bs.hpcre.2023.04.001.

[7]   J.M.D. Coey, Perspective and prospects for rare earth permanent magnets, Engineering (Beijing). 6 (2020) 119–131. https://doi.org/10.1016/j.eng.2018.11.034.

[8]   B.D. Cullity, C.D. Graham, Introduction to magnetic materials, John Wiley & Sons, Inc., Hoboken, NJ, USA, 2008. https://doi.org/10.1002/9780470386323.

[9]   S. Gupta, K.G. Suresh, Review on magnetic and related properties of RTX compounds, J. Alloys Compd. 618 (2015) 562–606. https://doi.org/10.1016/j.jallcom.2014.08.079.

[10]  Nikitin, S.A. ; Ovtchenkova, I.A. ; Skourski, Y.;. Morozkin, A. V., Magnetic properties of ternary scandium rare earth silicides and germanides, J. Alloys Compd. 345 (2002) 50–53. https://doi.org/10.1016/s0925-8388(02)00407-3.

[11]  Couillaud, S. Gaudin, E. Franco, V. Conde, A. ; Pöttgen, R. Heying, B. Rodewald, U. Ch. ; Chevalier, Bernard, The magnetocaloric properties of GdScSi and GdScGe, Intermetallics. 19 (2011) 1573–1578. https://doi.org/10.1016/j.intermet.2011.06.001.

[12]  S.D. Singh S. K. ; Mitra Chiranjib; Paulose P. L.; Manfrinetti Pietro; Palenzona A., The nature of magnetism in CeScSi and CeScGe, Journal of Physics: Condensed Matter. 13 (2001) 3753–3766. https://doi.org/10.1088/0953-8984/13/16/306.

[13]  S. Singh, S.K. Dhar, P. Manfrinetti, A. Palenzona, D. Mazzone, High magnetic transition temperatures in RScT (R=Pr, Nd and Sm; T=Si and Ge) compounds: multiple spin reorientations in PrScGe, Journal of Magnetism and Magnetic Materials. 269 (2004) 113–121. https://doi.org/10.1016/S0304-8853(03)00583-3.

[14]  I.A. Tskhadadze, V.V. Chernyshev, A.N. Streletskii, V.K. Portnoy, A.V. Leonov, I.A. Sviridov, I.V. Telegina, V.N. Verbetskii, Y.D. Seropegin, A.V. Morozkin, GdTiGe (CeScSi-type structure) and GdTiGe (CeFeSi-type structure) as the coherent phases with different magnetic and hydrogenation properties, Materials Research Bulletin. 34 (1999) 1773–1787. https://doi.org/10.1016/S0025-5408(99)00159-2.

[15]  A.V. Morozkin, S.A. Nikitin, Y.D. Seropegin, I.A. Sviridov, I.A. Tskhadadze, Structural and magnetic properties of new RRuGe compounds, Journal of Alloys and Compounds. 268 (1998) L1–L2. https://doi.org/10.1016/S0925-8388(97)00603-8.

[16]  T. Del Rose, A.K. Pathak, Y. Mudryk, V.K. Pecharsky, Distinctive exchange bias and unusual memory effects in magnetically compensated Pr $_{0.75}$ Gd $_{0.25}$ ScGe, Journal of Materials Chemistry C. 9 (2021) 181–188. https://doi.org/10.1039/D0TC05087C.

[17]  G.R. Bhimanapati, Z. Lin, V. Meunier, Y. Jung, J. Cha, S. Das, D. Xiao, Y. Son, M.S. Strano, V.R. Cooper, L. Liang, S.G. Louie, E. Ringe, W. Zhou, S.S. Kim, R.R. Naik, B.G. Sumpter, H. Terrones, F. Xia, Y. Wang, J. Zhu, D. Akinwande, N. Alem, J.A. Schuller, R.E. Schaak, M. Terrones, J.A. Robinson, Recent advances in two-dimensional

materials beyond graphene, ACS Nano. 9 (2015) 11509–11539.
https://doi.org/10.1021/acsnano.5b05556.

[18] S. Gupta, R. Ohshima, Y. Ando, T. Endo, Electrical transport properties of atomically thin WSe2 using perpendicular magnetic anisotropy metal contacts, J. Phys. D Appl. Phys. (2022). https://aip.scitation.org/doi/abs/10.1063/5.0079223.

[19] S. Gupta, F. Rortais, R. Ohshima, Y. Ando, T. Endo, Y. Miyata, M. Shiraishi, Approaching barrier-free contacts to monolayer MoS2 employing [Co/Pt] multilayer electrodes, NPG Asia Mater. 13 (2021). https://doi.org/10.1038/s41427-021-00284-1.

[20] Y. Uwatoko, M. Kosaka, T. Sigeoka, Magnetic properties of single crystals CeScGe and CeScSi, Physica B: Condensed Matter. 259–261 (1999) 114–115. https://doi.org/10.1016/S0921-4526(98)00968-5.

[21] A.V. Morozkin, New ternary compounds with CeFeSi-type structure (LuTiSi, LuTiGe) and CeScSi-type structure (ZrVGe and HfVGe), J. Alloys Compd. 289 (1999) L10–L11. https://doi.org/10.1016/s0925-8388(99)00179-6.

[22] V. Klosek, A. Vernière, B. Ouladdiaf, B. Malaman, Magnetic properties of CeFeSi-type RTiSi compounds (R=Gd–Tm, Lu, Y) from magnetic measurements and neutron diffraction, J. Magn. Magn. Mater. 246 (2002) 233–242. https://doi.org/10.1016/s0304-8853(02)00060-4.

[23] Morozkin, A.V. ; Seropegin, Y.D. ; Leonov, A.V. ; Sviridov, I.A. ; Tskhadadze, I. A; Nikitin, S. A., Crystallographic data of new ternary CeFeSi-type RTiGe (R=Y,Gd-Tm) compounds, J. Alloys Compd. 267 (1998) L14–L15. https://doi.org/10.1016/s0925-8388(97)00473-8.

[24] R.;. V. Welter A.;. Venturini, High rare earth sublattice ordering temperatures in new CeFeSi-type RTiGe (R≡La–Nd, Sm) compounds, J. Alloys Compd. 283 (1999) 54–58. https://doi.org/10.1016/s0925-8388(98)00904-9.

[25] A. Guzik, The magnetic and transport properties of the new GdTiSb compound, J. Alloys Compd. 423 (2006) 40–42.

[26] B. Chevalier, W. Hermes, E. Gaudin, R. Pöttgen, New high temperature modification of CeTiGe: structural characterization and physical properties, J. Phys. Condens. Matter. 22 (2010) 146003. https://doi.org/10.1088/0953-8984/22/14/146003.

[27] S.A. Nikitin, I.A. Tskhadadze, I.V. Telegina, A.V. Morozkin, Y.D. Seropegin, Magnetic properties of RTiGe compounds, J. Magn. Magn. Mater. 182 (1998) 375–380. https://doi.org/10.1016/s0304-8853(97)01036-6.

[28] Tencé, S. Gaudin, E. Isnard, O. Chevalier, Bernard, Magnetic and magnetocaloric properties of the high-temperature modification of TbTiGe, J. Phys. Condens. Matter. 24 (2012) 296002–296002. https://doi.org/10.1088/0953-8984/24/29/296002.

[29] M Deppe, N Caroca-Canales, S Hartmann, N Oeschler and C Geibel, New non-magnetically ordered heavy-fermion system CeTiGe, J. Phys. Condens. Matter. 21 (2009) 206001-NA. https://doi.org/10.1088/0953-8984/21/20/206001.

[30] M. Deppe, S. Lausberg, F. Weickert, M. Brando, Y. Skourski, N. Caroca-Canales, C. Geibel, and F. Steglich, Pronounced first-order metamagnetic transition in the paramagnetic heavy-fermion system CeTiGe, Phys. Rev. B: Condens. Matter Mater. Phys. 85 (2012) 060401-1-060401–5. https://doi.org/10.1103/physrevb.85.060401.

[31] A. Vernière, V. Klosek, R. Welter, G. Venturini, O. Isnard and B. Malaman, Neutron diffraction study of the CeFeSi-type RTiGe compounds (R=Pr, Nd, Tb–Er), J. Magn. Magn. Mater. 234 (2001) 261–273. https://doi.org/10.1016/s0304-8853(01)00352-3.

[32] Spatz, P. ; Gross, K.J. ; Züttel, A. Fauth, F. Fischer, P. Schlapbach, Louis, CeMnA1Hx, a new metal hydride, J. Alloys Compd. 261 (1997) 263–268. https://doi.org/10.1016/s0925-8388(97)00188-6.

[33] Dhar, S.K. ; Mitra, C. ; Manfrinetti, P. Palenzona, R. ; Palenzona, A., Synthesis and Magnetic Properties of NdMnAl and Some RMnIn (R = Rare Earth), J. Phase Equilib. Diffus. 23 (2002) 79–82. https://doi.org/10.1361/105497102770332252.

[34] H. Oesterreicher, Structural, magnetic and neutron diffraction studies on ErMn2ErAl2 and TbMnAl, J. Phys. Chem. Solids. 33 (1972) 1031–1039. https://doi.org/10.1016/s0022-3697(72)80263-4.

[35] J. Sakurai, K. Inaba, J. Schweizer, Spin glass states in compounds RMnGa (R; rare earth metals), Solid State Commun. 87 (1993) 1073–1076. https://doi.org/10.1016/0038-1098(93)90563-3.

[36] Welter, R. Venturini, G. ; Malaman, B., High rare earth sublattice ordering temperatures in RMnSi compounds (R LaSm, Gd) studied by susceptibility measurements and neutron diffraction, J. Alloys Compd. 206 (1994) 55–71. https://doi.org/10.1016/0925-8388(94)90011-6.

[37] Welter, R. ; Venturini, G. ; Ijjaali, I. ; Malaman, Bernard, Magnetic study of the new TiNiSi-type TbMnSi, DyMnSi and NdMnGe compounds, J. Magn. Magn. Mater. 205 (1999) 221–233. https://doi.org/10.1016/s0304-8853(99)00512-0.

[38] A.V. Morozkin, Y.D. Seropegin, I.A. Sviridov, V.A. Moskalev, I.A. Tskhadadze, I.G. Ryabinkin, Phase equilibria in the Sm,Tb,Tm–Mn–Si systems at 870/1070 K, J. Alloys Compd. 280 (1998) 178–187. https://doi.org/10.1016/s0925-8388(98)00742-7.

[39] Welter, R. ; Venturini, G. ; Ressouche, E. ; Malaman, Bernard, Crystallographic data and magnetic properties of new CeFeSi-type RMnGe compounds (R = FaSm) studied by magnetization and neutron diffraction measurements, J. Alloys Compd. 228 (1995) 59–74. https://doi.org/10.1016/0925-8388(95)01659-7.

[40] Klosek, V. ; Vernière, A. ; Ouladdiaf, B. ; Malaman, Bernard, Crystal and magnetic structures of the R(=Y, Dy–Tm)MnGe compounds, J. Magn. Magn. Mater. 256 (2003) 69–92. https://doi.org/10.1016/s0304-8853(02)00383-9.

[41] Ivanova, T.I. ; Nikitin, S.A. ; Morozkin, A.V. ; Ovchenkova, I.A. ; Bogdanov, A.E. ; Suski, W. ; Warchulska, J. ; Gilewski, A., A magnetic study of TiNiSi-type GdMn1−xTixGe alloys, J. Alloys Compd. 365 (2004) 15–20. https://doi.org/10.1016/s0925-8388(03)00645-5.

[42] Chevalier, B. Bobet, J.-L.;. Etourneau, Jean, Spin-glass behaviour in GdMnAl obtained by mechanical grinding, J. Alloys Compd. 339 (2002) 35–39. https://doi.org/10.1016/s0925-8388(01)01994-6.

[43] M. Klimczak, E. Talik, Magnetocaloric effect of GdTX (T = Mn, Fe, Ni, pd, X=Al, in) and GdFe6Al6ternary compounds, J. Phys. Conf. Ser. 200 (2010) 092009. https://doi.org/10.1088/1742-6596/200/9/092009.

[44] G.A. Cabrera-Pasca, J.F. Magno, W.L. Ferreira, A.C. Campos, B. Bosch-Santos, T.S.N. Sales, L.F.D. Pereira, A. Burimova, R.N. Saxena, R.S. Freitas, A.W. Carbonari, Local inspection of magnetic properties in GdMnIn by measuring hyperfine interactions, AIP Adv. 11 (2021) 015322. https://doi.org/10.1063/9.0000037.

[45] Nikitin, S.A. ; Tskhadadze, I. A; Makarova, M. Morozkin, A. V., Negative magnetic moment induced by a magnetic field in the region of the magnetic phase transition in SmMnSi compound, J. Phys. D Appl. Phys. 32 (1999) L23–L25. https://doi.org/10.1088/0022-3727/32/6/001.

[46] M.K. Ray, B. Maji, K. Motla, Sajilesh, R.P. Singh, Multiple magnetization reversal and field induced orbital moment switching in intermetallic SmMnSi compound, J. Appl. Phys. 128 (2020) 073909. https://doi.org/10.1063/5.0017821.

[47] S. Gupta, F. Matsukura, H. Ohno, Properties of sputtered full Heusler alloy Cr2MnSb and its application in a magnetic tunnel junction, J. Phys. D Appl. Phys. 52 (2019)

495002. https://doi.org/10.1088/1361-6463/ab3fc6.

[48] Y. Venkateswara, S. Gupta, S.S. Samatham, M.R. Varma, Enamullah, K.G. Suresh, A. Alam, Competing magnetic and spin-gapless semiconducting behavior in fully compensated ferrimagnetic CrVTiAl: Theory and experiment, Phys. Rev. B. 97 (2018). https://doi.org/10.1103/physrevb.97.054407.

[49] E. Hovestreydt, A three-dimensional structure-stability diagram for ternary equiatomic RTM intermetallic compounds, Journal of the Less Common Metals. 143 (1988) 25–30. https://doi.org/10.1016/0022-5088(88)90026-4.

[50] Ventutini, G. ; ljjaali, I. ; Ressouche, E. ; Malaman, Bernard, Neutron diffraction study of the HoMnSi, LuMnSi and Sc0.9Lu0.1MnSi compounds, J. Alloys Compd. 256 (1997) 65–75. https://doi.org/10.1016/s0925-8388(96)03009-5.

[51] Venturini, G. ; Malaman, B. Ressouche, E., Neutron diffraction study of the TbMnGe compound, J. Alloys Compd. 243 (1996) 98–105. https://doi.org/10.1016/s0925-8388(96)02388-2.

[52] H. Oesterreicher, Structural studies of rare-earth compounds RFeAl, Journal of the Less Common Metals. 25 (1971) 341–342. https://doi.org/10.1016/0022-5088(71)90160-3.

[53] A.V. Morozkin, V.K. Genchel, A.V. Garshev, V.O. Yapaskurt, R. Nirmala, S. Quezado, S.K. Malik, MgZn2-type Ho, Er, TmFeGa rare earth compounds: Crystal structure and magnetic properties, J. Solid State Chem. 253 (2017) 238–241. https://doi.org/10.1016/j.jssc.2017.06.003.

[54] O.I. Bodak, E.I. Gladyshevskii, P.I. Kripyakevich, Crystal structure of CeFeSi and related compounds, J. Struct. Chem. 11 (1970) 283–288. https://doi.org/10.1007/bf00745235.

[55] Welter, R. Venturini, G. ; Malaman, B., Magnetic properties of RFeSi (RLa-Sm, Gd-Dy) from susceptibility measurements and neutron diffraction studies, J. Alloys Compd. 189 (1992) 49–58. https://doi.org/10.1016/0925-8388(92)90045-b.

[56] H. Oesterreicher, Magnetic Properties of Scatter Order Compounds RFeA1 (R = Gd, Tb, Dy, Ho, Er, Tm, Lu, and Y), Phys. Status Solidi . 40 (1977) K139–K143. https://doi.org/10.1002/pssa.2210400250.

[57] J. Kaštil, P. Javorský, J. Kamarád, L.V.B. Diop, O. Isnard, Z. Arnold, Magnetic and magnetocaloric properties of partially disordered RFeAl (R = Gd, Tb) intermetallic, Intermetallics (Barking). 54 (2014) 15–19. https://doi.org/10.1016/j.intermet.2014.05.008.

[58] Dong, Q. Shen, B.-G. Chen, J. Shen, J. Zhang, H.-W. Sun, Jirong, Magnetic entropy change and refrigerant capacity in GdFeAl compound, J. Appl. Phys. 105 (2009) 7A305-NA. https://doi.org/10.1063/1.3059372.

[59] H.S. Nair, A.M. Strydom, Exchange bias-like effect in TbFeAl intermetallic induced by atomic disorder, ArXiv [Cond-Mat.Str-El]. (2016). http://arxiv.org/abs/1605.02381.

[60] Y. Zhang, G. Wilde, X. Li, Z. Ren, L. Li, Magnetism and magnetocaloric effect in the ternary equiatomic REFeAl (RE = Er and Ho) compounds, Intermetallics (Barking). 65 (2015) 61–65. https://doi.org/10.1016/j.intermet.2015.06.003.

[61] Mulders, A. M; Kraan, W.H. ; Gubbens, P.C. M; Buschow, K.H. ; Stüßer, N. Hofmann, Michael, Observation of magnetic nanodomains in TmFeAl, J. Alloys Compd. 299 (2000) 88–93. https://doi.org/10.1016/s0925-8388(99)00803-8.

[62] Zhang, H. Sun, Y. Niu, E. ; Yang, L. Shen, J. Hu, F. Sun, J.R. ; Shen, B. G., Large magnetocaloric effects of RFeSi (R = Tb and Dy) compounds for magnetic refrigeration in nitrogen and natural gas liquefaction, Appl. Phys. Lett. 103 (2013) 202412-NA. https://doi.org/10.1063/1.4832218.

[63] R.K. Chouhan, T.J. Del Rose, Y. Mudryk, V.K. Pecharsky, Inducing Fe moment in

LaFeSi with p-block element substitution, AIP Adv. 12 (2022) 035130. https://doi.org/10.1063/9.0000334.

[64] Zhang, H.-W. Sun, Y. Yang, L. Niu, E. ; Wang, H.S. Hu, F.X. ; Sun, J.R. ; Shen, B. G., Successive inverse and normal magnetocaloric effects in HoFeSi compound, J. Appl. Phys. 115 (2014) 063901-NA. https://doi.org/10.1063/1.4865297.

[65] Zhang, H. Shen, B.G. ; Xu, Z.Y. ; Shen, J. Hu, F. Sun, J.R. ; Long, Yi, Large reversible magnetocaloric effects in ErFeSi compound under low magnetic field change around liquid hydrogen temperature, Appl. Phys. Lett. 102 (2013) 092401-NA. https://doi.org/10.1063/1.4794415.

[66] H. Oesterreicher, Structural studies of rare-earth (R) compounds RCoAl, Journal of the Less Common Metals. 25 (1971) 228–230. https://doi.org/10.1016/0022-5088(71)90134-2.

[67] O. Niehaus, R.-D. Hoffmann, S. Tencé, B. Chevalier, R. Pöttgen, Intermediate-valent CeCoAl – a commensurate modulated structure with short Ce–Co distances, Z. Kristallogr. Cryst. Mater. 230 (2015) 579–591. https://doi.org/10.1515/zkri-2015-1856.

[68] Goraus, J. Ślebarski, A. Fijałkowski, Marcin, Electronic and magnetic properties of CeCoGa, Intermetallics. 32 (2013) 219–224. https://doi.org/10.1016/j.intermet.2012.08.011.

[69] Welter, R. Venturini, G. ; Ressouche, E. ; Malaman, B., Magnetic properties of RCoSi (R La—Sm, Gd, Tb) compounds from susceptibility measurements and neutron diffraction studies, J. Alloys Compd. 210 (1994) 279–286. https://doi.org/10.1016/0925-8388(94)90150-3.

[70] Welter, R. Venturini, G. ; Malaman, B. ; Ressouche, E., Crystallographic data and magnetic properties of new RCoGe (RLa-Nd) compounds from susceptibility measurements and neutron diffraction study, J. Alloys Compd. 201 (1993) 191–196. https://doi.org/10.1016/0925-8388(93)90883-o.

[71] A.E. Dwight, P.P. Vaishnava, C.W. Kimball, J.L. Matykiewicv, Crystal structure and Mössbauer effect study of evuiatomic (Sc,Y,Ln)-Co-(Si,Ge,Sn) ternary compounds (Ln • Gd — Tm, Lu), J. Less-Common Met. 119 (1986) 319–326. https://doi.org/10.1016/0022-5088(86)90692-2.

[72] X.X. Zhang, F.W. Wang, G.H. Wen, Magnetic entropy change in RCoAl (R = Gd, Tb, Dy, and Ho) compounds: candidate materials for providing magnetic refrigeration in the temperature range 10 K to 100 K, J. Phys. Condens. Matter. 13 (2001) L747–L752. https://doi.org/10.1088/0953-8984/13/31/102.

[73] Z.-J. Mo, J. Shen, L.-Q. Yan, C.-C. Tang, L.-C. Wang, J.-F. Wu, J.-R. Sun, B.-G. Shen, Magnetic property and magnetocaloric effect in TmCoAl compound, Intermetallics (Barking). 56 (2015) 75–78. https://doi.org/10.1016/j.intermet.2014.08.006.

[74] Chevalier, B. Matar, S. F., Effect of H insertion on the magnetic, electronic, and structural properties of CeCoSi, Phys. Rev. B: Condens. Matter Mater. Phys. 70 (2004) 174408-NA. https://doi.org/10.1103/physrevb.70.174408.

[75] Gupta, S. Suresh, K. G., Observation of giant magnetocaloric effect in HoCoSi, Mater. Lett. 113 (2013) 195–197. https://doi.org/10.1016/j.matlet.2013.09.076.

[76] Y.V. Knyazev, A.V. Lukoyanov, Y.I. Kuz'min, S. Gupta, K.G. Suresh, Electronic structure and optical properties of the HoCoSi and ErNiSi compounds, J. Exp. Theor. Phys. 123 (2016) 638–642. https://doi.org/10.1134/s1063776116090132.

[77] J.W. Xu, X.Q. Zheng, S.X. Yang, L. Xi, J.Y. Zhang, Y.F. Wu, S.G. Wang, J. Liu, L.C. Wang, Z.Y. Xu, B.G. Shen, Giant low field magnetocaloric effect in TmCoSi and TmCuSi compounds, J. Alloys Compd. 843 (2020) 155930. https://doi.org/10.1016/j.jallcom.2020.155930.

[78] Y. Zhang, Q. Dong, X. Zheng, Y. Liu, S. Zuo, J. Xiong, B. Zhang, X. Zhao, R. Li, D. Liu, F.-X. Hu, J. Sun, T. Zhao, B. Shen, Complex magnetic properties and large magnetocaloric effects in RCoGe (R=Tb, Dy) compounds, AIP Adv. 8 (2018) 056418. https://doi.org/10.1063/1.5007114.

[79] Chevalier, B. Gaudin, E. Weill, F. Bobet, Jean-Louis, Hydrogenation and physical properties of the ternary germanide CeCoGe: an anisotropic expansion of the unit cell, Intermetallics. 12 (2004) 437–442. https://doi.org/10.1016/j.intermet.2003.12.007.

[80] Leciejewicz, J. ; Stüsser, N. ; Kolenda, M. ; Szytuła, A. ; Zygmunt, A., Magnetic ordering in HoCoSi and TbCoGe, J. Alloys Compd. 240 (1996) 164–169. https://doi.org/10.1016/0925-8388(96)02307-9.

[81] Görlich, E.A. ; Kmieć, R. ; łątka, K. ; Szytuła, A. ; Zygmunt, A., Magnetic properties and 119Sn hyperfine interactions investigated in RCoSn (R=Tb, Dy, Ho, Er) compounds, J. Phys. Condens. Matter. 6 (1994) 11127–11139. https://doi.org/10.1088/0953-8984/6/50/020.

[82] Bażela, W. ; Leciejewicz, J. ; Stuesser, N. ; Szytu[la, A. ; Zygmunt, A., Magnetic ordering in RCoSn (R = Dy, Ho, Er) compounds, J. Magn. Magn. Mater. 137 (1994) 219–223. https://doi.org/10.1016/0304-8853(94)90208-9.

[83] André, G. Bażela, W. ; Bourée, F. Guillot, M. ; Oleś, A.M. ; Sikora, W. Szytuła, A. ; Zygmunt, A, Magnetic ordering of TbCoSn and HoCoSn studied by neutron diffraction and magnetic measurements, J. Alloys Compd. 221 (1995) 254–263. https://doi.org/10.1016/0925-8388(94)01469-8.

[84] H. Oesterreicher, Structural and magnetic studies on rare-earth compounds RNiAl and RCuAl, Journal of the Less Common Metals. 30 (1973) 225–236. https://doi.org/10.1016/0022-5088(73)90109-4.

[85] Prchal, J. Javorský, P. Rusz, J. de Boer, F.S. ; Diviš, M. Kitazawa, H. Dönni, A. Daniš, S. Sechovský, Vladimír, Structural discontinuity in the hexagonalRTAlcompounds: Experiments and density-functional theory calculations, Phys. Rev. B: Condens. Matter Mater. Phys. 77 (2008) 134106-NA. https://doi.org/10.1103/physrevb.77.134106.

[86] F. Merlo, S. Cirafici, F. Canepa, Structural anomaly in GdNiAl: a crystallographic, electric and magnetic investigation, J. Alloys Compd. 266 (1998) 22–25. https://doi.org/10.1016/s0925-8388(97)00505-7.

[87] Jarosz, J. ; Talik, E. Mydlarz, T. Kusz, J. Böhm, H. ; Winiarski, Antoni, Crystallographic, electronic structure and magnetic properties of the GdTAl; T=Co, Ni and Cu ternary compounds, J. Magn. Magn. Mater. 208 (2000) 169–180. https://doi.org/10.1016/s0304-8853(99)00592-2.

[88] Kaštil, J. Klicpera, M. Prchal, J. Míšek, M. Prokleška, J. Javorský, Pavel, Effect of hydrostatic and uniaxial pressure on structural and magnetic transitions in TbNiAl, J. Alloys Compd. 585 (2014) 98–102. https://doi.org/10.1016/j.jallcom.2013.09.129.

[89] Chevalier, B. Bobet, J.-L. Gaudin, E. Pasturel, M. Etourneau, Jean, The ternary gallide CeNiGa: polymorphism and hydrogen absorption, J. Solid State Chem. 168 (2002) 28–33. https://doi.org/10.1006/jssc.2002.9671.

[90] F. Canepa, S. Cirafici, F. Merlo, A. Palenzona, Physical properties of GdNiGa, J. Alloys Compd. 279 (1998) L11–L12. https://www.scopus.com/inward/record.uri?eid=2-s2.0-0032475693&partnerID=40&md5=331db114c976c1108358b13d67b2dbea.

[91] P.A. Kotsanidis, J.K. Yakinthos, Classical antiferromagnetism in the TbNiGa compound, J. Magn. Magn. Mater. 81 (1989) 159–162. https://doi.org/10.1016/0304-8853(89)90245-x.

[92] P.A. Kotsanidis, J.K. Yakinthos, E. Ressouche, V.N. Nguyen, Transversal sine-modulated magnetic structure of DyNiGa, J. Magn. Magn. Mater. 131 (1994) 135–138.

https://doi.org/10.1016/0304-8853(94)90020-5.

[93] Kotsanidis, P.A. ; Yakinthos, J.K. ; Semutelou, I. ; Roudaut, E., Transversal modulated magnetic structure of ErNiGa, J. Magn. Magn. Mater. 116 (1992) 95–98. https://doi.org/10.1016/0304-8853(92)90149-i.

[94] Y.-X. Wang, H. Zhang, M.-L. Wu, K. Tao, Y.-W. Li, T. Yan, K.-W. Long, T. Long, Z. Pang, Y. Long, Large reversible magnetocaloric effect induced by metamagnetic transition in antiferromagnetic HoNiGa compound, Chin. Physics B. 25 (2016) 127104. https://doi.org/10.1088/1674-1056/25/12/127104.

[95] Kotsanidis, P.A. ; Papathanassiou, G.F. ; Yakinthos, J.K. ; Ressouche, E. ; Nguyen, V. N., Crystal and magnetic structure of TmNiGa, J. Magn. Magn. Mater. 176 (1997) 255–260. https://doi.org/10.1016/s0304-8853(97)00467-8.

[96] Trovarelli, O. ; Geibel, C. ; Cardoso, R. ; Mederle, S. ; Borth, R. ; Buschinger, B. ; Grosche, F.M. ; Grin, Y. Sparn, G. ; Steglich, Frank, Low-temperature properties of the Yb-based heavy-fermion antiferromagnets YbPtIn, YbRhSn, and YbNiGa, Phys. Rev. B: Condens. Matter Mater. Phys. 61 (2000) 9467–9474. https://doi.org/10.1103/physrevb.61.9467.

[97] Canepa, F. Napoletano, M. ; Palenzona, A. ; Merlo, F. Cirafici, Salvino, Magnetocaloric properties of GdNiGa and GdNiIn intermetallic compounds, J. Phys. D Appl. Phys. 32 (1999) 2721–2725. https://doi.org/10.1088/0022-3727/32/21/303.

[98] I.I. Bulyk, V.A. Yartys, R.V. Denys, Y.M. Kalychak, I.R. Harris, Hydrides of the RNiIn (R=La, Ce, Nd) intermetallic compounds: crystallographic characterisation and thermal stability, J. Alloys Compd. 284 (1999) 256–261. https://doi.org/10.1016/s0925-8388(98)00953-0.

[99] A.L. Lapolli, R.N. Saxena, J. Mestnik-Filho, D.M.T. Leite, A.W. Carbonari, Local investigation of magnetism at R and In sites in RNiIn (R=Gd, Tb, Dy, Ho) compounds, J. Appl. Phys. 101 (2007) 09D510. https://doi.org/10.1063/1.2709421.

[100] H. Zhang, Z.Y. Xu, X.Q. Zheng, J. Shen, F.X. Hu, J.R. Sun, B.G. Shen, Magnetocaloric effects in RNiIn (R = Gd-Er) intermetallic compounds, J. Appl. Phys. 109 (2011) 123926. https://doi.org/10.1063/1.3603044.

[101] Baran, S. Kaczorowski, D. Arulraj, A. ; Penc, B. ; Szytuła, A., Investigation of thermodynamic properties and magnetic ordering in TmNiIn, J. Magn. Magn. Mater. 323 (2011) 833–837. https://doi.org/10.1016/j.jmmm.2010.11.028.

[102] A. Szytuła, M. Bałanda, M. Hofmann, J. Leciejewicz, M. Kolenda, B. Penc, A. Zygmunt, Antiferromagnetic properties of ternary silicides RNiSi (R=Tb–Er), J. Magn. Magn. Mater. 191 (1999) 122–132. https://doi.org/10.1016/s0304-8853(98)00330-8.

[103] Pasturel, M. Bobet, J.-L.,;. Isnard, O. Chevalier, Bernard, Unusual increase of the Kondo effect by hydrogenation: case of the ternary silicide CeNiSi, J. Alloys Compd. 384 (2004) 39–43. https://doi.org/10.1016/j.jallcom.2004.03.126.

[104] Lee, W.H. ; Yang, F.A. ; Shih, C.R. ; Yang, H. D., Crystal structure and superconductivity in the Ni-based ternary compound LaNiSi, Phys. Rev. B Condens. Matter. 50 (1994) 6523–6525. https://doi.org/10.1103/physrevb.50.6523.

[105] Pasturel, M. Weill, F. Bourée, F. Bobet, J.-L. Chevalier, Bernard, Hydrogenation of the ternary silicides RENiSi (RE = Ce, Nd) crystallizing in the tetragonal LaPtSi-type structure, J. Alloys Compd. 397 (2005) 17–22. https://doi.org/10.1016/j.jallcom.2005.01.015.

[106] S. Gupta, K.G. Suresh, A.V. Lukoyanov, Y.V. Knyazev, Y.I. Kuz'min, Understanding the magnetic, electronic and optical properties of ternary rare earth intermetallic compound HoNiSi, J. Alloys Compd. 650 (2015) 542–546. https://doi.org/10.1016/j.jallcom.2015.08.036.

[107]   Kotsanidis, P.A. ; Yakinthos, J.K. ; Gamari-Seale, E., Susceptibility of the rare earth ternary equiatomic germanides RNiGe (R ≡ Gd, Tb, Dy, Ho, Er, Tm and Y), Journal of the Less Common Metals. 157 (1990) 295–300. https://doi.org/10.1016/0022-5088(90)90184-l.

[108]   Weill, F. Pasturel, M. Bobet, J.-L. Chevalier, Bernard, Ordering phenomena in intermetallic CeMX (M = Ni, Cu and X = Si, Ge, Sn) upon hydrogenation, J. Phys. Chem. Solids. 67 (2006) 1111–1116. https://doi.org/10.1016/j.jpcs.2006.01.032.

[109]   A.E. Dwight, Crystal structure of RENiSn and REPdsn (RE rare earth) equiatomic compounds, Journal of the Less Common Metals. 93 (1983) 411–413. https://doi.org/10.1016/0022-5088(83)90195-9.

[110]   Kasaya, M. Tani, T. ; Iga, F. ; Kasuya, Tadao, Electrical resistivity, hall constant and magnetic susceptibility in the Kondo states Ce(Ni1−xPdx)Sn and YbNiSn, J. Magn. Magn. Mater. 76 (1988) 278–280. https://doi.org/10.1016/0304-8853(88)90395-2.

[111]   Hartjes, K. Jeitschko, Wolfgang, Crystal structures and magnetic properties of the lanthanoid nickel antimonides LnNiSb (Ln = LaNd, Sm, GdTm, Lu), J. Alloys Compd. 226 (1995) 81–86. https://doi.org/10.1016/0925-8388(95)01573-6.

[112]   L. Menon, A. Agarwal, S.K. Malik, Magnetic and transport measurements on CeNiAl, Physica B Condens. Matter. 230–232 (1997) 201–203. https://doi.org/10.1016/s0921-4526(96)00590-x.

[113]   Bobet, J.-L.;. Chevalier, B. Darriet, B. Nakhl, M. Weill, F. Etourneau, Jean, Hydrogen absorption properties of CeNiAl: influence on its crystal structure and magnetic behaviour, J. Alloys Compd. 317 (2001) 67–70. https://doi.org/10.1016/s0925-8388(00)01358-x.

[114]   P. Javorský, V. Sechovský, R.R. Arons, P. Burlet, E. Ressouche, P. Svoboda, G. Lapertot, Neutron diffraction study of magnetic ordering in NdNiAl and PrNiAl, J. Magn. Magn. Mater. 164 (1996) 183–186. https://doi.org/10.1016/s0304-8853(96)00381-2.

[115]   C. Schank, G. Olesch, J. Köhler, U. Tegel, U. Klinger, J. Diehl, S. Klimm, G. Sparn, S. Horn, C. Geibel, F. Steglich, YbNiAl: A new Yb-based heavy-fermion antiferromagnet, J. Magn. Magn. Mater. 140–144 (1995) 1237–1238. https://doi.org/10.1016/0304-8853(94)01423-x.

[116]   Ehlers, G. Maletta, H., Frustrated magnetic moments in RNiAl intermetallic compounds, Physica B Condens. Matter. 234 (1997) 667–669. https://doi.org/10.1016/s0921-4526(96)01081-2.

[117]   G. Ehlers, H. Casalta, R.E. Lechner, H. Maletta, Dynamics of frustrated magnetic moments in TbNiAl, Appl. Phys. A Mater. Sci. Process. 74 (2002) s613–s615. https://doi.org/10.1007/s003390101126.

[118]   L. He, Zero-magnetization ferromagnet induced by hydrogenation, Solid State Commun. 151 (2011) 985–987. https://doi.org/10.1016/j.ssc.2011.05.004.

[119]   A.V. Andreev, N.V. Mushnikov, T. Goto, J. Prchal, Magnetic anisotropy of a DyNiAl single crystal, Physica B Condens. Matter. 346–347 (2004) 201–205. https://doi.org/10.1016/j.physb.2004.01.050.

[120]   P. Javorský, H. Sugawara, D. Rafaja, F. Bourdarot, H. Sato, Magnetic ordering in HoNiAl-single crystal study, J. Alloys Compd. 323–324 (2001) 472–476. https://doi.org/10.1016/s0925-8388(01)01128-8.

[121]   Y.-H. Ding, F.-Z. Meng, L.-C. Wang, R.-S. Liu, J. Shen, Metamagnetic transition and reversible magnetocaloric effect in antiferromagnetic DyNiGa compound, Chin. Physics B. 29 (2020) 077501. https://doi.org/10.1088/1674-1056/ab90f3.

[122]   P. Kotsanidis, I. Semitelou, J.K. Yakinthos, E. Roudaut, Sine modulated magnetic

structure of HoNiGa, J. Magn. Magn. Mater. 102 (1991) 67–70. https://doi.org/10.1016/0304-8853(91)90267-e.

[123]   Gondek, L. ; Szytuła, A. ; Baran, S. Hernandez-Velasco, J., Neutron diffraction studies of the hexagonal RTIn (R=rare earth, T=Au or Ni) compounds, J. Magn. Magn. Mater. 272 (2004) E443–E444. https://doi.org/10.1016/j.jmmm.2003.12.641.

[124]   K.T. Matsumoto, M. Kobayashi, N. Morioka, K. Hiraoka, Specific heat of Cu substituted GdNiSi, Jpn. J. Appl. Phys. (2008). 60 (2021) 043001. https://doi.org/10.35848/1347-4065/abeb8c.

[125]   H. Zhang, Y. Li, E. Liu, Y. Ke, J. Jin, Y. Long, B. Shen, Giant rotating magnetocaloric effect induced by highly texturing in polycrystalline DyNiSi compound, Sci. Rep. 5 (2015) 11929. https://doi.org/10.1038/srep11929.

[126]   Gupta, S. Rawat, R. Suresh, K. G., Field induced large magnetocaloric effect and magnetoresistance in ErNiSi, Appl. Phys. Lett. 105 (2014) 012403-NA. https://doi.org/10.1063/1.4887336.

[127]   Chevalier, B. Pasturel, M. Bobet, J.-L.;. Decourt, R. Etourneau, J. Isnard, O. Marcos, J. Sanchez; Fernández, J. Rodríguez, Hydrogenation of the ternary compounds CeNiX (X=Al, Ga, In, Si, Ge and Sn): influence on the valence state of cerium, J. Alloys Compd. 383 (2004) 4–9. https://doi.org/10.1016/j.jallcom.2004.04.006.

[128]   G. André, F. Bourée, M. Kolenda, A. Oleś, A. Pacyna, M. Pinot, W. Sikora, A. Szytuła, Crystal and magnetic structure of TbNiGe and DyNiGe compounds, J. Magn. Magn. Mater. 116 (1992) 375–385. https://doi.org/10.1016/0304-8853(92)90119-9.

[129]   André, G. Bourée, F. Bombik, A. ; Oles, A. ; Sikora, W. Kolenda, M. ; Szytuła, A., Neutron diffraction and magnetic study of the HoNiGe and ErNiGe compounds, J. Magn. Magn. Mater. 127 (1993) 83–92. https://doi.org/10.1016/0304-8853(93)90200-l.

[130]   Routsi, C.D. ; Yakinthos, J.K. ; Gamari-Seale, E., Magnetic characteristics of some RNiSn ternary equiatomic alloys, J. Magn. Magn. Mater. 98 (1991) 257–260. https://doi.org/10.1016/0304-8853(91)90240-b.

[131]   Echizen, Y. ; Umeo, K. Takabatake, Toshiro, Superconductivity and magnetoresistance in a single-crystal LaNiSn, Solid State Commun. 111 (1999) 153–157. https://doi.org/10.1016/s0038-1098(99)00160-x.

[132]   Routsi, C.D. ; Yakinthos, J.K. ; Gamari-Seale, H., Magnetic characteristics of some RNiSn (R = Ce, Pr, Nd, Sm) and RRhSn (R = Ce, Pr, Nd) compounds, J. Magn. Magn. Mater. 117 (1992) 79–82. https://doi.org/10.1016/0304-8853(92)90294-x.

[133]   M. Kurisu, Y. Andoh, Magnetic properties of a SmNiSn single crystal, Physica B Condens. Matter. 327 (2003) 393–396. https://doi.org/10.1016/s0921-4526(02)01754-4.

[134]   Szytuła, A. ; Penc, B. ; Stüsser, N., Incommensurate magnetic ordering in NdNiSn, J. Magn. Magn. Mater. 265 (2003) 94–97. https://doi.org/10.1016/s0304-8853(03)00229-4.

[135]   Y. Öner, O. Kamer, J.H. Ross Jr, Magnetic and electrical properties of NdNiSn, J. Alloys Compd. 460 (2008) 69–73. https://doi.org/10.1016/j.jallcom.2007.06.040.

[136]   Takabatake, T. Nakamoto, G. Tanaka, H. Bando, Y. ; Fujii, H. Nishigori, S. ; Goshima, H. ; Suzuki, T. ; Fujita, T. Oguro, I. Hiraoka, T. ; Malik, S. K., Coherence Kondo gap in CeNiSn and CeRhSb, Physica B Condens. Matter. 199 (1994) 457–462. https://doi.org/10.1016/0921-4526(94)91868-6.

[137]   Izawa, K. ; Suzuki, T. Fujita, T. Takabatake, T. Nakamoto, G. Fujii, H. Maezawa, Kunihiko, Anomalous field dependence of specific heat of CeNiSn below 1 K, J. Magn. Magn. Mater. 177 (1998) 395–396. https://doi.org/10.1016/s0304-8853(97)00408-3.

[138]   Andoh, Y. ; Kurisu, M. Kawano, S. Oguro, Isamu, Magnetic properties and magnetic anisotropy of RNiSn (R = Gd, Ho and Er) single crystals, J. Magn. Magn. Mater. 177 (1998) 1063–1064. https://doi.org/10.1016/s0304-8853(98)80008-5.

[139]   Karla, I. ; Pierre, J. ; Skolozdra, R. V., Physical properties and giant magnetoresistance in RNiSb compounds, J. Alloys Compd. 265 (1998) 42–48. https://doi.org/10.1016/s0925-8388(97)00419-2.

[140]   S.K.;. R. Dhar S.;. Vijayaraghavan, MAGNETIC BEHAVIOR OF YBNISB, Phys. Rev. B Condens. Matter. 49 (1994) 641–643. https://doi.org/10.1103/physrevb.49.641.

[141]   Bielemeier, B. ; Wortmann, G. Casper, F. Ksenofontov, V. Felser, Claudia, Magnetic properties of GdPdSb and GdNiSb studied by 155Gd-Mössbauer spectroscopy, J. Alloys Compd. 480 (2009) 117–119. https://doi.org/10.1016/j.jallcom.2008.09.206.

[142]   Javorský, P. Havela, L. Sechovský, V. Michor, H. Jurek, Karel, Magnetic behaviour of RCuAl compounds, J. Alloys Compd. 264 (1998) 38–42. https://doi.org/10.1016/s0925-8388(97)00198-9.

[143]   Chevalier, B. Bobet, J.-L. Pasturel, M. Gaudin, E. Etourneau, Jean, Structure and magnetic properties of the ternary gallides CeMGa (M=Mn, Co and Cu) and their hydrides, J. Alloys Compd. 356 (2003) 147–150. https://doi.org/10.1016/s0925-8388(02)01223-9.

[144]   A.E. Dwight, Crystal structure of EuCuGa and related compounds, Journal of the Less Common Metals. 127 (1987) 175–178. https://doi.org/10.1016/0022-5088(87)90375-4.

[145]   Szytuła, A. ; Tyvanchuk, Y.;. Kalychak, Y.M. ; Winiarski, A. Penc, B. ; Zarzycki, Arkadiusz, Electronic structure and magnetic properties of RCuIn (R = La, Ce, Pr, Nd and Lu) compounds, J. Alloys Compd. 442 (2007) 279–281. https://doi.org/10.1016/j.jallcom.2006.07.144.

[146]   A. Szytuła, S. Baran, T. Jaworska-Gołąb, B. Penc, A. Zarzycki, N. Stűsser, A. Arulraj, Y. Tyvanchuk, Magnetic structure of RCuIn (R = Nd, Tb, ho, er), Acta Phys. Pol. A. 113 (2008) 1185–1192. https://doi.org/10.12693/aphyspola.113.1185.

[147]   A. Iandelli, A low temperature crystal modification of the rare earth ternary compounds RCuSi, Journal of the Less Common Metals. 90 (1983) 121–126. https://doi.org/10.1016/0022-5088(83)90123-6.

[148]   Y.V. Knyazev, A.V. Lukoyanov, Y.I. Kuz'min, S. Gupta, K.G. Suresh, Electronic structure and spectral properties of RCuSi (R=Nd,Gd) compounds, Physica B Condens. Matter. 487 (2016) 85–89. https://doi.org/10.1016/j.physb.2016.02.005.

[149]   A. Mugnoli, A. Albinati, A.W. Hewat, A neutron powder diffraction study of the crystal structure of LaCuSi, J. Less-Common Met. 97 (1984) L1–L3. https://doi.org/10.1016/0022-5088(84)90041-9.

[150]   Rieger, W. Parthé, Erwin, Ternäre Erdalkali-und Seltene Erd-Silicide und-Germanide mit AlB2-Struktur, Monatshefte F�r Chemie. 100 (1969) 439–443. https://doi.org/10.1007/bf00904085.

[151]   Baran, S. Szytuła, A. ; Leciejewicz, J. ; Stüsser, N. ; Zygmunt, A. ; Tomkowicz, Z. Guillot, M., Magnetic structures of RCuGe (R = Pr, Nd, Tb, Dy, Ho and Er) compounds from neutron diffraction and magnetic measurements, J. Alloys Compd. 243 (1996) 112–119. https://doi.org/10.1016/s0925-8388(96)02399-7.

[152]   A. Iandelli, The structure of ternary phases of rare earths with RCuGe composition, J. Alloys Compd. 198 (1993) 141–142. https://doi.org/10.1016/0925-8388(93)90157-i.

[153]   Y.V. Knyazev, A.V. Lukoyanov, Y.I. Kuzmin, S. Gupta, K.G. Suresh, Electronic and optical properties of RCuGe compounds (R = Dy, ho), Bull. Russ. Acad. Sci. Phys. 84 (2020) 1152–1155. https://doi.org/10.3103/s1062873820090191.

[154]   A.V. Lukoyanov, L.N. Gramateeva, Y.V. Knyazev, Y.I. Kuz'min, S. Gupta, K.G. Suresh, Effect of electronic correlations on the electronic structure, magnetic and optical properties of the ternary RCuGe compounds with R = Tb, Dy, Ho, Er, Materials (Basel). 13 (2020) 3536. https://doi.org/10.3390/ma13163536.

[155]   S. Gupta, K.G. Suresh, Variations of magnetocaloric effect and magnetoresistance across RCuGe (R=Tb, Dy, Ho, Er) compounds, J. Magn. Magn. Mater. 391 (2015) 151–155. https://doi.org/10.1016/j.jmmm.2015.04.116.

[156]   C.P. Sebastian, C. Fehse, H. Eckert, R.-D. Hoffmann, R. Pöttgen, Structure, 119Sn solid state NMR and Mössbauer spectroscopy of RECuSn (), Solid State Sci. 8 (2006) 1386–1392. https://doi.org/10.1016/j.solidstatesciences.2006.07.006.

[157]   Baran, S. Hofmann, M. Leciejewicz, J. ; Slaski, M. Szytuła, A., Antiferromagnetic ordering in PrCuSn and NdCuSn, J. Phys. Condens. Matter. 10 (1998) 2107–2114. https://doi.org/10.1088/0953-8984/10/9/014.

[158]   Baran, S. Leciejewicz, J. ; Szytuła, A. ; Stüsser, N. ; Tomkowicz, Zbigniew, Incommensurate-commensurate magnetic phase transition in TbCuSn, Solid State Commun. 101 (1997) 631–634. https://doi.org/10.1016/s0038-1098(96)00641-2.

[159]   S.I. Baran V; Leciejewicz, Magnetism of ternary stannides RCuSn (R=Gd-Er), J. Alloys Compd. 257 (1997) 5–13. https://doi.org/10.1016/s0925-8388(96)02731-4.

[160]   R. Pöttgen, The stannides EuCuSn and EuAgSn, J. Alloys Compd. 243 (1996) L1–L4. https://doi.org/10.1016/s0925-8388(96)02354-7.

[161]   Mishra, T. Schellenberg, I. Eul, M. Pöttgen, Rainer, Structure and properties of EuTSb (T= Cu, Pd, Ag, Pt, Au) and YbIrSb, Zeitschrift Für Kristallographie. 226 (2011) 590–601. https://doi.org/10.1524/zkri.2011.1387.

[162]   J. Tong, J. Parry, Q. Tao, G.-H. Cao, Z.-A. Xu, H. Zeng, Magnetic properties of EuCuAs single crystal, J. Alloys Compd. 602 (2014) 26–31. https://doi.org/10.1016/j.jallcom.2014.02.157.

[163]   Kaczorowski, D. Leithe-Jasper, A. ; Rogl, P. Flandorfer, H. Cichorek, T. Pietri, R. ; Andraka, Bohdan, Magnetic, thermodynamic, and electrical transport properties of ternary equiatomic ytterbium compounds Yb TM ( T =transition metal, M = Sn and Bi), Phys. Rev. B: Condens. Matter Mater. Phys. 60 (1999) 422–433. https://doi.org/10.1103/physrevb.60.422.

[164]   P.S. Javorský Vladimír; Havela, Magnetic behaviour of PrCuAl and NdCuAl, J. Magn. Magn. Mater. 177 (1998) 1052–1053. https://doi.org/10.1016/s0304-8853(97)00243-6.

[165]   Mo, Z.-J. Shen, J. Yan, L. Wu, J. Wang, L.-C. Lin, J. Tang, C. Shen, Bao-gen, Low-field induced giant magnetocaloric effect in TmCuAl compound, Appl. Phys. Lett. 102 (2013) 052409-NA.

[166]   Chevalier, B. Bobet, J.-L., On the synthesis and physical properties of the intermetallic CeCuAl, Intermetallics. 9 (2001) 835–838. https://doi.org/10.1016/s0966-9795(01)00070-x.

[167]   Dong, Q.Y. ; Shen, B.G. ; Chen, J. Shen, J. Wang, F. Zhang, H. Sun, J. R., Magnetic properties and magnetocaloric effects in amorphous and crystalline GdCuAl ribbons, Solid State Commun. 149 (2009) 417–420. https://doi.org/10.1016/j.ssc.2008.12.006.

[168]   Pöttgen, R. Johrendt, Dirk, Equiatomic Intermetallic Europium Compounds: Syntheses, Crystal Chemistry, Chemical Bonding, and Physical Properties, Chem. Mater. 12 (2000) 875–897. https://doi.org/10.1021/cm991183v.

[169]   D. Vries, J.W. ; Thiel, R.C. ; Buschow, H.J. K., 155Gd Mössbauer effect and magnetic properties of some ternary gadolinium intermetallic compounds, Journal of the Less Common Metals. 111 (1985) 313–320. https://doi.org/10.1016/0022-5088(85)90203-6.

[170]   H. Oesterreicher, Magnetic studies on compounds RCuSi, R6Cu8Si8, and RCu2Si2 (R = Pr, Gd, Tb), Phys. Status Solidi . 34 (1976) 723–728. https://doi.org/10.1002/pssa.2210340237.

[171]   Gignoux, D. ; Schmitt, D. ; Zerguine, M., Magnetic properties of CeCuSi, Solid State

Commun. 58 (1986) 559–562. https://doi.org/10.1016/0038-1098(86)90796-9.

[172]   B.M. Sondezi-Mhlungu, D.T. Adroja, A.M. Strydom, S. Paschen, E.A. Goremychkin, Crystal electric field excitations in ferromagnetic Ce TX compounds, Physica B Condens. Matter. 404 (2009) 3032–3034. https://doi.org/10.1016/j.physb.2009.07.014.

[173]   Strydom, A.M. ; Sondezi-Mhlungu, B M, Magnetic ordering in hexagonal PrCuSi, J. Phys. Conf. Ser. 200 (2010) 032071-NA. https://doi.org/10.1088/1742-6596/200/3/032071.

[174]   H.S. Nair, C.M.N. Kumar, D.T. Adroja, C. Ritter, A.S. Wills, W.A. Kockelmann, P.P. Deen, A. Bhattacharyya, A.M. Strydom, Magnetic structure and field-dependent magnetic phase diagram of Ni2In-type PrCuSi, J. Phys. Condens. Matter. 30 (2018) 435803. https://doi.org/10.1088/1361-648X/aae28d.

[175]   S. Gupta, K.G. Suresh, A.V. Lukoyanov, Effect of complex magnetic structure on the magnetocaloric and magneto-transport properties in GdCuSi, J. Mater. Sci. 50 (2015) 5723–5728. https://doi.org/10.1007/s10853-015-9116-8.

[176]   K. Łątka, R. Kmieć, A.W. Pacyna, T. Fickenscher, R.-D. Hoffmann, R. Pöttgen, Magnetic ordering in GdAuCd, J. Magn. Magn. Mater. 280 (2004) 90–100. https://doi.org/10.1016/j.jmmm.2004.02.025.

[177]   R.-D. Hoffmann, R. Pöttgen, T. Fickenscher, C. Felser, K. Łątka, R. Kmieć, Ferromagnetic ordering in GdPdCd, Solid State Sci. 4 (2002) 609–617. https://doi.org/10.1016/s1293-2558(02)01304-3.

[178]   Gupta, S. Suresh, K.G. ; Nigam, A.K. ; Mudryk, Y. Paudyal, D. Pecharsky, V.K. ; Gschneidner, K. A., The nature of the first order isostructural transition in GdRhSn, J. Alloys Compd. 613 (2014) 280–287. https://doi.org/10.1016/j.jallcom.2014.06.027.

[179]   S. Gupta, K.G. Suresh, A. Das, A.K. Nigam, A. Hoser, Effects of antiferro-ferromagnetic phase coexistence and spin fluctuations on the magnetic and related properties of NdCuSi, APL Mater. 3 (2015) 066102. https://doi.org/10.1063/1.4922387.

[180]   Oleś, A. ; Duraj, R. ; Kolenda, M. ; Penc, B. ; Szytuła, A., Magnetic properties of DyCuSi and HoCuSi studied by neutron diffraction and magnetic measurements, J. Alloys Compd. 363 (2004) 63–67. https://doi.org/10.1016/s0925-8388(03)00481-x.

[181]   Chen, J. Shen, B.G. ; Dong, Q.Y. ; Sun, Jirong, Giant magnetic entropy change in antiferromagnetic DyCuSi compound, Solid State Commun. 150 (2010) 1429–1431. https://doi.org/10.1016/j.ssc.2010.05.017.

[182]   Chen, J. Shen, B.G. ; Dong, Q. Hu, F. Sun, Jirong, Giant reversible magnetocaloric effect in metamagnetic HoCuSi compound, Appl. Phys. Lett. 96 (2010) 152501-NA. https://doi.org/10.1063/1.3386536.

[183]   P. Schobinger-Papamantellos, K.H.J. Buschow, N.P. Duong, C. Ritter, Magnetic phase diagram of ErCuSi studied by neutron diffraction and magnetic measurements, J. Magn. Magn. Mater. 223 (2001) 203–214. https://doi.org/10.1016/s0304-8853(00)01347-0.

[184]   Schobinger-Papamantellos, P. Ritter, C. Buschow, K.H. ; Duong, N. P., Crossover from antiferromagnetic to ferromagnetic ordering in TmCuSi by neutron diffraction, J. Magn. Magn. Mater. 247 (2002) 207–214. https://doi.org/10.1016/s0304-8853(02)00180-4.

[185]   Gupta, S. Suresh, K.G. ; Nigam, A. K., Magnetocaloric and magnetotransport properties in RRhSn (R = Tb-Tm) series, J. Appl. Phys. 112 (2012) 103909-NA. https://doi.org/10.1063/1.4766900.

[186]   S. Gupta, K.G. Suresh, A.K. Nigam, Magnetic, magnetocaloric and magnetotransport properties of RSn1+xGe1−x compounds (R=Gd, Tb, and Er; x=0.1), J. Magn. Magn. Mater. 342 (2013) 61–68. https://doi.org/10.1016/j.jmmm.2013.04.004.

[187]   S. Gupta, K.G. Suresh, A.V. Lukoyanov, Y.V. Knyazev, Y.I. Kuz'min, Theoretical and

experimental investigations on the magnetic and related properties of RAgSn2 (R=Ho, Er) compounds, J. Mater. Sci. 51 (2016) 6341–6347. https://doi.org/10.1007/s10853-016-9930-7.

[188] D.T. Adroja, B.D. Rainford, A.J. Neville, Crystal fields and spin dynamics of hexagonal CeTSn compounds (T = Cu, Ag and Au), J. Phys. Condens. Matter. 9 (1997) L391–L395. https://doi.org/10.1088/0953-8984/9/27/002.

[189] Müllmann, R. Ernet, U. Mosel, B.D. ; Eckert, H. Kremer, R.K. ; Hoffmann, R.-D. Pöttgen, Rainer, A 119Sn and 151Eu Mössbauerspectroscopic, magnetic susceptibility, and electrical conductivity investigationof the stannides EuTSn (T = Cu, Pd, Ag, Pt), J. Mater. Chem. 11 (2001) 1133–1140. https://doi.org/10.1039/b100055l.

[190] C. Tomuschat, H.-U. Schuster, Magnetische Eigenschaften der Verbindungsreihe EuBX mit B = Element der ersten Neben- und X = Element der f�nften Hauptgruppe, Z. Anorg. Allg. Chem. 518 (1984) 161–167. https://doi.org/10.1002/zaac.19845181116.

[191] Gribanov, A. Tursina, A. Grytsiv, A. Murashova, E. Bukhan'ko, N.G. ; Rogl, P. Seropegin, Y.D. ; Giester, Gerald, Crystal structures of isotypic aluminides CeRuAl and CeRhAl, J. Alloys Compd. 454 (2008) 164–167. https://doi.org/10.1016/j.jallcom.2006.12.087.

[192] Welter, R. Venturini, G. ; Malaman, B. ; Ressouche, E., Crystallographic data and magnetic properties of new RTX compounds (RLa-Sm, Gd; TRu, Os; XSi, Ge). Magnetic structure of NdRuSi, J. Alloys Compd. 202 (1993) 165–172. https://doi.org/10.1016/0925-8388(93)90536-v.

[193] A.V. Morozkin, Y.D. Seropegin, I.A. Sviridov, I.G. Riabinkin, Crystallographic data of new ternary Co2Si-type RTSi (R=Y, Tb–Tm, T=Mn, Ru) compounds, J. Alloys Compd. 282 (1999) L4–L5. https://doi.org/10.1016/s0925-8388(98)00784-1.

[194] Riecken, J.F. ; Hermes, W. Chevalier, B. Hoffmann, R.-D. Schappacher, F.M. ; Pöttgen, Rainer, Trivalent-intermediate valent cerium ordering in CeRuSn - A static intermediate valent cerium compound with a superstructure of the CeCoAl type, Z. Anorg. Allg. Chem. 633 (2007) 1094–1099. https://doi.org/10.1002/zaac.200700066.

[195] A.G. Kuchin, S.P. Platonov, R.D. Mukhachev, A.V. Lukoyanov, A.S. Volegov, V.S. Gaviko, M.Y. Yakovleva, Large magnetic entropy change in GdRuSi optimal for magnetocaloric liquefaction of nitrogen, Metals (Basel). 13 (2023) 290. https://doi.org/10.3390/met13020290.

[196] S.B. Gupta, K.G. Suresh, A.K. Nigam, Magnetic, magnetocaloric and transport properties of HoRuSi compound, (2013). https://doi.org/10.48550/ARXIV.1301.3670.

[197] Gupta, S. Suresh, K. G., Giant low field magnetocaloric effect in soft ferromagnetic ErRuSi, Appl. Phys. Lett. 102 (2013) 022408-NA. https://doi.org/10.1063/1.4775690.

[198] S. Gupta, A. Das, K.G. Suresh, A. Hoser, Y.V. Knyazev, Y.I. Kuz'min, A.V. Lukoyanov, Experimental and theoretical investigations on magnetic and related properties of ErRuSi, Mater. Res. Express. 2 (2015) 046101. https://doi.org/10.1088/2053-1591/2/4/046101.

[199] Penc, B. Hofmann, M. Kolenda, M. Ślaski, M. Szytuła, A, Magnetic properties of RRuGe (R=Gd-Er) compounds, J. Alloys Compd. 267 (1998) L4–L8. https://doi.org/10.1016/s0925-8388(97)00523-9.

[200] Schappacher, F.M. ; Rayaprol, S. Pöttgen, Rainer, Structure and magnetism of GdRuGe, Solid State Commun. 148 (2008) 326–330. https://doi.org/10.1016/j.ssc.2008.08.033.

[201] Penc, B. ; Hofmann, M. Szytuła, A., Magnetic structure of HoRuGe and ErRuGe, J. Alloys Compd. 287 (1999) 48–50. https://doi.org/10.1016/s0925-8388(99)00052-3.

[202] D. Kaczorowski, A. Szytuła, Magnetic and related properties of ternary TmTX

intermetallics, Acta Phys. Pol. A. 127 (2015) 620–622. https://doi.org/10.12693/aphyspola.127.620.

[203]   Schwer, H. ; Hulliger, F., On new rare-earth compounds LnRhAl, J. Alloys Compd. 259 (1997) 249–253. https://doi.org/10.1016/s0925-8388(97)00125-4.

[204]   F. Hulliger, On new rare-earth compounds LnIrGa and LnRhGa, J. Alloys Compd. 239 (1996) 131–134. https://doi.org/10.1016/0925-8388(96)02262-1.

[205]   B. Penc, A. Szytuła, J. Hernández-Velasco, A. Zygmunt, Antiferromagnetic properties of ternary galides RRhGa (R=Tb, Ho and Er), J. Magn. Magn. Mater. 256 (2003) 373–380. https://doi.org/10.1016/s0304-8853(02)00966-6.

[206]   M.Z. Lukachuk Vasyl' I.;. Pöttgen, Synthesis and crystal structures of RERhIn (RE=Sm, Tb, Ho, Er, Tm, Yb, Lu), Intermetallics. 11 (2003) 581–587. https://doi.org/10.1016/s0966-9795(03)00042-6.

[207]   R. Ferro, R. Marazza, G. Rambaldi, On some Ternary Alloys of the Rare Earths having the Fe2P-type structure, Z. Anorg. Allg. Chem. 410 (1974) 219–224. https://doi.org/10.1002/zaac.19744100215.

[208]   D. Rossi, D. Mazzone, R. Marazza, R. Ferro, A contribution to the crystallochemistry of Ternary Rare Earth Intermetallic Phases, Z. Anorg. Allg. Chem. 507 (1983) 235–240. https://doi.org/10.1002/zaac.19835071230.

[209]   V.I. Zaremba, Y.M. Kalychak, V.P. Dubenskiy, R.-D. Hoffmann, R. Pöttgen, Indides LnNiIn2 (Ln=pr, Nd, Sm) and ferromagnetic PrRhIn, J. Solid State Chem. 152 (2000) 560–567. https://doi.org/10.1006/jssc.2000.8731.

[210]   Adroja, D.T. ; Malik, S.K. ; Padalia, B.D. ; Vijayaraghavan, R., CeRhIn: A new mixed-valent cerium compound, Phys. Rev. B Condens. Matter. 39 (1989) 4831–4833. https://doi.org/10.1103/physrevb.39.4831.

[211]   R. Pöttgen, R.-D. Hoffmann, M.H. Möller, G. Kotzyba, B. Künnen, C. Rosenhahn, B.D. Mosel, Syntheses, crystal structures, and properties of EuRhIn, EuIr2, and EuIrSn2, J. Solid State Chem. 145 (1999) 174–181. https://doi.org/10.1006/jssc.1999.8236.

[212]   S.B. Gupta, K.G. Suresh, Study of magnetocaloric effect in GdRhIn compound, in: AIP, 2013. https://doi.org/10.1063/1.4791425.

[213]   Hovestreydt, E. ; Engel, N. ; Klepp, K. ; Chabot, B. ; Parthé, Erwin, Equiatomic ternary rare earth-transition metal silicides, germanides and gallides, Journal of the Less Common Metals. 85 (1982) 247–274. https://doi.org/10.1016/0022-5088(82)90075-3.

[214]   Chevalier, B. Cole, A. Lejay, P. ; Vlasse, M. ; Etourneau, J. Hagenmuller, P. Georges, Roland, Crystal structure and magnetic properties of new rare earth ternary equiatomic silicides RERhSi, Mater. Res. Bull. 17 (1982) 251–258. https://doi.org/10.1016/0025-5408(82)90153-2.

[215]   Chevalier, B. Lejay, P. ; Cole, A. ; Vlasse, M. ; Etourneau, Jean, Crystal structure, superconducting and magnetic properties of new ternary silicides LaRhSi, LaIrSi and NdIrSi, Solid State Commun. 41 (1982) 801–804. https://doi.org/10.1016/0038-1098(82)91252-2.

[216]   A.V. Morozkin, Y.D. Seropegin, I.A. Sviridov, Crystallographic data of the ternary compounds RRhGe (R=Gd-Tm) of the TiNiSi-type structure, J. Alloys Compd. 270 (1998) L4–L6. https://doi.org/10.1016/s0925-8388(98)00249-7.

[217]   Katoh, K. ; Mano, Y. ; Nakano, K. ; Terui, G. ; Niide, Y. ; Ochiai, A., Magnetic properties of YbTGe (T=Rh, Cu, Ag), J. Magn. Magn. Mater. 268 (2004) 212–218. https://doi.org/10.1016/s0304-8853(03)00501-8.

[218]   Y.V.;. N. Kochetkov V.N.;. Klestov, An investigation of the samarium compounds Sm1Ru(Rh)1Ge(Si)1 and Sm1Ru(Rh)2Ge(Si)2, J. Magn. Magn. Mater. 157 (1996) 665–

666. https://doi.org/10.1016/0304-8853(95)01001-7.

[219]   Gupta, S. Suresh, K.G. Nigam, A. K., The magnetic, electronic and optical properties of HoRhGe, J. Phys. D Appl. Phys. 47 (2014) 365002. https://doi.org/10.1088/0022-3727/47/36/365002.

[220]   Y.V. Knyazev, A.V. Lukoyanov, Y.I. Kuz'min, S. Gupta, K.G. Suresh, Optical spectroscopy and electronic structure of TmRhGe compound, Phys. Solid State. 57 (2015) 2357–2360. https://doi.org/10.1134/s1063783415120185.

[221]   Y.V. Knyazev, A.V. Lukoyanov, Y.I. Kuz'min, S. Gupta, K.G. Suresh, Ab initio simulation of the electron structure and optical spectroscopy of ErRhGe compound, Phys. Solid State. 59 (2017) 1275–1278. https://doi.org/10.1134/s1063783417070101.

[222]   Łątka, K. Kmieć, R. Pacyna, A.W. ; Pöttgen, Rainer, Electronic and magnetic properties of ternary stannides RERhSn (RE=light rare-earth metals), J. Magn. Magn. Mater. 320 (2008) L18–L20. https://doi.org/10.1016/j.jmmm.2007.05.031.

[223]   Y.V. Knyazev, A.V. Lukoyanov, Y.I. Kuz'min, S. Gupta, K.G. Suresh, Electronic structure of DyRhSn and HoRhSn compounds: band calculations and optical study, Eur. Phys. J. B. 92 (2019). https://doi.org/10.1140/epjb/e2019-100096-y.

[224]   Y.V. Knyazev, A.V. Lukoyanov, Y.I. Kuz'min, S. Gupta, K.G. Suresh, Electronic and spectral properties of RRhSn (R = Gd, Tb) intermetallic compounds, Phys. Solid State. 60 (2018) 225–229. https://doi.org/10.1134/s1063783418020130.

[225]   Gurgul, J. Łątka, K. Pacyna, A.W. ; Pöttgen, Rainer, Probing the SmRhSn magnetic state by AC/DC magnetic measurements and 119Sn Mössbauer spectroscopy, Intermetallics. 22 (2012) 154–159. https://doi.org/10.1016/j.intermet.2011.11.006.

[226]   Z.Y. Nie, J.W. Shu, A. Wang, H. Su, W.Y. Duan, A.D. Hillier, D.T. Adroja, P.K. Biswas, T. Takabatake, M. Smidman, H.Q. Yuan, Nodeless superconductivity in noncentrosymmetric LaRhSn, ArXiv [Cond-Mat.Supr-Con]. (2023). https://doi.org/10.1103/PhysRevB.105.134523.

[227]   Haase, M.G. ; Schmidt, T. Richter, C.G. ; Block, H. Jeitschko, Wolfgang, Equiatomic Rare Earth ( Ln) Transition Metal Antimonides LnTSb ( T=Rh, lr) and Bismuthides LnTBi ( T=Rh, Ni, Pd, Pt), J. Solid State Chem. 168 (2002) 18–27. https://doi.org/10.1006/jssc.2002.9670.

[228]   Malik, S.K. ; Takeya, H. Gschneidner, K. A., The low temperature behavior of some light lanthanide RRhSb and RPdSb compounds, J. Alloys Compd. 207–208 (1994) 237–240. https://doi.org/10.1016/0925-8388(94)90211-9.

[229]   B. Chevalier, R. Decourt, B. Heying, F.M. Schappacher, U.C. Rodewald, R.-D. Hoffmann, R. Poettgen, R. Eger, A. Simon, Inducing magnetism in the Kondo semiconductor CeRhSb through hydrogenation: Antiferromagnetic behavior of the new hydride CeRhSbH0.2, ChemInform. 38 (2007). https://doi.org/10.1002/chin.200715010.

[230]   Yoshii, S. Tazawa, D. Kasaya, Mitsuo, Kondo effect in CeRhBi and superconductivity in LaRhBi, Physica B Condens. Matter. 230 (1997) 380–382. https://doi.org/10.1016/s0921-4526(96)00728-4.

[231]   Dong, C.L. ; Asokan, K. ; Chen, C. L; Chang, C.L. ; Pong, W.F. ; Kumar, N. Harish; Malik, S. K., X-ray absorption studies of RRhAl (R ¼ La and Ce) compounds, Physica B Condens. Matter. 325 (2003) 235–239. https://doi.org/10.1016/s0921-4526(02)01530-2.

[232]   Kumar, N. Harish; Malik, S. K., Magnetic behaviour of RRhAl (R=La, Ce, Pr, Nd and Gd) compounds, Solid State Commun. 114 (2000) 223–226. https://doi.org/10.1016/s0038-1098(00)00034-x.

[233]   Ślebarski, A. Goraus, J. Hackemer, A. ; Sołyga, Monika, Electronic structure and thermodynamic properties of CeRhAl, Phys. Rev. B: Condens. Matter Mater. Phys. 70

(2004) 195123-NA. https://doi.org/10.1103/physrevb.70.195123.

[234]  Kundaliya, D.C. ; Malik, S. K., Superconductivity in the intermetallic compound YRhAl, Solid State Commun. 131 (2004) 489–491. https://doi.org/10.1016/j.ssc.2004.06.024.

[235]  F. Hulliger, On new rare-earth iridium aluminides LnIrAl, J. Alloys Compd. 229 (1995) 265–267. https://doi.org/10.1016/0925-8388(95)01926-x.

[236]  Baran, S. Hoser, A. Szytuła, A., Antiferromagnetic order in TmRhGa, J. Magn. Magn. Mater. 335 (2013) 97–100. https://doi.org/10.1016/j.jmmm.2013.01.039.

[237]  Goraus, J. Ślebarski, A. Fijałkowski, Marcin, Electronic and thermal properties of non-magnetic CeRhGa, J. Alloys Compd. 509 (2011) 3735–3739. https://doi.org/10.1016/j.jallcom.2010.12.183.

[238]  Higaki, H. Ishii, I. Hirata, D. ; Kim, M.-S. Takabatake, T. Suzuki, Takashi, Elastic, Thermal, Magnetic and Transport Properties of Kondo Compounds CeRhIn and CeRhSn, J. Phys. Soc. Jpn. 75 (2006) 024709–024709. https://doi.org/10.1143/jpsj.75.024709.

[239]  Gondek, Ł. ; Penc, B. ; Stüsser, N. ; Szytuła, A. ; Zygmunt, A., Magnetic structures and phase transitions in TbRhSi, DyRhSi and HoRhSi, Phys. Status Solidi . 196 (2003) 305–308. https://doi.org/10.1002/pssa.200306413.

[240]  Bażela, W. ; Leciejewicz, J. ; SzytuŁa, A., Magnetic structures of RERhSi (RE = Tb, Ho, Er) compounds, J. Magn. Magn. Mater. 50 (1985) 19–26. https://doi.org/10.1016/0304-8853(85)90081-2.

[241]  S. Gupta, A.V. Lukoyanov, Y.V. Knyazev, Y.I. Kuz'min, K.G. Suresh, Field induced metamagnetism and large magnetic entropy change in RRhSi (R = Tb, Dy, Ho) rare earth intermetallics, J. Alloys Compd. 888 (2021) 161493. https://doi.org/10.1016/j.jallcom.2021.161493.

[242]  Ueda, T. Honda, D. Shiromoto, T. Metoki, N. Honda, F. Kaneko, K. Haga, Y. Matsuda, T.D. ; Takeuchi, T. Thamizhavel, A. Sugiyama, K. Kindo, K. ; Settai, R. Onuki, Yoshichika, Magnetic Property and Pressure Effect of a Single Crystal CeRhGe, J. Phys. Soc. Jpn. 74 (2005) 2836–2842. https://doi.org/10.1143/jpsj.74.2836.

[243]  Bażela, W. ; Zygmunt, A. Szytuła, A. ; Ressouche, E. ; Leciejewicz, J. ; Sikora, Wieslawa, Magnetic properties of CeRhGe and NdRhGe compounds, J. Alloys Compd. 243 (1996) 106–111. https://doi.org/10.1016/s0925-8388(96)02393-6.

[244]  S. Gupta, K.G. Suresh, A.K. Nigam, Observation of large positive magnetoresistance and its sign reversal in GdRhGe, J. Alloys Compd. 586 (2014) 600–604. https://doi.org/10.1016/j.jallcom.2013.10.123.

[245]  A.V. Lukoyanov, Y.V. Knyazev, Y.I. Kuz'min, S. Gupta, K.G. Suresh, Electronic structure and optical spectroscopy of the GdRhGe compound, Opt. Spectrosc. 122 (2017) 574–579. https://doi.org/10.1134/s0030400x17040154.

[246]  Bażela, W. ; Hofmann, M. Penc, B. ; Szytuła, A., Neutron diffraction studies of the magnetic structures of the HoRhGe and ErRhGe compounds, J. Phys. Condens. Matter. 10 (1998) 2233–2239. https://doi.org/10.1088/0953-8984/10/10/007.

[247]  Bażela, W. ; Hofmann, M. Baran, S. Penc, B. ; Szytuła, A. ; Zygmunt, A., Magnetic Properties of RRhGe (R = Dy and Tm) Compounds, Acta Phys. Pol. A. 97 (2000) 819–822. https://doi.org/10.12693/aphyspola.97.819.

[248]  S. Gupta, K.G. Suresh, A.K. Nigam, A.V. Lukoyanov, Magnetism in RRhGe (R=Tb, Dy, Er, Tm): An experimental and theoretical study, J. Alloys Compd. 640 (2015) 56–63. https://doi.org/10.1016/j.jallcom.2015.02.126.

[249]  Y.V. Knyazev, A.V. Lukoyanov, Y.I. Kuz'min, S. Gupta, K.G. Suresh, A comparative study of the optical properties of TbRhGe and DyRhGe, Solid State Sci. 44 (2015) 22–

26. https://doi.org/10.1016/j.solidstatesciences.2015.03.020.

[250]  K. Łątka, R. Kmieć, A.W. Pacyna, R. Mishra, R. Pöttgen, Magnetism and hyperfine interactions in Gd2Ni2Mg, Solid State Sci. 3 (2001) 545–558. https://doi.org/10.1016/s1293-2558(01)01172-4.

[251]  S. Gupta, V.R. Reddy, G.S. Okram, K.G. Suresh, A comparative study of HoSn1.1Ge0.9 and DySn1.1Ge0.9 compounds using magnetic, magneto-thermal and magneto-transport measurements, J. Alloys Compd. 625 (2015) 107–112. https://doi.org/10.1016/j.jallcom.2014.11.123.

[252]  Andraka, B. Pietri, R. ; Kaczorowski, D. Leithe-Jasper, A. Rogl, Peter, Magnetism and heavy fermions in YbRhSn and YbPtSn, J. Appl. Phys. 87 (2000) 5149–5151. https://doi.org/10.1063/1.373278.

[253]  Yoshino, T. Takabatake, Toshiro, Crystal Growth and Magnetoresistance of a Superconductor LaRhSb, J. Phys. Soc. Jpn. 68 (1999) 1456–1457. https://doi.org/10.1143/jpsj.68.1456.

[254]  F. Hulliger, On the rare-earth palladium aluminides LnPdAl, J. Alloys Compd. 218 (1995) 44–46. https://doi.org/10.1016/0925-8388(94)01386-1.

[255]  Talik, E. Skutecka, M. Kusz, J. Böhm, H. ; Jarosz, J. ; Mydlarz, T. Winiarski, Antoni, Magnetic properties of GdPdAl single crystals, J. Alloys Compd. 325 (2001) 42–49. https://doi.org/10.1016/s0925-8388(01)01366-4.

[256]  Penc, B. ; Hofmann, M. ; Leciejewicz, J. ; Szytuła, A. ; Zygmunt, A., Magnetic properties of RPdGa (R=Gd–Er) compounds, J. Alloys Compd. 305 (2000) 24–31. https://doi.org/10.1016/s0925-8388(00)00742-8.

[257]  Gondek, Ł. Szytuła, A. ; Kaczorowski, D. Nenkov, K., Electronic structure and magnetism of RPdIn compounds (R=La, Ce, Pr, Nd), Solid State Commun. 142 (2007) 556–560. https://doi.org/10.1016/j.ssc.2007.04.015.

[258]  Gondek, Ł. ; Baran, S. Szytuła, A. ; Kaczorowski, D. Hernandez-Velasco, J., Crystal and magnetic structures of RPdIn (R=Nd, Ho, Er) compounds, J. Magn. Magn. Mater. 285 (2005) 272–278. https://doi.org/10.1016/j.jmmm.2004.08.002.

[259]  M. Bałanda, A. Szytuła, M. Guillot, Magnetic properties of RPdIn (R=Gd–Er) compounds, J. Magn. Magn. Mater. 247 (2002) 345–354. https://doi.org/10.1016/s0304-8853(02)00293-7.

[260]  D.X. Li, T. Yamamura, S. Nimori, K. Koyama, Y. Shiokawa, Magnetic properties of rare-earth compounds GdPdIn and TmPdIn, J. Alloys Compd. 418 (2006) 151–154. https://doi.org/10.1016/j.jallcom.2005.07.081.

[261]  Ito, T. ; Ohkubo, K. ; Hirasawa, T. ; Takeuchi, J. Hiromitsu, I. Kurisu, Makio, Magnetic properties of SmPdIn single crystals, J. Magn. Magn. Mater. 140 (1995) 873–874. https://doi.org/10.1016/0304-8853(94)01273-3.

[262]  Cirafici, S. Palenzona, A. ; Canepa, Fabio, Thermodynamic and physical properties of mixed-valence YbPdIn and of MPdIn intermetallic compounds (M Ca, Sr, Er, Eu), Journal of the Less Common Metals. 107 (1985) 179–187. https://doi.org/10.1016/0022-5088(85)90253-x.

[263]  R. Pöttgen, Metamagnetism in EuPdIn and EuAuIn, J. Mater. Chem. 6 (1996) 63–67. https://doi.org/10.1039/jm9960600063.

[264]  A. Mukhopadhyay, K. Singh, S. Sen, K. Mukherjee, A.K. Nayak, N. Mohapatra, Anomalous magnetoresistance and magneto-thermal properties of the half-Heuslers,RPdSi (R=Y, Gd-Er), J. Phys. Condens. Matter. 33 (2021) 435804. https://doi.org/10.1088/1361-648X/ac1880.

[265]  Y.M. Prots', R. Pöttgen, W. Jeitschko, The crystal structure of YPdSi, the isotypic compounds LnPdSi (Ln = Gd–Lu), and their structural relation to some other equiatomic

compounds of the rare earth and transition metals with main group elements, Z. Anorg. Allg. Chem. 624 (1998) 425–432. https://doi.org/10.1002/(sici)1521-3749(199803)624:3<425::aid-zaac425>3.0.co;2-a.

[266] Szytuła, A. ; Kolenda, M. ; Ressouche, E. ; Zygmunt, A., Magnetic properties of RPdGe (RCe, Pr and Tb) compounds, J. Alloys Compd. 259 (1997) 36–41. https://doi.org/10.1016/s0925-8388(97)00115-1.

[267] Penc, B. ; Hofmann, M. Szytuła, A. ; Zygmunt, A., Magnetic properties of RPdGe (R=Gd,Dy,Ho,Er) compounds, J. Alloys Compd. 282 (1999) 52–57. https://doi.org/10.1016/s0925-8388(98)00841-x.

[268] D. Niepmann, Y.M. Prots', R. Pöttgen, W. Jeitschko, The order of the palladium and germanium atoms in the germanides LnPdGe (Ln=la–Nd, Sm, Gd, Tb) and the new compound Yb3Pd4Ge4, J. Solid State Chem. 154 (2000) 329–337. https://doi.org/10.1006/jssc.2000.8789.

[269] Kotsanidis, P.A. ; Yakinthos, J.K. ; Roudaut, E. ; Gamari-Seale, H., Magnetic properties of RPdGe ternary compounds, J. Magn. Magn. Mater. 131 (1994) 139–147. https://doi.org/10.1016/0304-8853(94)90021-3.

[270] Adroja, D.T. ; Malik, S. K., Magnetic-susceptibility and electrical-resistivity measurements on RPdSn (R=Ce-Yb) compounds, Phys. Rev. B Condens. Matter. 45 (1992) 779–785. https://doi.org/10.1103/physrevb.45.779.

[271] D. Rossi, R. Marazza, R. Ferro, Ternary 1:1:1 alloys of nickel and palladium with tin and rare earths, J. Less-Common Met. 107 (1985) 99–104. https://doi.org/10.1016/0022-5088(85)90245-0.

[272] S.K. Malik, D.T. Adroja, Magnetic behaviour of RPdSb (R = rare earth) compounds, J. Magn. Magn. Mater. 102 (1991) 42–46. https://doi.org/10.1016/0304-8853(91)90262-9.

[273] Rainer, P. Johrendt, D. Kußmann, Dirk, Chapter 207 Structure-property relations of ternary equiatomic YbTX intermetallics, Handbook on the Physics and Chemistry of Rare Earths. 32 (2001) 453‑513. https://doi.org/10.1016/s0168-1273(01)32006-8.

[274] O. Pavlosiuk, D. Kaczorowski, P. Wiśniewski, Magnetic and transport properties of possibly topologically nontrivial half-Heusler bismuthides RMBi (R = Y, Gd, Dy, ho, Lu; M = pd, pt), Acta Phys. Pol. A. 130 (2016) 573–576. https://doi.org/10.12693/aphyspola.130.573.

[275] O. Pavlosiuk, M. Kleinert, P. Wiśniewski, D. Kaczorowski, Antiferromagnetic Order in the Half-Heusler Phase TbPdBi, Acta Phys. Pol. A. 133 (2018) 498–500. https://doi.org/10.12693/aphyspola.133.498.

[276] K. Gofryk, D. Kaczorowski, T. Plackowski, A. Leithe-Jasper, Y. Grin, Magnetic and transport properties of the rare-earth-based Heusler phasesRPdZandRPd2Z(Z=Sb,Bi), Phys. Rev. B Condens. Matter Mater. Phys. 72 (2005). https://doi.org/10.1103/physrevb.72.094409.

[277] Oyamada, A. Maegawa, S. Nishiyama, M. Kitazawa, H. Isikawa, Yosikazu, Ordering mechanism and spin fluctuations in a geometrically frustrated heavy-fermion antiferromagnet on the Kagome-like lattice CePdAl: AAl27NMR study, Phys. Rev. B: Condens. Matter Mater. Phys. 77 (2008) 064432-NA. https://doi.org/10.1103/physrevb.77.064432.

[278] R. Ballou, Geometric frustration in Rare Earth antiferromagnetic compounds, J. Alloys Compd. 275–277 (1998) 510–517. https://doi.org/10.1016/s0925-8388(98)00382-x.

[279] Dönni, A. Ehlers, G. Maletta, H. ; Fischer, P. Kitazawa, H. Zolliker, M., Geometrically frustrated magnetic structures of the heavy-fermion compound CePdAl

studied by powder neutron diffraction, J. Phys. Condens. Matter. 8 (1996) 11213–11229. https://doi.org/10.1088/0953-8984/8/50/043.

[280]   A. Oyamada, T. Kaibuchi, M. Nishiyama, T. Itou, S. Maegawa, Y. Isikawa, A. Dönni, A.H. Kitazawa, Critical behavior in a Kondo-screening partially-ordered antiferromagnet CePdAl, J. Phys. Conf. Ser. 320 (2011) 012067. https://doi.org/10.1088/1742-6596/320/1/012067.

[281]   S. Lucas, K. Grube, C.-L. Huang, A. Sakai, S. Wunderlich, E.L. Green, J. Wosnitza, V. Fritsch, P. Gegenwart, O. Stockert, H. Löhneysen V., Entropy evolution in the magnetic phases of partially frustrated CePdAl, Phys. Rev. Lett. 118 (2017) 107204. https://doi.org/10.1103/PhysRevLett.118.107204.

[282]   Z. Wang, H. Zhao, M. Lyu, J. Xiang, Y. Isikawa, S. Zhang, P. Sun, Frustrated antiferromagnetism and heavy-fermion-like behavior in PrPdAl, Phys. Rev. B. 105 (2022). https://doi.org/10.1103/physrevb.105.125113.

[283]   L. Keller, A. Dönni, H. Kitazawa, J. Tang, F. Fauth, M. Zolliker, Magnetic properties of PrPdAl and NdPdAl, Physica B Condens. Matter. 241–243 (1997) 660–662. https://doi.org/10.1016/s0921-4526(97)00684-4.

[284]   H. Kitazawa, A. Dönni, G. Kido, Magnetic properties of field oriented hexagonal TbPdAl, Physica B Condens. Matter. 281–282 (2000) 165–166. https://doi.org/10.1016/s0921-4526(99)01004-2.

[285]   Shen, J. Xu, Z.Y. ; Zhang, H. Zheng, X.-Q. Wu, J. Hu, F. Sun, J. Shen, Bao-gen, Metamagnetic transition and magnetocaloric effect in antiferromagnetic TbPdAl compound, J. Magn. Magn. Mater. 323 (2011) 2949–2952. https://doi.org/10.1016/j.jmmm.2011.05.042.

[286]   Talik, E. Skutecka, M. Mydlarz, T. Kusz, J. Böhm, H., Role of hybridization in magnetic properties of DyPdAl single crystals, J. Alloys Compd. 391 (2005) 1–7. https://doi.org/10.1016/j.jallcom.2004.08.063.

[287]   Talik, E. Skutecka, M. Kusz, J. Böhm, H. ; Mydlarz, Tadeusz, Electronic structure and magnetic properties of a HoPdAl single crystal, J. Alloys Compd. 359 (2003) 103–108. https://doi.org/10.1016/s0925-8388(03)00217-2.

[288]   Li, D. Shiokawa, Y. Nozue, T. ; Kamimura, T. Sumiyama, Kenji, Metastable characteristics in ferromagnetic TbPdIn and DyPdIn, J. Magn. Magn. Mater. 241 (2002) 17–24. https://doi.org/10.1016/s0304-8853(01)00950-7.

[289]   H. Su, Z.Y. Nie, F. Du, S.S. Luo, A. Wang, Y.J. Zhang, Y. Chen, P.K. Biswas, D.T. Adroja, C. Cao, M. Smidman, H.Q. Yuan, Fully gapped superconductivity with preserved time-reversal symmetry in noncentrosymmetric LaPdIn, Phys. Rev. B. 104 (2021). https://doi.org/10.1103/physrevb.104.024505.

[290]   E. Brück, M. van Sprang, J.C.P. Klaasse, F.R. de Boer, Magnetic order in the Kondo-lattice compound CePdIn, J. Appl. Phys. 63 (1988) 3417–3419. https://doi.org/10.1063/1.340751.

[291]   D.X. Li, S. Nimori, T. Shikama, Magnetic ordering and magnetocaloric effect in PrPdIn and NdPdIn, J. Phys. Conf. Ser. 400 (2012) 032045. https://doi.org/10.1088/1742-6596/400/3/032045.

[292]   Prots, Y.M. ; Jeitschko, W. Gerdes, M.H. ; Künnen, Bernd, Preparation, Crystal Structure, and Physical Properties of the Rare-Earth Metal Palladium Silicides LnPdSi (Ln = La, Ce, Pr) and the a-ThSi2 Type Compounds LaPd0.787(2)Si1.213(2) and CePd0.758(5)Si1.242(5), Z. Anorg. Allg. Chem. 624 (1998) 1855–1862. https://doi.org/10.1002/(sici)1521-3749(1998110)624:11<1855::aid-zaac1855>3.0.co;2-0.

[293]   Tsujii, N. Kitazawa, Hideaki, Ferromagnetic ordering with Heavy Femion behavior in

YbPdSi, Solid State Commun. 159 (2013) 65–69. https://doi.org/10.1016/j.ssc.2013.01.025.

[294] Itoh, Y. Kadomatsu, Hideoki, Electrical and magnetic properties of YbPdGe and YbPtGe, J. Alloys Compd. 280 (1998) 39–43. https://doi.org/10.1016/s0925-8388(98)00744-0.

[295] Zygmunt, A. ; Szytuła, A., Magnetic properties of RPdSn and RPdSb compounds, J. Alloys Compd. 219 (1995) 185–188. https://doi.org/10.1016/0925-8388(94)05036-8.

[296] Kolenda, M. ; Baran, S. Oleś, A.M. ; Stüsser, N. ; Szytuła, A., Magnetic structures of PrPdSn and NdPdSn, J. Alloys Compd. 269 (1998) 25–28. https://doi.org/10.1016/s0925-8388(98)00257-6.

[297] Y.-J. Li, Y. Andoh, M. Kurisu, G. Nakamoto, T. Tsutaoka, S. Kawano, Magnetic properties of DyPdSn, J. Alloys Compd. 692 (2017) 961–965. https://doi.org/10.1016/j.jallcom.2016.09.121.

[298] Baran, S. Leciejewicz, J. ; Stüsser, N. ; Szytuła, A. ; Zygmunt, A. Ivanov, V, Magnetic properties of PrPdSb and NdPdSb compounds, J. Phys. Condens. Matter. 8 (1996) 8397–8405. https://doi.org/10.1088/0953-8984/8/43/030.

[299] A. Mukhopadhyay, N. Lakshminarasimhan, N. Mohapatra, Multi-functional properties of non-centrosymmetric ternary half-Heuslers, RPdSb (R = Er and Ho), J. Phys. D Appl. Phys. 51 (2018) 265004. https://doi.org/10.1088/1361-6463/aac567.

[300] A. Mukhopadhyay, N. Lakshminarasimhan, N. Mohapatra, Magnetic and transport properties of half-Heuslers, RPdSb (R = Gd and Tb), J. Alloys Compd. 721 (2017) 712–720. https://doi.org/10.1016/j.jallcom.2017.06.014.

[301] C.-L. Zhang, S.-Y. Xu, I. Belopolski, Z. Yuan, Z. Lin, B. Tong, G. Bian, N. Alidoust, C.-C. Lee, S.-M. Huang, T.-R. Chang, G. Chang, C.-H. Hsu, H.-T. Jeng, M. Neupane, D.S. Sanchez, H. Zheng, J. Wang, H. Lin, C. Zhang, H.-Z. Lu, S.-Q. Shen, T. Neupert, M. Zahid Hasan, S. Jia, Signatures of the Adler-Bell-Jackiw chiral anomaly in a Weyl fermion semimetal, Nat. Commun. 7 (2016) 10735. https://doi.org/10.1038/ncomms10735.

[302] L. Fu, C.L. Kane, Superconducting proximity effect and majorana fermions at the surface of a topological insulator, Phys. Rev. Lett. 100 (2008) 096407. https://doi.org/10.1103/PhysRevLett.100.096407.

[303] A.R. Akhmerov, J. Nilsson, C.W.J. Beenakker, Electrically detected interferometry of Majorana fermions in a topological insulator, Phys. Rev. Lett. 102 (2009) 216404. https://doi.org/10.1103/PhysRevLett.102.216404.

[304] N.P. Butch, P. Syers, K. Kirshenbaum, A.P. Hope, J. Paglione, Superconductivity in the topological semimetal YPtBi, Phys. Rev. B Condens. Matter Mater. Phys. 84 (2011). https://doi.org/10.1103/physrevb.84.220504.

[305] B. Nowak, O. Pavlosiuk, D. Kaczorowski, Band inversion in topologically nontrivial half-Heusler bismuthides: $^{209}$bi NMR study, J. Phys. Chem. C Nanomater. Interfaces. 119 (2015) 2770–2774. https://doi.org/10.1021/jp5115493.

[306] A.M. Nikitin, Y. Pan, X. Mao, R. Jehee, G.K. Araizi, Y.K. Huang, C. Paulsen, S.C. Wu, B.H. Yan, A. de Visser, Magnetic and superconducting phase diagram of the half-Heusler topological semimetal HoPdBi, J. Phys. Condens. Matter. 27 (2015) 275701. https://doi.org/10.1088/0953-8984/27/27/275701.

[307] Goraus, J. Ślebarski, A. Fijałkowski, Marcin, Experimental and theoretical study of CePdBi, J. Phys. Condens. Matter. 25 (2013) 176002-NA. https://doi.org/10.1088/0953-8984/25/17/176002.

[308] O. Pavlosiuk, K. Filar, P. Wiśniewski, D. Kaczorowski, Magnetic order and SdH effect in half-Heusler phase ErPdBi, Acta Phys. Pol. A. 127 (2015) 656–658.

https://doi.org/10.12693/aphyspola.127.656.

[309]   O. Pavlosiuk, D. Kaczorowski, X. Fabreges, A. Gukasov, P. Wiśniewski, Antiferromagnetism and superconductivity in the half-Heusler semimetal HoPdBi, Sci. Rep. 6 (2016) 18797. https://doi.org/10.1038/srep18797.

[310]   Pan, Y. Nikitin, A.M. ; Bay, T.V. ; Huang, Y. Paulsen, C. Yan, B. de Visser, A., Superconductivity and magnetic order in the noncentrosymmetric half-Heusler compound ErPdBi, EPL. 104 (2013) 27001-NA. https://doi.org/10.1209/0295-5075/104/27001.

[311]   H. Xiao, T. Hu, W. Liu, Y.L. Zhu, P.G. Li, G. Mu, J. Su, K. Li, Z.Q. Mao, Superconductivity in the half-Heusler compound TbPdBi, Phys. Rev. B. 97 (2018). https://doi.org/10.1103/physrevb.97.224511.

[312]   S.M.A. Radmanesh, C. Martin, Y. Zhu, X. Yin, H. Xiao, Z.Q. Mao, L. Spinu, Evidence for unconventional superconductivity in half-Heusler YPdBi and TbPdBi compounds revealed by London penetration depth measurements, Phys. Rev. B. 98 (2018). https://doi.org/10.1103/physrevb.98.241111.

[313]   S.M.A. Radmanesh, S.A.S. Ebrahimi, A. Diaconu, J.Y. Liu, Nontrivial paired states in novel topological superconductors, J. Alloys Compd. 848 (2020) 156498. https://doi.org/10.1016/j.jallcom.2020.156498.

[314]   A. Mukhopadhyay, N. Lakshminarasimhan, N. Mohapatra, Electronic, thermal and magneto-transport properties of the half-Heusler, DyPdBi, Intermetallics (Barking). 110 (2019) 106473. https://doi.org/10.1016/j.intermet.2019.106473.

[315]   V. Bhardwaj, S.P. Pal, L.K. Varga, M. Tomar, V. Gupta, R. Chatterjee, Weak antilocalization and quantum oscillations of surface states in topologically nontrivial DyPdBi(110)half Heusler alloy, Sci. Rep. 8 (2018). https://doi.org/10.1038/s41598-018-28382-1.

[316]   V. Bhardwaj, A. Bhattacharya, L.K. Varga, A.K. Ganguli, R. Chatterjee, Thickness-dependent magneto-transport properties of topologically nontrivial DyPdBi thin films, Nanotechnology. 31 (2020) 384001. https://doi.org/10.1088/1361-6528/ab99f3.

[317]   M.L. Fornasini, A. Iandelli, F. Merlo, M. Pani, Crystal structure of the RCuZn, RAgZn and RAgAl intermetallic compounds (R = rare earth metals), Intermetallics (Barking). 8 (2000) 239–246. https://doi.org/10.1016/s0966-9795(99)00111-9.

[318]   A.E. Dwight, CRYSTAL STRUCTURE OF EQUIATOMIC TERNARY COMPOUNDS: LANTHANIDE-TRANSITION METAL ALUMINIDES, Journal of the Less Common Metals. 102 (1984) L9–L13. https://doi.org/10.1016/0022-5088(84)90401-6.

[319]   Rossi, D. ; Ferro, Riccardo, Ternary intermetallic RAgGa, RAuGa alloys (R=light rare earth and Yb), J. Alloys Compd. 317 (2001) 521–524. https://doi.org/10.1016/s0925-8388(00)01380-3.

[320]   A. Iandelli, The structure of ternary compounds of the rare earths: RAgSi, Journal of the Less Common Metals. 113 (1985) 25–27. https://doi.org/10.1016/0022-5088(85)90295-4.

[321]   Baran, S. Hofmann, M. Leciejewicz, J. ; Penc, B. ; Ślaski, M. ; Szytuła, A. ; Zygmunt, A., Magnetic properties and magnetic structures of RAgSi (R"Gd}Er) compounds, J. Magn. Magn. Mater. 222 (2000) 277–284. https://doi.org/10.1016/s0304-8853(00)00565-5.

[322]   V.K. Pecharsky, K.A. Gschneidner Jr, O.I. Bodak, A.S. Protsyk, The crystal structure, heat capacity (1.5–80 K) and magnetic susceptibility (1.6–300 K) of LaAgGe and CeAgGe, J. Less-Common Met. 168 (1991) 257–267. https://doi.org/10.1016/0022-5088(91)90307-p.

[323]   Morosan, E. Bud'ko, S.L. ; Canfield, P.C. ; Torikachvili, M.S. ; Lacerda, A. H., Thermodynamic and transport properties of RAgGe (R = Tb-Lu) single crystals, J. Magn. Magn. Mater. 277 (2004) 298–321. https://doi.org/10.1016/j.jmmm.2003.11.014.

[324]   R. Pöttgen, Preparation and Crystal Structure of EuAgGe, Zeitschrift Für Naturforschung B. 50 (1995) 1071–1074. https://doi.org/10.1515/znb-1995-0715.

[325]   Mazzone, D. Rossi, D. ; Marazza, R. Ferro, Riccardo, A contribution to the crystal chemistry of ternary 1:1:1 alloys: RAgSn and RCuTl compounds (R ≡ rare earth), Journal of the Less Common Metals. 80 (1981) P47–P52. https://doi.org/10.1016/0022-5088(81)90100-4.

[326]   D.;. I. Fus Vladimir; Jezierski, ELECTRONIC AND TRANSPORT PROPERTIES OF RAgSn (R=Ce, Pr, Nd, Dy) COMPOUNDS, Acta Phys. Pol. A. 98 (2000) 571–586. https://doi.org/10.12693/aphyspola.98.571.

[327]   S. Baran, J. Leciejewicz, N. Stüsser, A. Szytula, A. Zygmunt, Y. Ding, Neutron diffraction study of magnetic ordering in RAgSn (R = Ce, Pr, Nd, Gd, Tb, Dy, Ho, Er) compounds, J. Magn. Magn. Mater. 170 (1997) 143–154. https://doi.org/10.1016/s0304-8853(97)00017-6.

[328]   Suresh, K.G. ; Dhar, S.K. ; Nigam, A. K., Inhomogeneous magnetic state in RAgAl (R = rare earth), J. Magn. Magn. Mater. 288 (2005) 452–459. https://doi.org/10.1016/j.jmmm.2004.09.136.

[329]   Ślebarski, A. Kaczorowski, D. Głogowski, W. ; Goraus, Jerzy, The low-temperature magnetic and thermal properties and electronic structure of CeAgAl; experiment and calculations, J. Phys. Condens. Matter. 20 (2008) 315208-NA. https://doi.org/10.1088/0953-8984/20/31/315208.

[330]   J. Heimann, M. Kulpa, On the new ternary gadolinium silver aluminide GdAgAl, J. Alloys Compd. 296 (2000) L8–L10. https://doi.org/10.1016/s0925-8388(99)00530-7.

[331]   J. Heimann, D. Dunikowski, Magnetic state in RAgAl (R=Dy, Ho, Er), J. Alloys Compd. 423 (2006) 43–46. https://doi.org/10.1016/j.jallcom.2005.12.055.

[332]   Y. Zhang, L. Hou, Z. Ren, X. Li, G. Wilde, Magnetic properties and magnetocaloric effect in TmZnAl and TmAgAl compounds, J. Alloys Compd. 656 (2016) 635–639. https://doi.org/10.1016/j.jallcom.2015.10.026.

[333]   Goraus, J. Ślebarski, A. Fijałkowski, M. Hawełek, Ł., Inhomogenous magnetic ground state in CeAgGa, Eur. Phys. J. B. 80 (2011) 65–71. https://doi.org/10.1140/epjb/e2011-10726-9.

[334]   Adroja, D.T. ; Rainford, B.D. ; Malik, S. K., CeAgGa: a ferromagnetic Ce-based compound, Physica B Condens. Matter. 186 (1993) 566–568. https://doi.org/10.1016/0921-4526(93)90636-k.

[335]   Sill, L.R. ; Esau, E. D., Magnetic characteristics of RAgGa compounds, J. Appl. Phys. 55 (1984) 1844–1846. https://doi.org/10.1063/1.333496.

[336]   Guillot, M. ; Szytuła, A. ; Tomkowicz, Z. Zach, R., Magnetic properties of RAgSn (RNd, Tb, Ho), TbPdSn and TbAgGa compounds in high magnetic fields☆, J. Alloys Compd. 226 (1995) 131–135. https://doi.org/10.1016/0925-8388(95)01627-9.

[337]   D. Silva, L.M.;. dos Santos, A.O. ; Coelho, A.A. Cardoso, L. Pavie, Magnetic properties and magnetocaloric effect of the HoAgGa compound, Appl. Phys. Lett. 103 (2013) 162413-NA. https://doi.org/10.1063/1.4826440.

[338]   Zygmunt, A. ; Szytuła, A. ; Kolenda, M. ; Tomkowicz, Z. Stüsser, N. ; Leciejewicz, J., Magnetic properties of RAgGa (R Tb, Dy, Ho) compounds, J. Magn. Magn. Mater. 161 (1996) 127–132. https://doi.org/10.1016/s0304-8853(96)00035-2.

[339]   J.K.P. França, D.C. dos Reis, H. Fabrelli, E.M. Bittar, A.O. dos Santos, L.M. da Silva, Evidence of re-entrant spin-glass behavior in the GdAgGa compound, Appl. Phys. A

Mater. Sci. Process. 129 (2023). https://doi.org/10.1007/s00339-023-06719-6.

[340]  Baran, S. Kaczorowski, D. Hoser, A. Penc, B. ; SzytuŁa, A., Magnetic behavior in TmAgSi, J. Magn. Magn. Mater. 323 (2011) 222–225. https://doi.org/10.1016/j.jmmm.2010.09.005.

[341]  Öner, Y. ; Ross, J.H. ; Sologub, O. ; Salamakha, P., Magnetic phase transitions in intermetallic NdAgSi compound, J. Alloys Compd. 415 (2006) 38–42. https://doi.org/10.1016/j.jallcom.2005.07.049.

[342]  S. Baran, M. Hofmann, J. Leciejewicz, B. Penc, M. Ślaski, A. Szytuła, Magnetic order in RAgGe (R=Gd–Er) intermetallic compounds, J. Alloys Compd. 281 (1998) 92–98. https://doi.org/10.1016/s0925-8388(98)00721-x.

[343]  P.A. Goddard, J. Singleton, A.L. Lima, E. Morosan, S.J. Blundell, S.L. Bud'ko, P.C. Canfield, Magnetic-field-orientation dependence of the metamagnetic transitions in TmAgGe up to 55 T, J. Phys. Conf. Ser. 51 (2006) 219–226. https://doi.org/10.1088/1742-6596/51/1/050.

[344]  E. Morosan, S.L. Bud'ko, P.C. Canfield, Angular-dependent planar metamagnetism in the hexagonal compounds TbPtIn and TmAgGe, Phys. Rev. B Condens. Matter Mater. Phys. 71 (2005). https://doi.org/10.1103/physrevb.71.014445.

[345]  N. Li, Q. Huang, X.Y. Yue, S.K. Guang, K. Xia, Y.Y. Wang, Q.J. Li, X. Zhao, H.D. Zhou, X.F. Sun, Low-temperature transport properties of the intermetallic compound HoAgGe with a kagome spin-ice state, Phys. Rev. B. 106 (2022). https://doi.org/10.1103/physrevb.106.014416.

[346]  K. Zhao, H. Deng, H. Chen, K.A. Ross, V. Petříček, G. Günther, M. Russina, V. Hutanu, P. Gegenwart, Realization of the kagome spin ice state in a frustrated intermetallic compound, Science. 367 (2020) 1218–1223. https://doi.org/10.1126/science.aaw1666.

[347]  Bażela, W. ; Guillot, M. ; Leciejewicz, J. ; Małetka, K. ; Szytuła, A. ; Tomkowicz, Z., Magnetic properties of RAgSn (R Nd, Tb, Ho) compounds, J. Magn. Magn. Mater. 140 (1995) 1137–1138. https://doi.org/10.1016/0304-8853(94)01492-2.

[348]  D.H. Ryan, J.M. Cadogan, V.I. Krylov, A. Legros, R. Rejali, C.D. Boyer, The magnetic structures of GdCuSn, GdAgSn and GdAuSn, J. Phys. Condens. Matter. 29 (2017) 495804. https://doi.org/10.1088/1361-648X/aa977a.

[349]  S. Baran, A. Szytuła, D. Kaczorowski, F. Damay, Magnetic structures in TmPdIn and TmAgSn, J. Alloys Compd. 662 (2016) 11–15. https://doi.org/10.1016/j.jallcom.2015.11.200.

[350]  D. Kaczorowski, A. Szytuła, Long-range magnetic ordering in TmAgSn, J. Alloys Compd. 615 (2014) 1–3. https://doi.org/10.1016/j.jallcom.2014.06.153.

[351]  C. Tomuschat, H.-U. Schuster, ABX-Verbindungen mit modifizierter $Ni_2In$-Struktur / ABX-Compounds with a Modified $Ni_2In$ Structure, Z. Naturforsch. B J. Chem. Sci. 36 (1981) 1193–1194. https://doi.org/10.1515/znb-1981-0929.

END